\documentclass[a4paper,fleqn,usenatbib]{mnras}

\usepackage{upgreek}
\usepackage[T1]{fontenc}

\usepackage{graphicx}	
\usepackage{amsmath}	
\usepackage{amssymb}	

\newcommand{\MS}{\ifmmode{\,}\else\thinspace\fi{\rm M}\ifmmode_{\odot}\else$_{\odot}$\fi}
\newcommand{\LS}{\ifmmode{\,}\else\thinspace\fi{\rm L}\ifmmode_{\odot}\else$_{\odot}$\fi}
\newcommand{\RS}{\ifmmode{\,}\else\thinspace\fi{\rm R}\ifmmode_{\odot}\else$_{\odot}$\fi}
\newcommand{\sn}{\ifmmode {\rm S/N}\else${\rm S/N}$\fi}
\newcommand{\fx}{\ifmmode f_{\rm x}\else$f_{\rm x}$\fi}
\newcommand{\px}{\ifmmode P_{\rm x}\else$P_{\rm x}$\fi}
\newcommand{\Pm}{\ifmmode P_{\rm m}\else$P_{\rm m}$\fi}
\newcommand{\Pmj}{\ifmmode P_{\rm m1}\else$P_{\rm m1}$\fi}
\newcommand{\Pmd}{\ifmmode P_{\rm m2}\else$P_{\rm m2}$\fi}
\newcommand{\PF}{\ifmmode P_{\rm 0}\else$P_{\rm 0}$\fi}
\newcommand{\PO}{\ifmmode P_{\rm 1}\else$P_{\rm 1}$\fi}
\renewcommand{\fm}{\ifmmode f_{\rm m}\else$f_{\rm m}$\fi}
\newcommand{\fmj}{\ifmmode f_{\rm m1}\else$f_{\rm m1}$\fi}
\newcommand{\fmd}{\ifmmode f_{\rm m2}\else$f_{\rm m2}$\fi}
\newcommand{\fF}{\ifmmode f_{\rm 0}\else$f_{\rm 0}$\fi}

\newcommand{\fjd}{\ifmmode  f_{\rm 0}/2\else $f_{\rm 0}/2$\fi}
\newcommand{\ftd}{\ifmmode 3f_{\rm 0}/2\else$3f_{\rm 0}/2$\fi}
\newcommand{\fpd}{\ifmmode 5f_{\rm 0}/2\else$5f_{\rm 0}/2$\fi}
\newcommand{\fsd}{\ifmmode 7f_{\rm 0}/2\else$7f_{\rm 0}/2$\fi}

\newcommand{\fjc}{\ifmmode  f_{\rm 0}/4\else $f_{\rm 0}/4$\fi}
\newcommand{\ftc}{\ifmmode 3f_{\rm 0}/4\else$3f_{\rm 0}/4$\fi}
\newcommand{\fpc}{\ifmmode 5f_{\rm 0}/4\else$5f_{\rm 0}/4$\fi}
\newcommand{\fdc}{\ifmmode 9f_{\rm 0}/4\else$9f_{\rm 0}/4$\fi}
\newcommand{\ftrzc}{\ifmmode 13f_{\rm 0}/4\else$13f_{\rm 0}/4$\fi}

\newcommand{\fO}{\ifmmode f_{\rm 1}\else$f_{\rm 1}$\fi}
\newcommand{\cd}{\ifmmode {\rm c\,d}^{-1}\else c\,d$^{-1}$\fi}
\newif\iftech
\newif\ifoffi
\techtrue
\offifalse
\offitrue
\techfalse

\newcommand\tech[1]{\iftech #1\fi}
\newcommand\offi[1]{\ifoffi #1\fi}

\title[Dynamical phenomena in type II Cepheids]{Diversity of dynamical phenomena in type II Cepheids of the OGLE collection}

\author[Smolec et al.]
{
 R. Smolec,$^{1}$\thanks{E-mail: smolec@camk.edu.pl}
 P. Moskalik,$^{1}$
 E. Plachy,$^{2,\,3}$
 I. Soszy\'nski,$^{4}$
 A. Udalski$^{4}$
\\
$^{1}$Nicolaus Copernicus Astronomical Center of the Polish Academy of Sciences, ul. Bartycka 18, PL-00-716 Warszawa, Poland\\
$^{2}$Konkoly Observatory, MTA CSFK, Konkoly Thege Mikl\'os \'ut 15-17, H-1121 Budapest, Hungary\\
$^{3}$MTA CSFK Lend\"ulet Near-Field Cosmology Research Group\\
$^{4}$Warsaw University Observatory, Al. Ujazdowskie 4, PL-00-478 Warszawa, Poland\\
}

\date{Accepted XXX. Received YYY; in original form ZZZ}

\pubyear{2018}

\begin{document}
\label{firstpage}
\pagerange{\pageref{firstpage}--\pageref{lastpage}}
\maketitle

\begin{abstract}
We have analysed the photometry of 924 type II Cepheids from the Galactic bulge observed by the Optical Gravitational Lensing Experiment. Our goal was to detect and analyse various dynamical phenomena, including, e.g. multi-mode pulsation, period-doubled pulsation, period-$k$ pulsation, or quasi-periodic modulation of pulsation. First examples of double-mode BL~Her-type stars, pulsating simultaneously in the fundamental and first overtone modes are reported. Two double-mode pulsators are not isolated cases however, but form an extension of recently identified class of long-period RRd stars. Period doubling effect was detected in three BL~Her-type stars: two were identified previously and one is a new discovery. Through identification of numerous period-doubled W~Vir stars we show that the appearance of the period doubling effect, and hence the transition towards RV~Tau class, is a smooth process that starts at pulsation periods slightly above 15\,d. Interchange of the deep and shallow pulsation cycles is identified in tens of RV~Tau variables. One RV~Tau star is a strong candidate for period-4 pulsation. Quasi-periodic modulation of pulsation in all sub-classes of type II Cepheids is reported for the first time. Modulation was detected in 16 BL~Her, 9 W~Vir and in 7 RV~Tau stars. In the last case, modulation is the most likely scenario, but two other scenarios, excitation of non-radial modes, or period-4 pulsation, can also account for the observed pulsation pattern. Irregular changes of pulsation amplitude and period on various time-scales are common in all subgroups of type II Cepheids.
\end{abstract}




\section{Introduction}\label{sect:intro}

Type II Cepheids are low-mass, population II stars, pulsating in the radial fundamental mode. They are divided into three groups based on their pulsation period: BL~Her stars, with pulsation periods between $1$ and $5$\,d, W~Vir stars, with pulsation periods between $5$ and $20$\,d, and RV~Tau stars with pulsation periods above $20$\,d. The distinctive feature of the latter group, is a period doubling (PD) phenomenon. The light curve shows alternating deep and shallow brightness minima/maxima. The borderlines between the three groups are conventional and a bit arbitrary; different values can be found in the literature \citep[for a summary, see][]{csbook}. In the Magellanic Clouds, \cite{ogle3_lmc_t2,ogle3_smc_t2} put the borderline between the BL~Her and W~Vir classes at a shorter period of $4$\,d. The three classes are believed to correspond to different phases of the post-horizontal branch evolution \citep{gingold,csbook}, we note however, that evolutionary status of type II Cepheids and their sub-groups is still a poorly studied subject. Within the W~Vir class, a subgroup of peculiar W~Vir (pW~Vir) stars, which are brighter and bluer, and show a characteristic asymmetry in the light curve, was identified by \cite{ogle3_lmc_t2}. RV~Tau stars are also divided into two sub-classes: RVa stars, with roughly constant mean brightness, and RVb stars, which exhibit additional long-period variation of the mean brightness.

Type II Cepheids obey a period-luminosity relation. Since they are fainter than classical Cepheids, and their $P-L$ relation has a much higher dispersion, their use as distance indicators is not that common. On the other hand, type II Cepheids are extremely interesting in their own right, as complex dynamical systems. Pulsation of these low-mass giants is strongly non-adiabatic and non-linear. Their pulsation is rarely as regular as clockwork. The longer the pulsation period the more irregular their pulsation is. In addition, at periods above $20$\,d, the period doubling phenomenon becomes common.

The complex dynamical behaviour of type II Cepheids was a subject of several theoretical investigations. Large-amplitude radial pulsation can be modelled with non-linear pulsation codes. The early calculations were purely radiative. \cite{kb88} conducted a large survey of radiative, non-linear models of type II Cepheids. In a typical model sequence, three types of dynamical behaviour were detected as pulsation period was increasing. The shortest period models displayed a regular limit-cycle pulsation. For longer periods, period doubled pulsation was observed; first with two, then with four, eight etc., brightness minima/maxima per repetition period (hereafter period-2, period-4, period-8, etc., pulsation). Such PD cascade ultimately led to a third type of dynamical behaviour, the deterministic chaos. In the more luminous models, the transition to chaos was occurring through a tangent bifurcation. Using the amplitude equation formalism, \cite{mb90} showed, that period doubling may be triggered by the half-integer resonances between the radial pulsation modes. 

It is extremely challenging to detect chaotic dynamics in the variability of long-period pulsators, as long, uninterrupted and precise observations are necessary. Still, the evidence for chaotic dynamics was reported in three type II Cepheids of RV~Tau type \citep[R~Sct, AC~Her and DF~Cyg;][respectively]{bks96,kbsm98,plachyRRL17_chaos} as well as in several more luminous semi-regular variables \citep{bkc04}, and in Mira-type variable \citep{ks03}.

Period doubling on the other hand, is relatively easy to detect and is a well known, distinct feature of RV~Tau variables. However, in the radiative models of \cite{kb88}, the PD behaviour appears at a much shorter pulsation periods than observed (above $\sim\!7-10$\,d). PD was found in even shorter period, BL~Her-type models by \cite{bm92}. They found a robust PD behaviour in the period range between 2.0 and 2.6\,d, caused by the 3:2 resonance between the radial fundamental and radial first overtone modes. They concluded that PD effect should also be present in real stars. It took 20 years however, till their prediction was confirmed, when the first period-doubled BL~Her star was identified in the OGLE data \citep{blher_ogle,blherPD}.

The discovery of period doubled BL~Her star shows a great predictive power of non-linear pulsation models. The period doubled pulsation and chaotic dynamics are not the only predictions of non-linear models. The recent convective models of type II Cepheids show, that we may expect quasi-periodic modulation of pulsation \citep{blher_mod,s16}, or the full wealth of dynamical behaviours characteristic for deterministic chaos, including, e.g. type-I and type-III intermittency, or period-$(2k\!+\!1)$ pulsation \citep{blher_chaos}.

The aim of this paper is to search for various dynamical phenomena in the pulsation of type II Cepheids using the OGLE Galactic bulge data \citep{ogle4_gb_t2}. The OGLE collection is the largest in existence, and consists of the longest series of homogeneous photometric data for type II Cepheids. The OGLE data and adopted methods of data analysis are presented in Sect.~\ref{sec:data}, results are presented in Sect.~\ref{sec:results} and discussed in Sect.~\ref{sec:discussion}. Summary concludes the paper.

\section{Data analysis}\label{sec:data}

The OGLE collection of variable stars consists of 924 type II Cepheids \citep{ogle4_gb_t2}, of which 350 were included in the OGLE-III Catalog of Variable Stars \citep{blher_ogle}. During OGLE-IV, which has started in March, 2010 \citep{o4}, brightness of 873 type II Cepheids is monitored. Of these, 372 are of BL~Her type, 366 are of W~Vir type, and 135 are of RV~Tau type. The data sampling is typically much better than during OGLE-III. Hence, OGLE-IV $I$-band data are our primary data source. $V$-band data are also available for most of the stars but they are scarce. The data that we analyse cover seven observing seasons (more than $2400$\,d), hereafter abbreviated as s1,\ldots, s7. The corresponding, approximate Julian Date ranges are given in Tab.~\ref{tab:seasons}. The number of the data points strongly vary from field to field, from about $100$ points in the least observed fields, up to more than $13\,000$ points in the most observed fields.

\begin{table}
\centering
\caption{Approximate dates and Julian Date ranges for the first seven observing seasons of OGLE-IV in the Galactic bulge.}
\label{tab:seasons}
\begin{tabular}{lll}
season & start -- end date        & approximate JD range     \\
       & (dd-mm-yyyy)             & (${\rm JD}-2\,450\,000$) \\   
\hline
s1     & 05-03-2010 -- 07-11-2010 &  5261 -- 5508 \\
s2     & 03-02-2011 -- 22-11-2011 &  5596 -- 5888 \\
s3     & 30-01-2012 -- 11-11-2012 &  5957 -- 6243 \\
s4     & 03-02-2013 -- 04-11-2013 &  6327 -- 6601 \\
s5     & 29-01-2014 -- 01-11-2014 &  6687 -- 6963 \\
s6     & 06-02-2015 -- 08-11-2015 &  7088 -- 7335 \\
s7     & 25-01-2016 -- 30-10-2016 &  7413 -- 7692 \\
\hline
\end{tabular}
\end{table}

For 299 stars of the OGLE-IV sample, data from the earlier phase of the OGLE project are available \citep{blher_ogle}. We merge the data only exceptionally, however. This is because data are often of different quality, with typically much worse sampling during OGLE-III. In addition, the variability of type II Cepheids is often irregular, resulting in strong, broad, non-stationary signals in the frequency spectrum, that increase the noise level, and complicate the analysis. In many cases, increasing the time base only worsens the problem. We also note that there might be systematic differences between the OGLE-III and OGLE-IV photometry: the zero points and the pulsation amplitudes may slightly differ. The latter effect (amplitude change) is most likely due to the filter change and indeed in some stars we observe tiny amplitude differences between OGLE-III and OGLE-IV data (see, e.g. Fig.~\ref{fig:tdfd075}).

The initial analysis follows a standard consecutive prewhitening technique. Periodicities present in the data are identified with the help of the discrete Fourier transform (DFT). Next, the data are fitted with a Fourier series of the following form:
\begin{equation}
m(t)=m_0+\sum_{i}A_i\sin(2\uppi f_it+\phi_i)\,,\label{eq}
\end{equation}
in which $f_i$ are independent frequencies detected in the data and $A_i$ and $\phi_i$ are respective amplitudes and phases. All quantities are adjusted using non-linear least square procedure. In all cases, the solution includes frequency of the fundamental mode, $\fF$, and its harmonics, $k\fF$. Residuals from the fit are searched for additional periodicities with the DFT. In the equation above we include only resolved frequencies. The formal frequency resolution of the transform is $1/\Delta T$ ($\Delta T$ -- data length), but we adopt a more conservative criterion and regard two frequencies as resolved, if their separation is larger than $2/\Delta T$.

\begin{figure}
\includegraphics[width=\columnwidth]{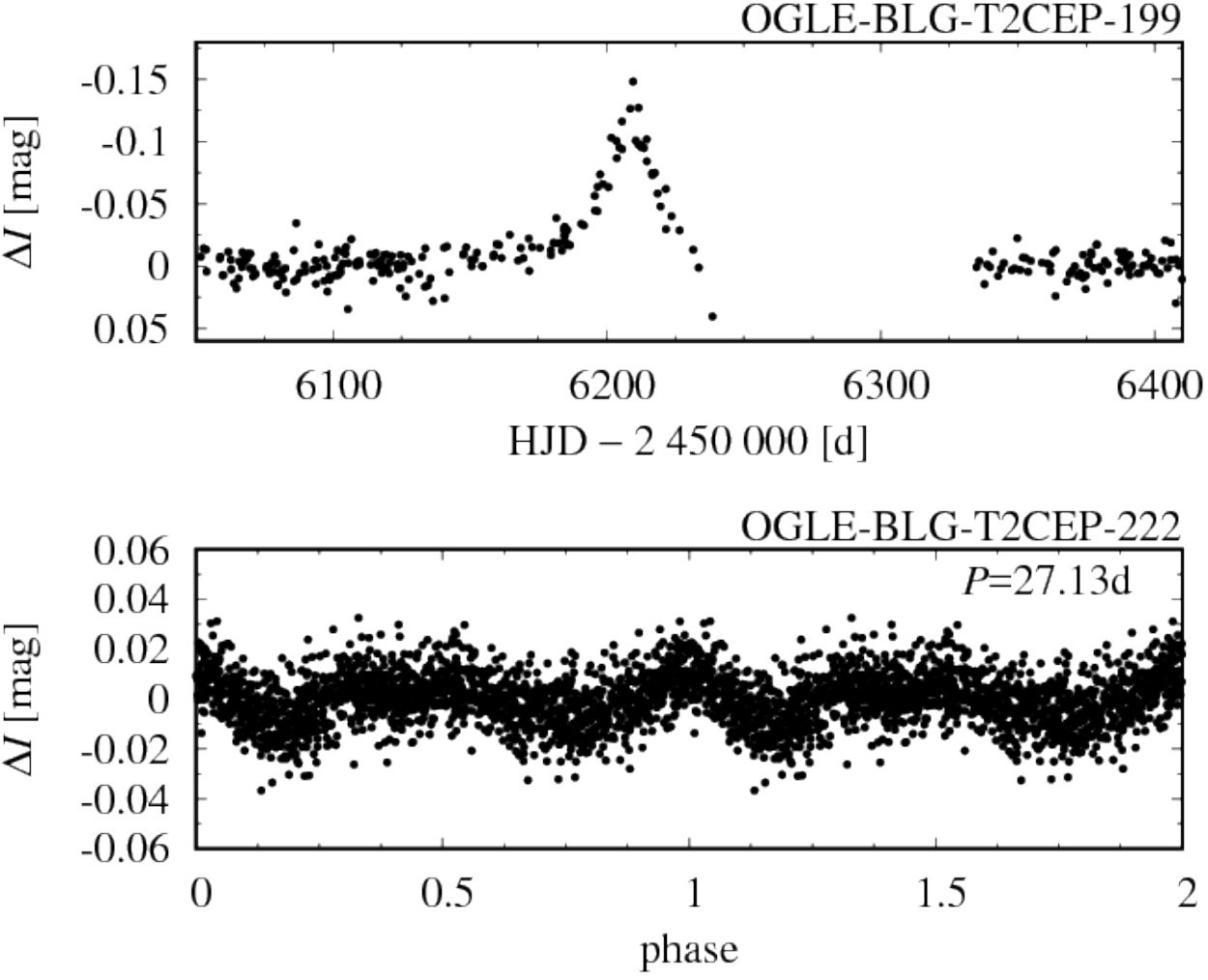}
\caption{Two examples of additional variability detected in the residual data: microlensing event in the residuals of BL~Her-type variable T2CEP-199 (top panel) and ellipsoidal variability in T2CEP-222 (bottom panel).}
\label{fig:resids}
\end{figure}

During the procedure, we remove the obvious outliers and model the slow trends with polynomial function. A care and constant inspection of the data is needed however, as quite often the data deviate from the fit due to, e.g. strong phase changes, while trends may in fact be connected to a star itself or its direct neighbourhood. This is the case for RVb variables, in which periodic large amplitude variation of the mean brightness is most likely due to obscuration by a surrounding disc \citep{kb17}. Other two examples of interesting variability detected in the residuals are illustrated in Fig.~\ref{fig:resids}. In the top panel we show a section of residuals for OGLE-BLG-T2CEP-199\footnote{From now on, we drop the first part of the star's ID.} (BL Her; $\PF=1.021186$\,d) with a clear symmetric brightening that is most likely due to microlensing event \citep[see][for similar examples]{lens_ex1,lens_ex2,wyrzyk}. Admittedly, due to large crowding within 1\,arcsec, we cannot rule out that, in fact, other nearby source experienced a gravitational microlensing episode. In the residuals of T2CEP-222 (BL Her; $\PF=1.147797$\,d) variability with a period of $P_{\rm orb}=27.13$\,d is present. Data folded with this period are plotted in the bottom panel of Fig.~\ref{fig:resids}. The light curve resembles ellipsoidal variability in systems with eccentric orbits \citep[see, e.g.][]{elips}. In the frequency spectrum we detect two combination frequencies, $\fF-f_{\rm orb}$ and $\fF+2f_{\rm orb}$, which indicate that additional variability is intrinsic to T2CEP-222 and not due to contamination.

Since we analyse ground-based and seasonal data, both daily and 1-yr aliases are prominent in the window function and sometimes may become a source of confusion during the analysis. A few such cases are explicitly discussed in the text.

The very common complication in the analysis of type II Cepheids is non-stationary nature of the dominant variability. The amplitude and phase of the fundamental mode slowly vary in an irregular way. As a result, after the prewhitening with the fundamental mode and its harmonics, strong residual power remains in the frequency spectrum. The shorter the time-scale of irregular variations, the broader the power excess observed at $\fF$. The highest peak within the power excess is typically placed at a location unresolved with $\fF$. Similar structures are observed at the harmonic frequencies. Such signals strongly increase the noise level in the Fourier transform and may easily hide other low-amplitude signals present in the data.

\begin{figure}
\includegraphics[width=\columnwidth]{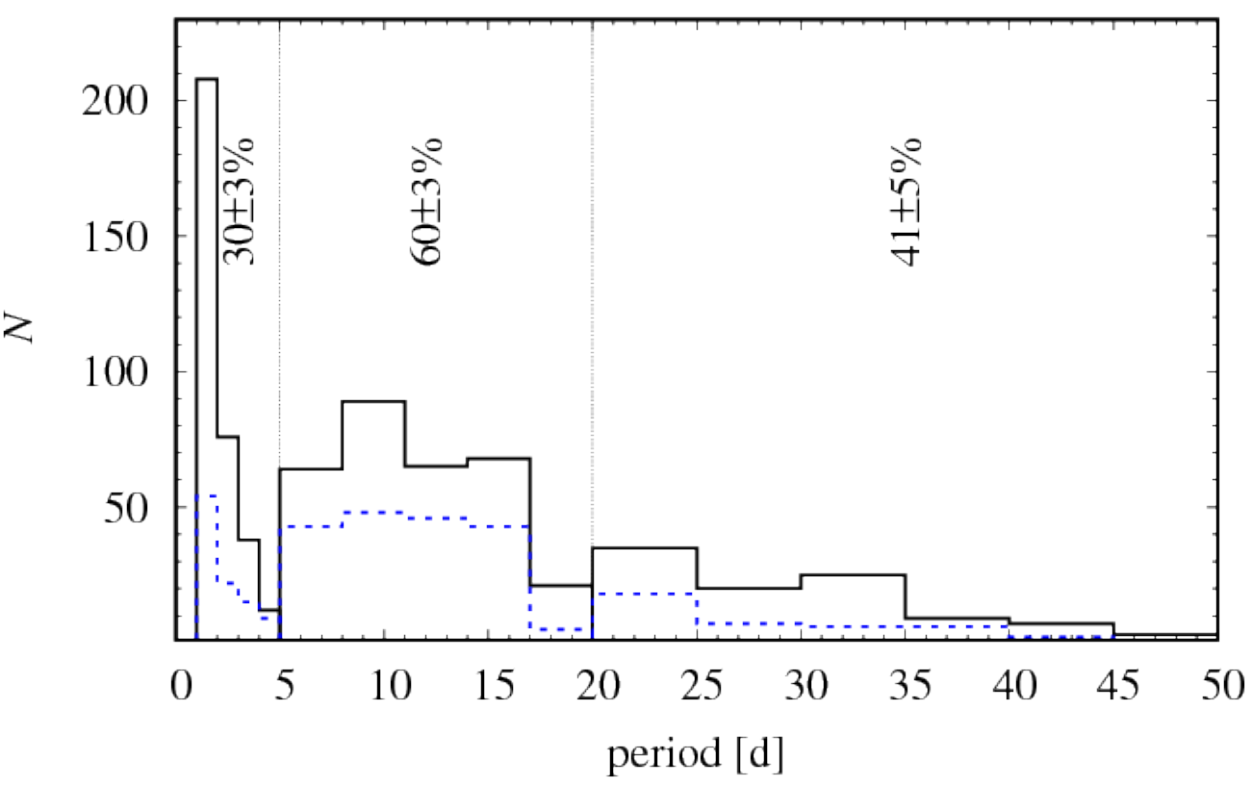}
\caption{Period distribution for all type II Cepheids for which data are available for seven observing seasons (solid line) and for stars in which unresolved remnant power is detected at $\fF$ after the prewhitening (dashed line). Incidence rates of such stars for the three sub-classes of type II Cepheids, separated with thin vertical lines, are given in the plot.}
\label{fig:unrhisto}
\end{figure}

To quantify the problem, in Fig.~\ref{fig:unrhisto} we plot the period distribution for type II Cepheids (solid line) and, separately, for those stars in which unresolved remnant power is detected at $\fF$ after the prewhitening (dashed line). Note that to construct the plot we have only used the stars for which data were collected during all observing seasons, s1,\ldots, s7 ($\sim$80\,per cent of the OGLE sample). Unresolved remnant power is detected in $30$\,per cent of BL~Her stars, $60$\,per cent of W~Vir stars and $41$\,per cent of RV~Tau stars. 

To study the amplitude/phase changes, and to get rid of the unwanted power excess at $k\fF$, we use the time-dependent Fourier analysis \citep{kbd87} followed by a time-dependent prewhitening \citep[][the Appendix]{pamsm15}. In a nutshell, the data are divided into chunks of roughly equal length, $\Delta t$. We fix the pulsation frequency, $\fF$, and fit the sine series to each of the chunks separately. Since $\fF$ is fixed, period changes are absorbed in a phase coefficient, $\phi_1$. In the next step, we conduct a time-dependent prewhitening. The individual sine series are subtracted from the corresponding data chunks. Amplitudes and phases of the sine series are kept constant within each data chunk. The residuals are inspected for the presence of additional signals. Obviously, the procedure is sensitive only to the amplitude/phase changes on a time scale longer than $\Delta t$. In the data prewhitened this way, any possible changes on a time scale longer than $\Delta t$ are filtered out.

\begin{figure}
\includegraphics[width=\columnwidth]{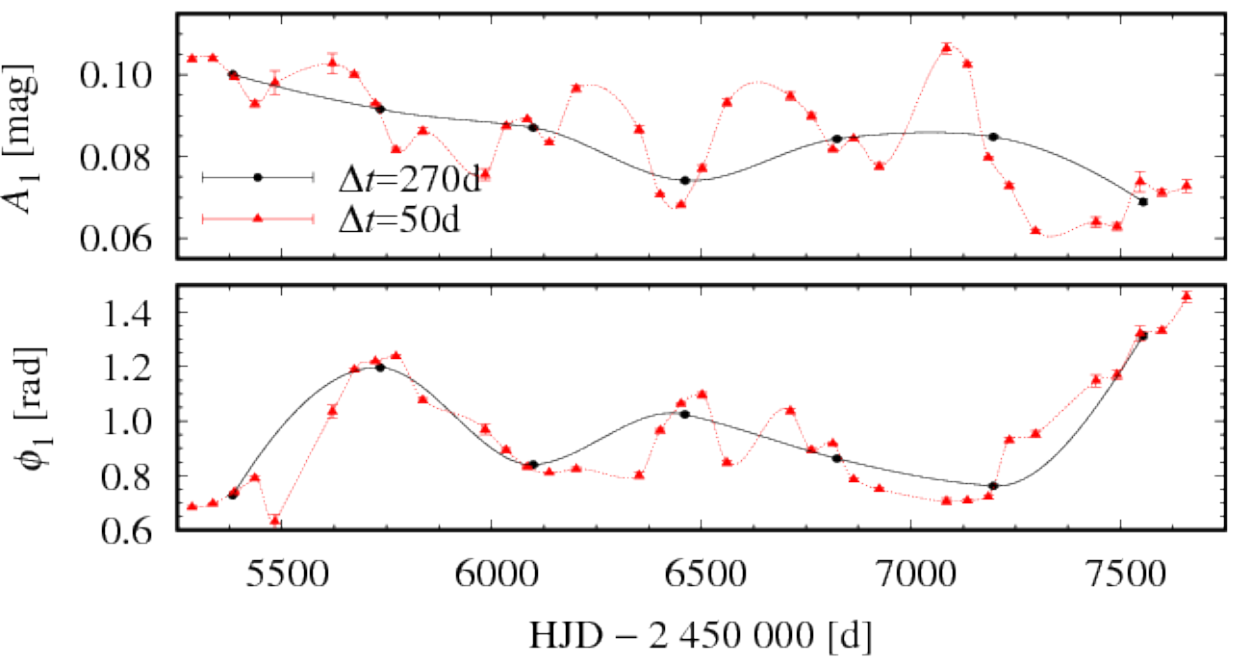}
\caption{Time-dependent Fourier analysis for W~Vir star T2CEP-126. Amplitude and phase changes are plotted in the top and bottom panels respectively. The data were divided into groups of approximately equal length of $\Delta t\approx270$\,d (one season; black circles) or $\Delta t\approx50$\,d (red triangles). Data were fitted with cubic splines to better visualise the changes.}
\label{fig:tdp_tdfd}
\end{figure}

\begin{figure}
\includegraphics[width=\columnwidth]{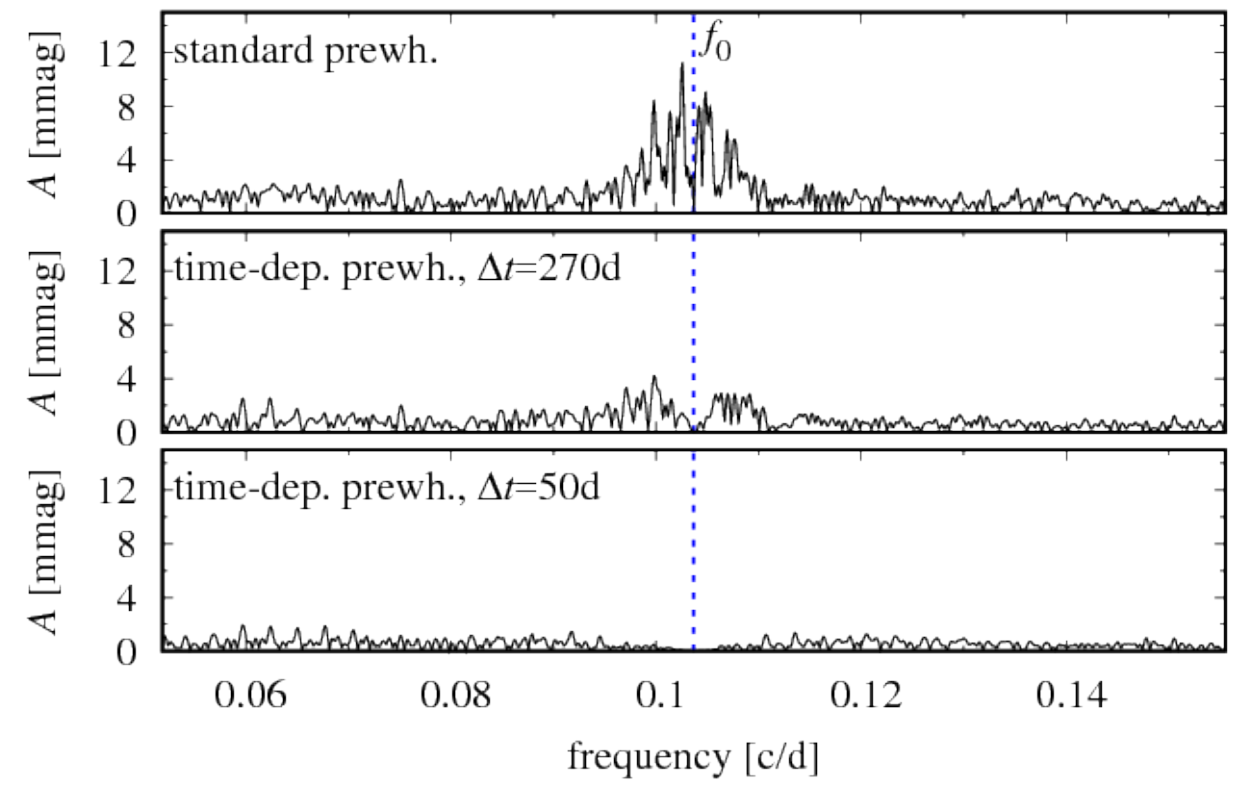}
\caption{Time-dependent prewhitening illustrated for T2CEP-126 with a section of frequency spectrum centred at $\fF$. Top panel: standard prewhitening. Middle panel: time-dependent prewhitening with $\Delta t\!\approx\!270$\,d. Bottom panel: time-dependent prewhitening with $\Delta t\!\approx\!50$\,d. Dashed vertical line in the panels marks the location of $\fF$.}
\label{fig:tdp_fsp}
\end{figure}

The procedure and the sensitivity of the results to $\Delta t$ are illustrated in Fig.~\ref{fig:tdp_tdfd} (time-dependent Fourier analysis) and in Fig.~\ref{fig:tdp_fsp} (frequency spectra) with the help of W~Vir-type star, T2CEP-126 ($\PF=9.64566$\,d). With $\Delta t=270$\,d (which is $\sim$observing season's length), amplitude and phase clearly vary in time and the changes are strong, and aperiodic. As a result, in the frequency spectrum after standard prewhitening with the fundamental mode and its harmonics,  we find a strong remnant power (top panel of Fig.~\ref{fig:tdp_fsp}). After the time-dependent prewhitening with $\Delta t=270$\,d, the power excess is reduced, but does not vanish. Significant power, centred at $\fF$ with a broad minimum in the middle, is still observed in the frequency spectrum (middle panel of Fig.~\ref{fig:tdp_fsp}). It indicates that there is a significant variability on a time scale shorter than $\sim$270\,d. Indeed, with $\Delta t=50$\,d, changes on a time scale of a few tens of days are unambiguously revealed (triangles and dashed lines in Fig.~\ref{fig:tdp_tdfd}). The changes are irregular. With $\Delta t=50$\,d, time-dependent prewhitening effectively cleans the frequency spectrum -- no additional significant signals are detected (bottom panel of Fig.~\ref{fig:tdp_fsp}).

In practice, there is a lower limit for $\Delta t$, motivated by the number of data points. In addition, $\Delta t$ should be longer than the pulsation period, so at least a few cycles are covered. Only then the amplitudes and phases can be determined reliably. The latter constraint becomes important, e.g. for RV~Tau stars in which the length of a period-doubled cycle can amount to a significant fraction of the observing season.

We stress that time-dependent prewhitening was conducted only when it was necessary, i.e. when strong remnant power was present in the frequency spectrum after standard prewhitening. The value of $\Delta t$ was adjusted for each star individually, based on the quality of the data (number of data points and their sampling) and the pulsation period. In many of the analysed stars, time-dependent analysis on a season-to-season basis was the only possibility. Quite often, the following time-dependent prewhitening led to a $\sim$symmetric structure around $\fF$, similar to that in the middle panel of Fig.~\ref{fig:tdp_fsp}. It may mimic the modulation side-peaks, but in fact, it reflects nothing more than irregular variability on a time-scale shorter than observing season.

In this study, the irregular variability of the fundamental mode, clearly worth a separate study (in prep.), is an unwanted complication. The phenomena we are particularly interested in are multi-mode pulsation, period-doubling phenomenon (in general, period-$k$ pulsation) and quasi-periodic modulation of pulsation.

\begin{figure}
\includegraphics[width=\columnwidth]{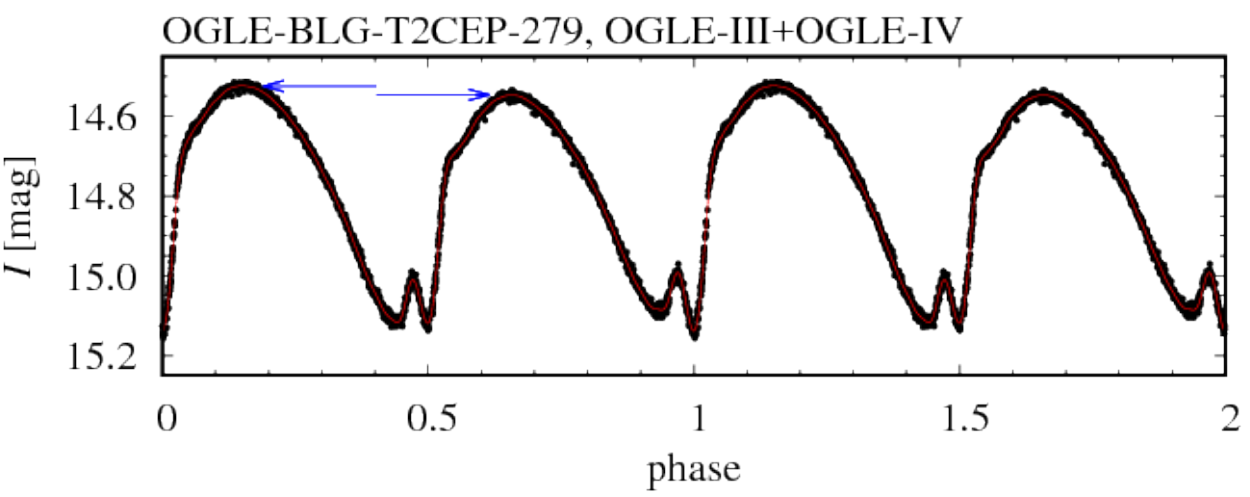}\\
\includegraphics[width=\columnwidth]{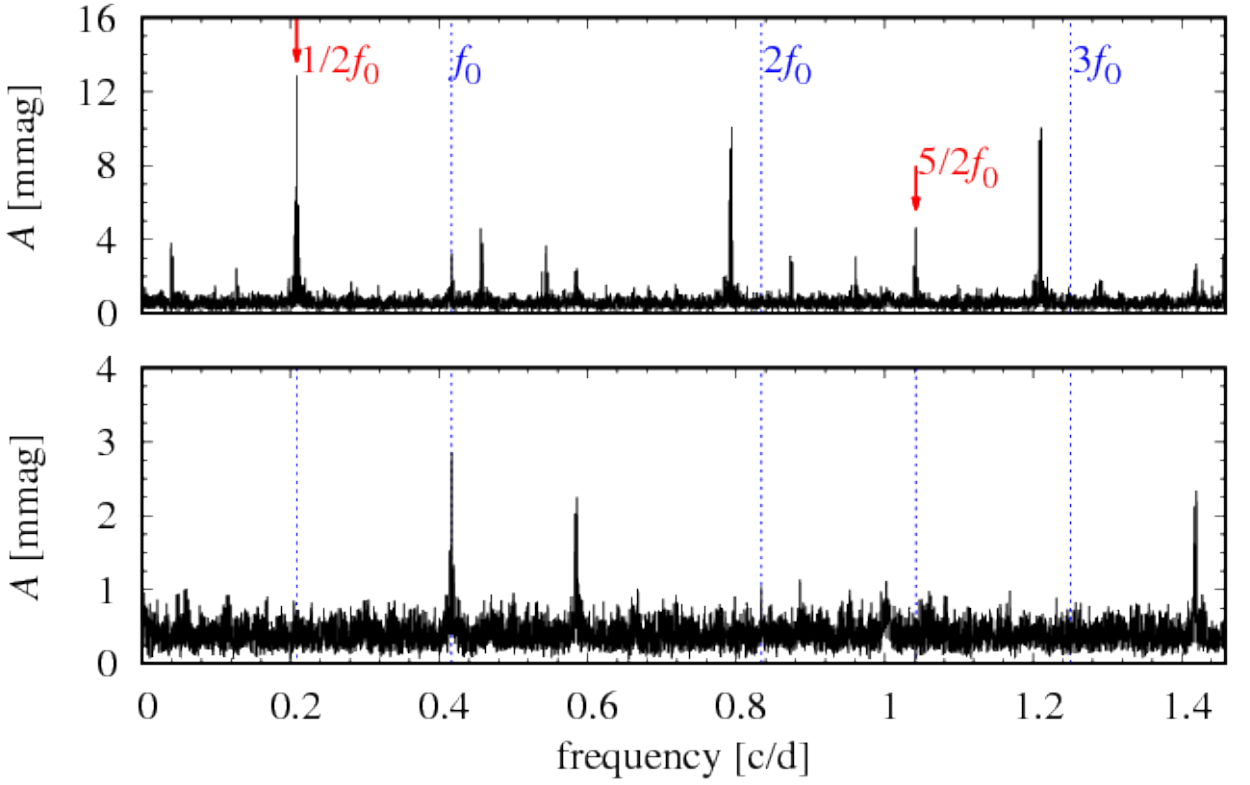}
\caption{Top panel: phased light curve for BL~Her-type star T2CEP-279 showing clear PD effect. Arrows visualize the differences in the brightness maxima of the consecutive cycles. Middle and bottom panels: frequency spectrum for T2CEP-279 after prewhitening with the fundamental mode and its harmonics (dashed lines). In the middle panel, two signals located at sub-harmonic frequencies ($\fjd$ and $\fpd$) are marked with arrows; other signals are daily aliases. In the bottom panel, all signals detected at sub-harmonic frequencies were prewhitened. The only remaining signal is unresolved with $\fF$ (its daily aliases are also present).}
\label{fig:pd279}
\end{figure}

In the time domain, the PD phenomenon manifests as alternating deep and shallow brightness maxima/minima. We illustrate it in the top panel of Fig.~\ref{fig:pd279} with the help of T2CEP-279 ($\PF=2.399260$\,d), the first BL~Her stars with period doubling effect. In the frequency spectrum, the PD phenomenon manifests as signals at sub-harmonic frequencies, i.e. at $(2n+1)\fF/2$. In the case when PD is a regular phenomenon, its signature in the Fourier transform is a set of coherent peaks located exactly at the sub-harmonic frequencies. These peaks can be prewhitened with sine waves of constant amplitude/phase. It is illustrated in the bottom panels of Fig.~\ref{fig:pd279}

\begin{figure}
\includegraphics[width=\columnwidth]{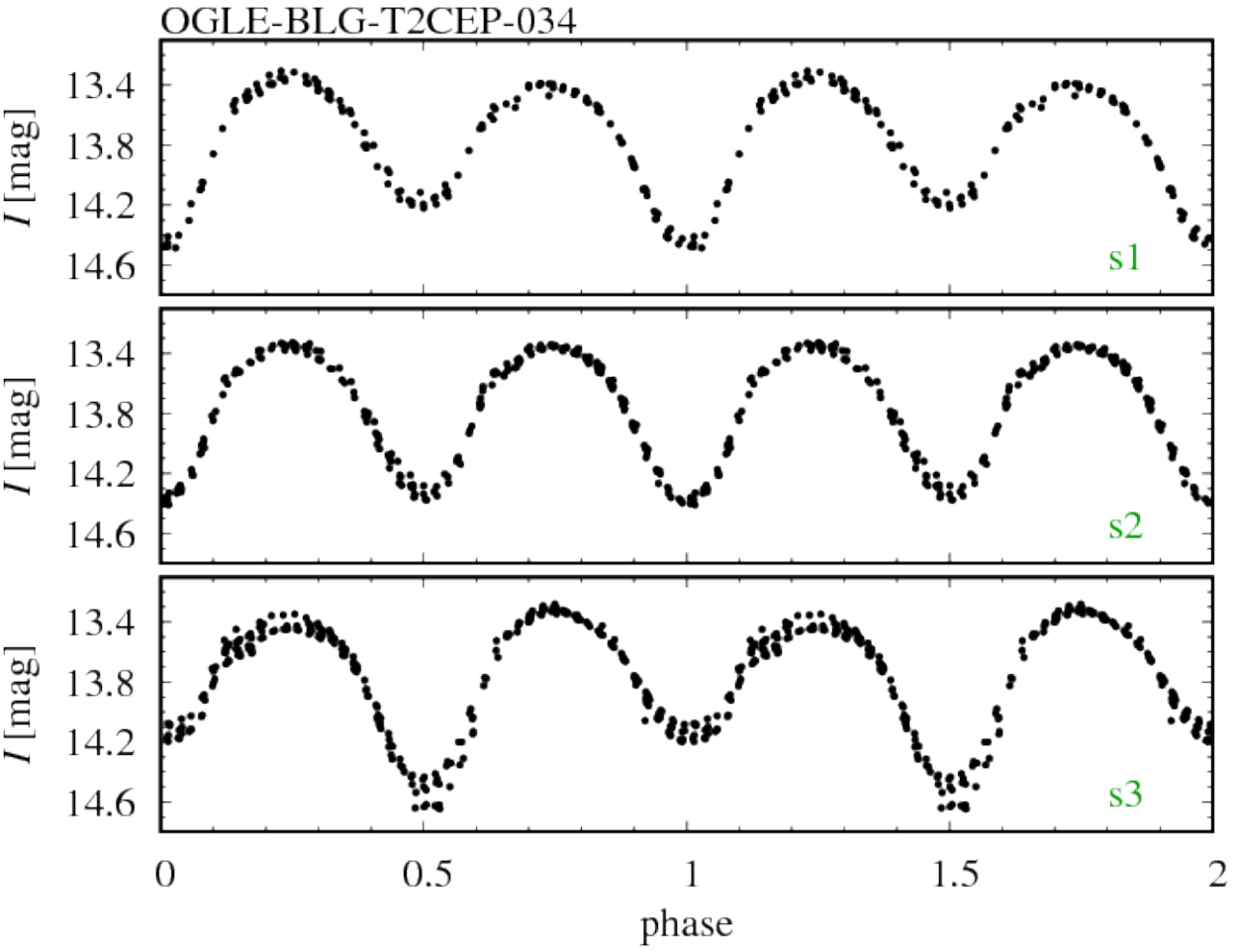}\\
\includegraphics[width=\columnwidth]{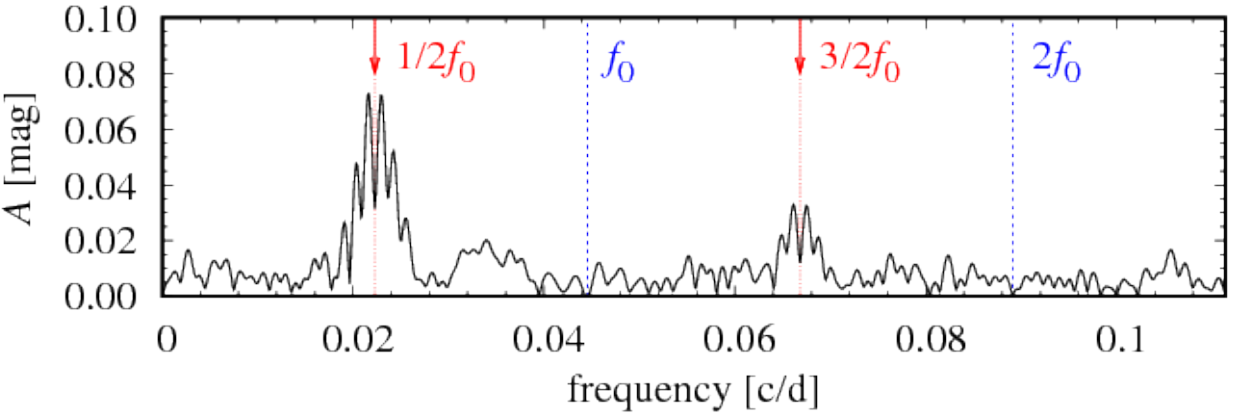}
\caption{Top panels: phased light curves for RV~Tau-type star T2CEP-034 plotted separately for three consecutive observing seasons. Interchange of the deep/shallow minima in between s1 and s3 is clear. Bottom panel: frequency spectrum for the three observing seasons of T2CEP-034, after prewhitening with the fundamental mode and its harmonics (dashed lines). Power excesses, centred at sub-harmonic frequencies, with minima in the middle, are clearly detected.}
\label{fig:pd034}
\end{figure}

Interchanges (switching) of the deep and shallow minima/maxima may also occur. We illustrate this in the top panels of Fig.~\ref{fig:pd034} in which we phase the data for RV~Tau variable T2CEP-034 ($\PF=22.4498$\,d), for three observing seasons separately. The deep/shallow minima clearly switch between s1 and s3; during s2 the PD effect is barely noticeable. The effect may also be strongly irregular or transient. Switching PD, or irregular effect, manifest as more complex structures in the frequency spectrum. It is illustrated in the bottom panel of Fig.~\ref{fig:pd034} for T2CEP-034. Power excesses, centred at $\fjd$ and $\ftd$ are clear, but exactly at the locations of the sub-harmonic frequencies, marked with arrows, minima are noticeable. Further examples will be presented in Sect.~\ref{ssec:rvt-switch}.

For period-$k$ pulsation, the light curve repeats after $k$ fundamental mode pulsation cycles. In the frequency spectrum, we expect to detect signals centred at $\fF/k$ and its harmonics.

Periodic modulation of pulsation manifests as equally spaced multiplets centred at the pulsation frequency and its harmonics. Frequency separation within the multiplet corresponds to the modulation frequency, $\fm$. The signal at the modulation frequency, which corresponds to the modulation of mean stellar brightness, may also be detected. Amplitudes of the multiplet components may be strongly asymmetric depending, e.g. on the phase relation between the amplitude and phase modulation, see \cite{benko11}. For low-amplitude modulation and ground-based data, modulation typically manifests as triplets at the pulsation frequency, i.e. peaks at $\fF$ and $\fF\pm\fm$, and also at the harmonics. In case one of the triplet components is missing, we detect doublets, see, e.g. \cite{rs15a} or \cite{s16} in which we analysed low-amplitude modulations in the OGLE data for RR~Lyrae stars and classical Cepheids, respectively. To claim the modulation we require the detection of at least two peaks related to a modulation with a common frequency.

\section{Results}\label{sec:results}

\subsection{Double-mode F+1O BL~Her stars}

Two short-period BL~Her stars have been identified as double-mode pulsators.

\begin{figure}
\includegraphics[width=\columnwidth]{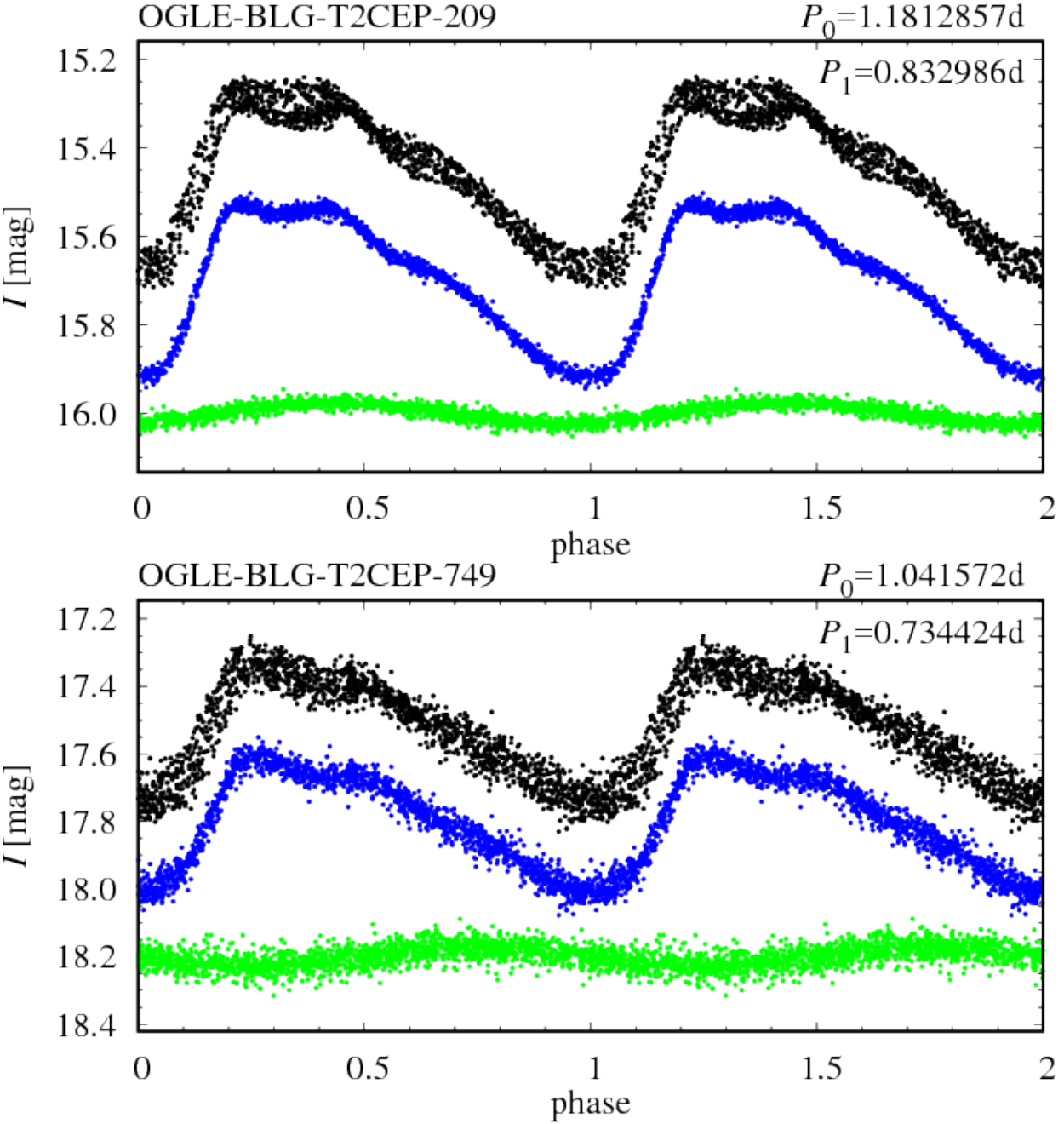}
\caption{Folded light curves for T2CEP-209 (top panel) and T2CEP-749 (bottom panel). Three curves in each panel correspond to: all data folded with fundamental mode period (top) and to disentangled light curves for radial fundamental mode (middle) and radial first overtone (bottom). The two latter curves are shifted in magnitude to allow comparison.}
\label{fig:749_lc}
\end{figure}

\offi{T2CEP-209}. Only OGLE-III data are available for the star. In addition to $\fF=0.8465353(4)$\thinspace\cd\ and its harmonics (up to $7\fF$), we detect $\fO=1.200501(3)$\thinspace\cd\ and several linear combination frequencies (11 in total). We note that three secondary periodicities were given already in the OGLE catalog: one corresponds to $\fO$, the other two correspond to $\fF\!+\!\fO$ and  $2\fF\!+\!\fO$.  The additional periodicity has low amplitude, $A_1=0.0237$\,mag ($A_1/A_0=0.13$). In the top panel of Fig.~\ref{fig:749_lc} we present the light curves for T2CEP-209. From top to bottom the light curves are: original data folded with $\PF$ and the disentangled light curves corresponding to the dominant mode (middle light curve) and to the low-amplitude variability (bottom light curve; sine wave).

\offi{T2CEP-749}\tech{BLG535.23.28416}. In addition to $\fF\!=\!0.9600873(9)$\thinspace\cd\ and its harmonics (up to $5\fF$), we clearly detect $\fO\!=\!1.361612(5)$\thinspace\cd\ and several linear combination frequencies ($\fF\!+\!\fO$, $2\fF\!+\!\fO$, $3\fF\!+\!\fO$, $4\fF\!+\!\fO$, $\fO\!-\!\fF$, $\fF\!+\!2\fO$ and $2\fF\!+\!2\fO$). The additional periodicity is of low amplitude, $A_1=0.0267$\,mag ($A_1/A_0=0.15$). In the bottom panel of Fig.~\ref{fig:749_lc} we present the light curves for T2CEP-749, organised in the same way as for T2CEP-209.

The period ratios, $\PO/\PF=0.7052$ for T2CEP-209, and nearly the same value, $0.7051$, for T2CEP-749, indicate that these might be the first double-mode BL~Her stars pulsating simultaneously in the radial fundamental mode and radial first overtone. As we discuss below, their properties fit well into other group of double-mode pulsators identified recently in the OGLE data, the extreme RRd stars \citep{eRRd}. With their periods, $\PF=1.1812857(5)$\,d (T2CEP-209) and $\PF=1.041572(1)$\,d (T2CEP-749), very close to the borderline between RR~Lyrae and BL~Her stars, they would be the longest-period members of this group, now spanning across both classes. 

\begin{figure}
\includegraphics[width=\columnwidth]{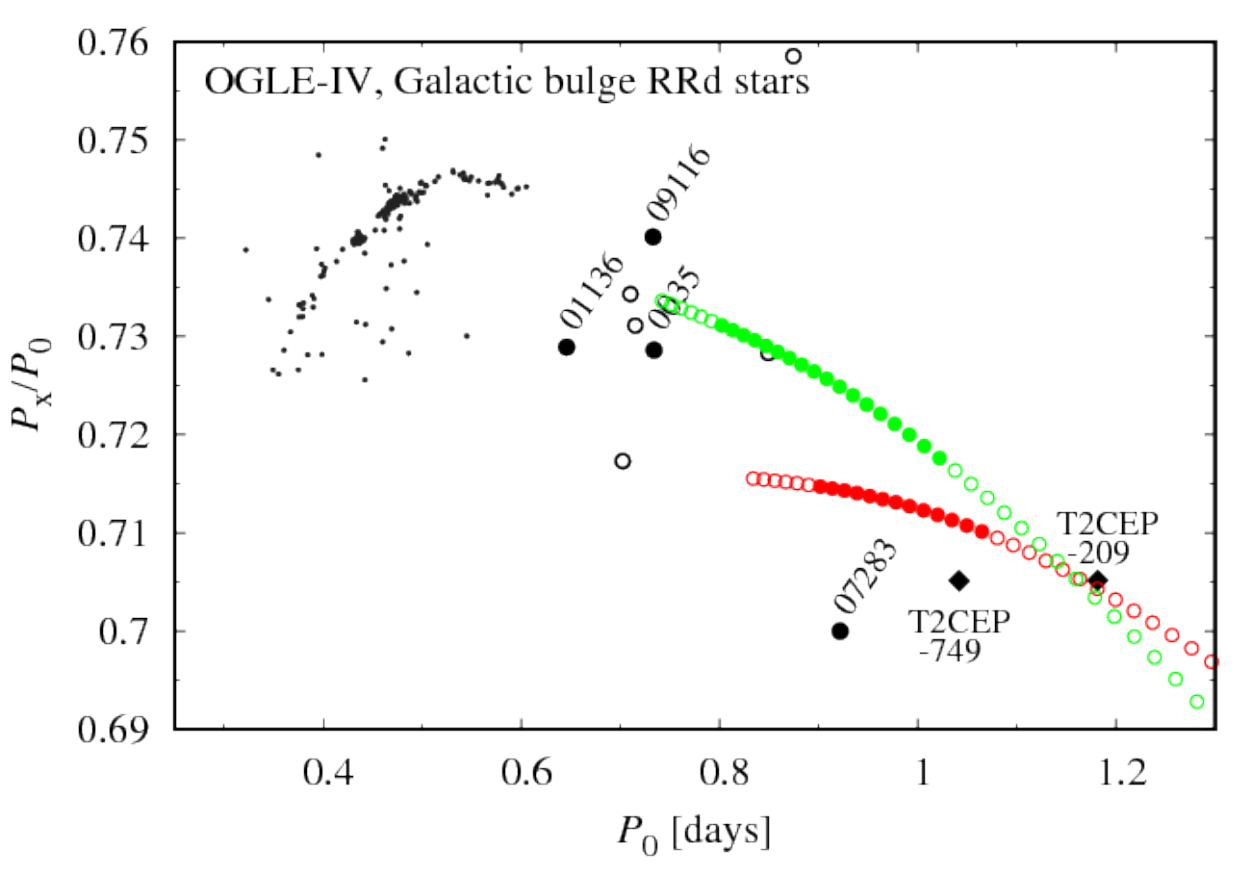}
\caption{Petersen diagram for classical RRd stars (small dots) and long-period double-mode pulsators identified in the OGLE Galactic bulge data: RR~Lyrae stars (circles) and BL~Her stars (diamonds). Two sequences of linear pulsation models, both with $M=0.5\MS$ and $L=70\LS$, but with different metallicities, ${\rm [Fe/H]}=0$ and ${\rm [Fe/H]}=-1.5$, ares plotted with circles. Higher metallicity models are plotted with red circles and lower metallicity ones with green circles.}
\label{fig:749_pet}
\end{figure}

In Fig.~\ref{fig:749_pet} we show the Petersen diagram in which we marked the location of classical RRd stars (dots), extreme RRd stars identified by \cite{eRRd} -- four firm stars (large filled circles) and a few candidates (open circles), and of T2CEP-209 and T2CEP-749 (diamonds). Both BL~Her stars have slightly longer periods than OGLE-BLG-RRLYR-07283, the longest-period double-mode pulsator identified by \cite{eRRd}, and a slightly larger period ratios. \cite{eRRd} calculated a grid of linear pulsation models and showed that for majority of these long-period double-mode pulsators suitable models, in which radial fundamental and radial first overtone are simultaneously unstable, can be found. Here we just overplot two sequences of luminous models that run close to the locations of BL~Her stars (green and red circles). These models have $M=0.5\MS$, $L=70\LS$ and ${\rm [Fe/H]}=0$ (red circles) or ${\rm [Fe/H]}=-1.5$ (green circles). Consecutive models in a sequence cross the instability strip horizontally with $25$\,K-step in effective temperature. Further details of the models are given in \cite{eRRd}, in particular period ratios for the full model grid are plotted in their fig. 12. Filled circles correspond to models in which both radial modes are simultaneously unstable. Match with the location of both BL~Her stars is reasonable. For T2CEP-209 both model sequences match the position of the star very well, but in these models only the fundamental mode is unstable. For higher metallicity, the domain in which the two radial modes are simultaneously unstable is closer. For T2CEP-749 the higher metallicity models are clearly better. We note that only for metal rich models, with high $L/M$ ratio, the two radial modes can be simultaneously unstable at long fundamental mode periods and low period ratios. In all discussed long-period double-mode variables it is the fundamental mode that dominates the pulsation. Amplitude of the first overtone is only a small fraction of the fundamental mode's amplitude.

\begin{figure}
\includegraphics[width=\columnwidth]{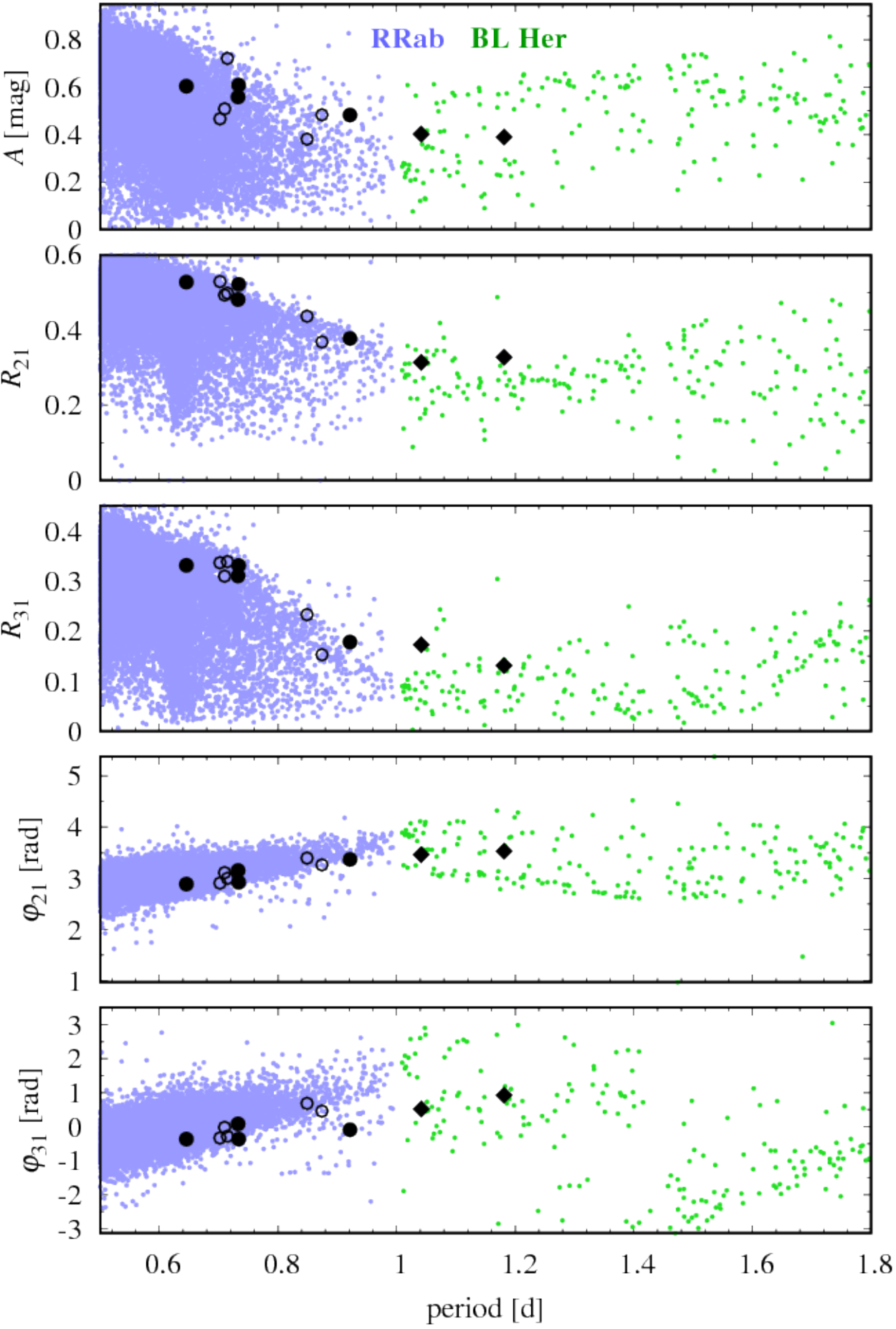}
\caption{Peak-to-peak amplitude (top panel) and Fourier parameters, $R_{21}$, $R_{31}$, $\varphi_{21}$, and $\varphi_{31}$, plotted against the pulsation period for RRab stars (bluish dots), BL~Her variables (green dots) and double-mode variables discussed in the text (large symbols).}
\label{fig:749_fou}
\end{figure}

T2CEP-209, T2CEP-749 and extreme RRd stars, cover a rather large area in the Petersen diagram. All these stars have long periods and low period-ratios (as compared to RRd stars), and they have a similar shape of the light curve of the dominant fundamental mode. This is best quantified with the help of the Fourier decomposition parameters \citep{sl81}, which we plot in Fig.~\ref{fig:749_fou}. The consecutive panels show peak-to-peak amplitude, amplitude ratios, $R_{21}$ and $R_{31}$, and Fourier phases $\varphi_{21}$ and $\varphi_{31}$, plotted vs. the pulsation period. Parameters for fundamental mode RR~Lyrae stars (RRab stars) are plotted with small bluish dots, while parameters for BL~Her stars are plotted with green dots. The other discussed stars are plotted with the same symbols as in Fig.~\ref{fig:749_pet}, and they form a well defined group in the Fourier parameter plots. As compared with RRab stars, we note, that at a given period, the peak-to-peak amplitude and amplitude ratios are among the highest observed for RRab stars. The Fourier phases, on the other hand, are among the lowest. The two BL~Her star fall well within the progression formed by shorter-period double-mode RR~Lyrae stars.

A noticeable feature in the light curves of T2CEP-209 and T2CEP-749 is a bump located on the descending branch (Fig.~\ref{fig:749_lc}). In T2CEP-209, which is of longer period, the bump actually distorts the maximum light. The bump may also be noticed on the descending branch of extreme RRd stars -- see fig.~2 in \cite{eRRd}. Considering the whole group, the bump location depends on the pulsation period -- the longer the period the higher the bump location on the descending branch. This bump progression may be caused by the 2:1 resonance between the fundamental mode and the second overtone as inferred from fig.~13 in \cite{s16}. The same resonance may also be crucial for excitation of the double-mode pulsations in these stars. We plan to investigate this possibility in more detail in a separate study.  

We conclude that T2CEP-209 and T2CEP-749 are new members of the class first identified by \cite{eRRd}, which now spans across the RR~Lyrae and BL~Her classes. We note that the commonly adopted boundary between RR~Lyrae and BL~Her stars, at pulsation period of 1\,d is just a convention. Physically, both groups are low-mass, population II stars that evolve away the ZAHB.

\subsection{Period doubling in BL~Her variables}\label{ssec:pdinblher}

In total, three BL~Her-type variables show a PD phenomenon. Two stars were already reported in the past, based on the analysis of OGLE-III data. T2CEP-279 was the first BL~Her star with PD effect, reported by \cite{blher_ogle} and \cite{blherPD}. In the same papers, T2CEP-257 was announced as a candidate. For both stars new OGLE-IV data are available. The third star, T2CEP-820, briefly announced in \cite{rrl17}, is analysed in detail for the first time. Basic data for these stars are collected in Tab.~\ref{tab:blherpd}.

\begin{table*}
\centering
\caption{Basic characteristics of BL~Her-type variables with PD effect. Consecutive columns contain: star's ID, period and amplitude of the fundamental mode, amplitude of the highest peak detected at $\fjd$, $A_{1/2}$, as determined from the analysis of all OGLE-IV data, and list of all sub-harmonic frequencies detected in the data.}
\label{tab:blherpd}
\begin{tabular}{lllll}
star's ID & $\PF$\,(d) & $A_0$\,(mag) & $A_{1/2}$\,(mag) & sub-harmonic frequencies \\
\hline
\offi{T2CEP-279} \tech{BLG514.29.119264-EP} & 2.399260(1) & 0.2810 & 0.0134 & $\fjd$, $5\fF/2-13\fF/2$ \\
\offi{T2CEP-257} \tech{BLG513.15.94723-RS}  & 2.245698(1) & 0.2438 & 0.0029 & $\fjd$, $5\fF/2-9\fF/2$  \\
\offi{T2CEP-820} \tech{BLG505.24.151399-RS} & 2.4005048(4)& 0.2612 & 0.0014 & $\fjd$    \\
\hline
\end{tabular}
\end{table*}

{\bf T2CEP-279} ($\PF\!=\!2.399260$\,d) is the only BL~Her star in which PD is a persistent phenomenon and is clearly visible in the phased light curve, which we present in Fig.~\ref{fig:pd279}. Currently, the OGLE data for this star span more than $5550$\,d, which corresponds to more than $2300$ fundamental mode pulsation cycles. The phenomenon is very stable, no interchanges of the deep/shallow minima/maxima are observed. This is also confirmed by the analysis of the frequency spectrum, section of which is plotted in the middle and bottom panels of Fig.~\ref{fig:pd279}. In addition to the fundamental mode and its harmonics (detected up to $17\fF$, but without $13\fF$ and $14\fF$), sub-harmonics at $\fjd$ and from $\fpd$ to $1\ftd$ are clearly detected. The peak at $\fjd$ is the highest with $A_{1/2}/A_0=0.048$ (Tab.~\ref{tab:blherpd}). Interestingly, there is no significant signal at $\ftd$. The peaks at sub-harmonic frequencies are coherent, see the bottom panel of Fig.~\ref{fig:pd279}. Residual, unresolved power is detected at $\fF$, which results from: (i) slow phase drift and, (ii) a slight ($\approx 7$\,mmag) amplitude change between OGLE-III and OGLE-IV observations. As the amplitude change occurred in between two phases of the OGLE, we cannot rule out the instrumental origin of this change (see Sect.~\ref{sec:data}).

\begin{figure*}
\includegraphics[width=2\columnwidth]{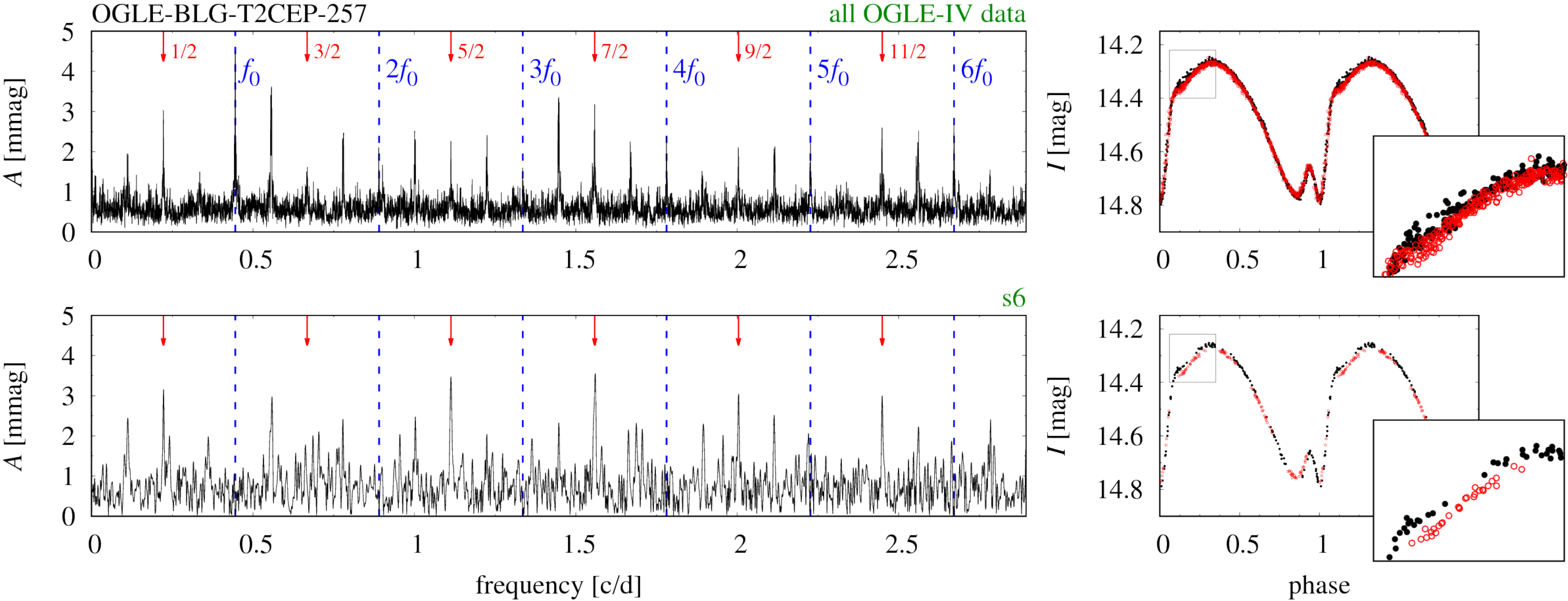}
\caption{Transient PD phenomenon in BL~Her-type star T2CEP-257. The top row shows the analysis of all OGLE-IV data, while the bottom row shows the analysis for s6 only. Left panels show the frequency spectrum after prewhitening with the fundamental mode and its harmonics (dashed lines). Locations of the sub-harmonic frequencies are indicated with arrows. Right panels show the light curve folded with the pulsation period, $\PF$. Even and odd pulsation cycles are plotted with different colours to visualise the PD effect. Insets show zooms around the brightness maximum.}
\label{fig:257}
\end{figure*}

OGLE-III data for {\bf T2CEP-257} ($\PF\!=\!2.245698$\,d) were analysed in detail by \cite{blherPD}. Due to a period change, only a section of the data was analysed. The only detected sub-harmonic ($\fjd$) was very weak, with amplitude of $2.9$\,mmag. The effect was barely visible in the folded light curve. In Fig.~\ref{fig:257} we show the frequency spectrum for OGLE-IV data after prewhitening with the fundamental mode and its harmonics. First we note that $\fF$ and its harmonics are non-stationary; residual, unresolved power remains in the frequency spectrum at their location. Time-dependent Fourier analysis clearly shows that the pulsation period slowly changes. The expected locations of sub-harmonic frequencies are marked with arrows, and indeed, several significant peaks are present there. T2CEP-257 has a somewhat unfortunate pulsation frequency, however. As $4.5\fF\approx2$\,\cd, the location of the sub-harmonics overlap with the location of daily aliases of the residual signals at $k\fF$. Indeed, at the interesting locations we observe a broadened peaks with multiple maxima. In the folded light curve (top-right panel of Fig.~\ref{fig:257}) a signature of the PD effect is barely noticeable.

Due to above mentioned unfortunate pulsation frequency, we cannot use time-dependent prewhitening to get rid of the period changes that complicate the analysis -- this technique would remove the signals at sub-harmonic frequencies as well. Instead, we conducted a seasonal analysis. The bottom row of Fig.~\ref{fig:257} shows the result for s6. During one season period changes are insignificant and there is no remnant power at $k\fF$. Sub-harmonics are now clearly identified; for peaks at $\fjd$, $\fpd$ and $\fsd$ we have a detection well above the $\sn\!=\!4.0$. The peak at $\fjd$ is the highest with $A_{1/2}/A_0=0.012$ (Tab.~\ref{tab:blherpd}). Folded light curve also clearly shows the PD effect. Interestingly, the effect seems to be restricted to s6. Only in the preceding season, s5, we have a significant detection of only one sub-harmonic (at $\fjd$, $\sn\!=\!4.3$). Hence, PD in T2CEP-257 is a transient phenomenon, present in some of the observing seasons only. It is also the case for the earlier OGLE-III observations.

\begin{figure*}
\includegraphics[width=2\columnwidth]{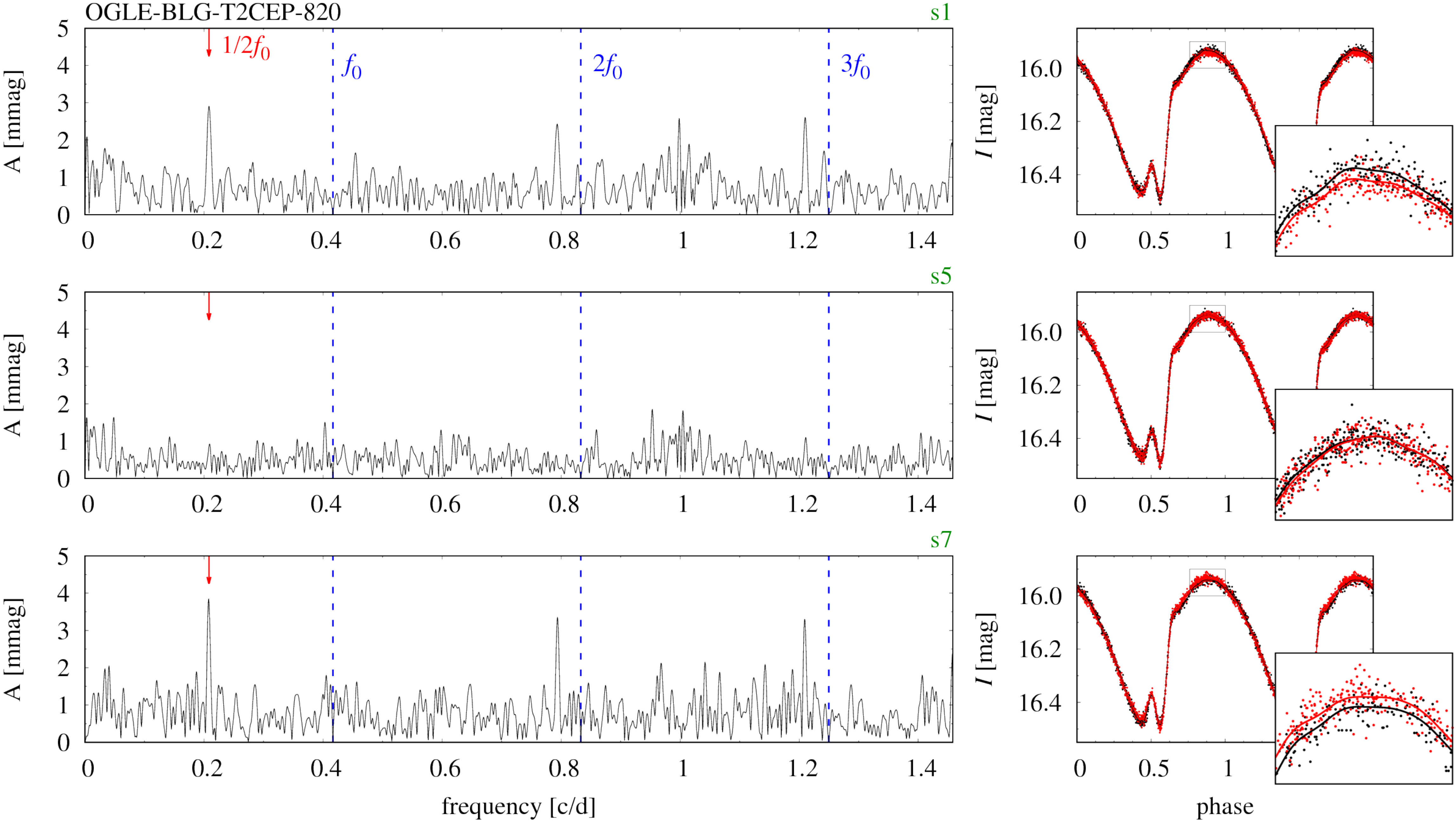}
\caption{Transient and switching PD phenomenon in BL~Her-type variable T2CEP-820. Separate analysis for three observing seasons, s1, s5 and s7 is presented in three consecutive rows. Left panels show the frequency spectrum after prewhitening with the fundamental mode and its harmonics (dashed lines). Location of the $\fjd$ sub-harmonic frequency is indicated with an arrow. Right panels show the light curve folded with the pulsation period, $\PF$. Even and odd pulsation cycles, and respective Fourier fits, are plotted with different colours to visualise the PD effect. Insets show zooms around the brightness maximum.}
\label{fig:820}
\end{figure*}

{\bf T2CEP-820} ($\PF\!=\!2.4005048$\,d) was observed only during the fourth phase of the OGLE project. Signature of PD was detected in the frequency spectrum as a signal at sub-harmonic frequency, $\fjd$. The signal is pronounced ($\sn\!=\!6.0$); still its amplitude is only $\approx 1.4$\,mmag ($A_{1/2}/A_0=0.005$; Tab.~\ref{tab:blherpd}). We note that in this star season-to-season zero point offsets are significant and must be removed in the analysis. Residual power at $\fF$ is also detected and time-dependent Fourier analysis indicates it is due to period change. In addition, much weaker signals (all with $3.0\!<\!\sn\!<\!3.5$) centred at $\fpd$, $\fsd$ and $9/2\fF$ are present in the spectrum. All these signals have a complex structure, similar to that displayed in the bottom panel of Fig.~\ref{fig:pd034}, which suggests that the PD effect is irregular. It is confirmed by seasonal analysis of the data, illustrated in Fig.~\ref{fig:820}. Signatures of PD (signals at $\fjd$) are clear only during s1 (top panel of Fig.~\ref{fig:820}), s6 and s7 (bottom panel of Fig.~\ref{fig:820}). In between, no significant power at sub-harmonic frequencies is detected, as illustrated for s5 -- middle panel of Fig.~\ref{fig:820}. Folded light curves presented in the right panels of Fig.~\ref{fig:820} confirm the frequency analysis. The effect is noticeable close to the brightness maximum during s1 and s7, and is not present in s5. Interestingly, the deep/shallow maxima interchange. 

We postpone a more detailed discussion of the PD effect in BL~Her-type stars to Sect.~\ref{sec:discussion}. Here we just note that all discussed stars occupy a narrow period range and have very similar light curve shapes. In all stars the sub-harmonic at $\fjd$ is the highest, while there is no significant detection at $\ftd$, which appears surprising, as signals at higher-order sub-harmonics are detected (at least in T2CEP-279 and T2CEP-257).

\subsection{Period doubling in W~Vir stars}\label{ssec:pdinwvir}

The $20$\,d borderline between W~Vir stars and RV~Tau variables is a convention. Not surprisingly, single stars that do show the PD effect and have period slightly shorter than $20$\,d were reported, e.g. by \cite{ogle3_lmc_t2,blher_ogle,ogle4_gb_t2,plachy17,plachyRRL17_PD}. W~Vir itself ($\PF\approx17.27$\,d), the prototype of the class, also shows a weak PD effect \citep{th07}. On the other hand, there are type II Cepheids with no PD alternations and periods above $20$\,d. They are formally classified as RV~Tau stars in the OGLE catalogues but, as noted by \cite{ogle4_gb_t2}, such stars may also be classified as yellow semi-regular variables (SRd stars).

\begin{figure*}
\includegraphics[width=.66\columnwidth]{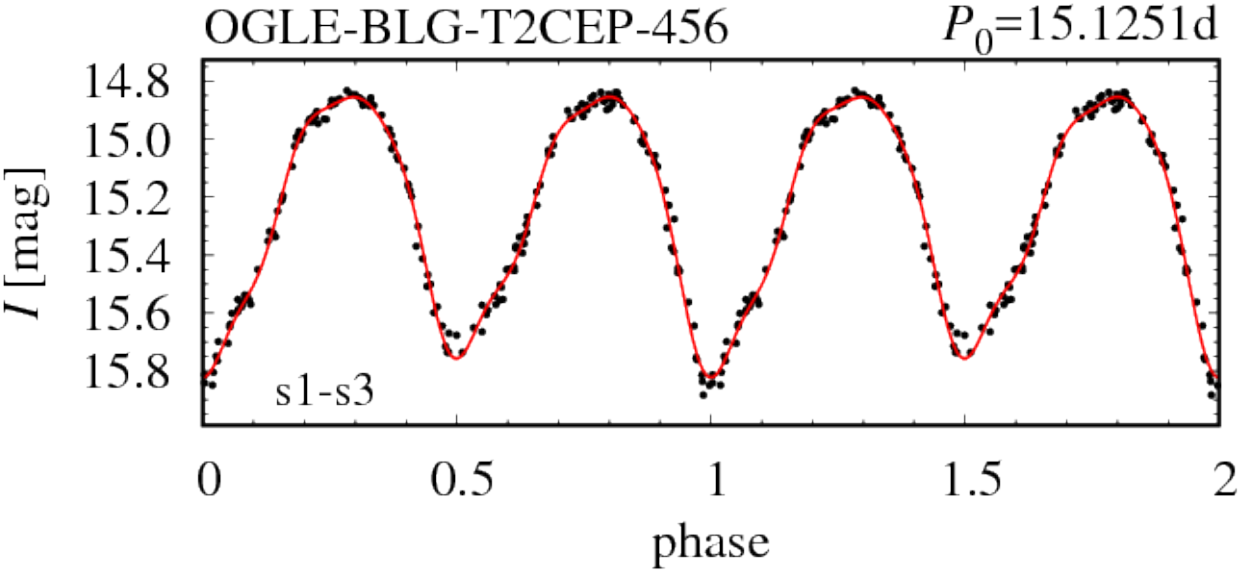}    
\includegraphics[width=.66\columnwidth]{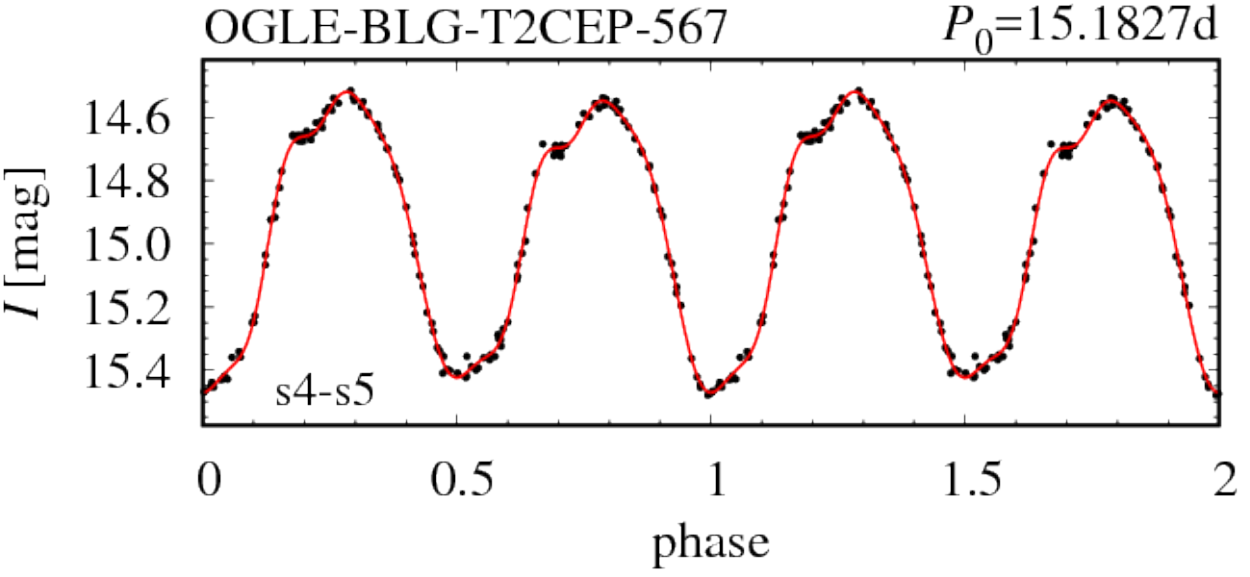}   
\includegraphics[width=.66\columnwidth]{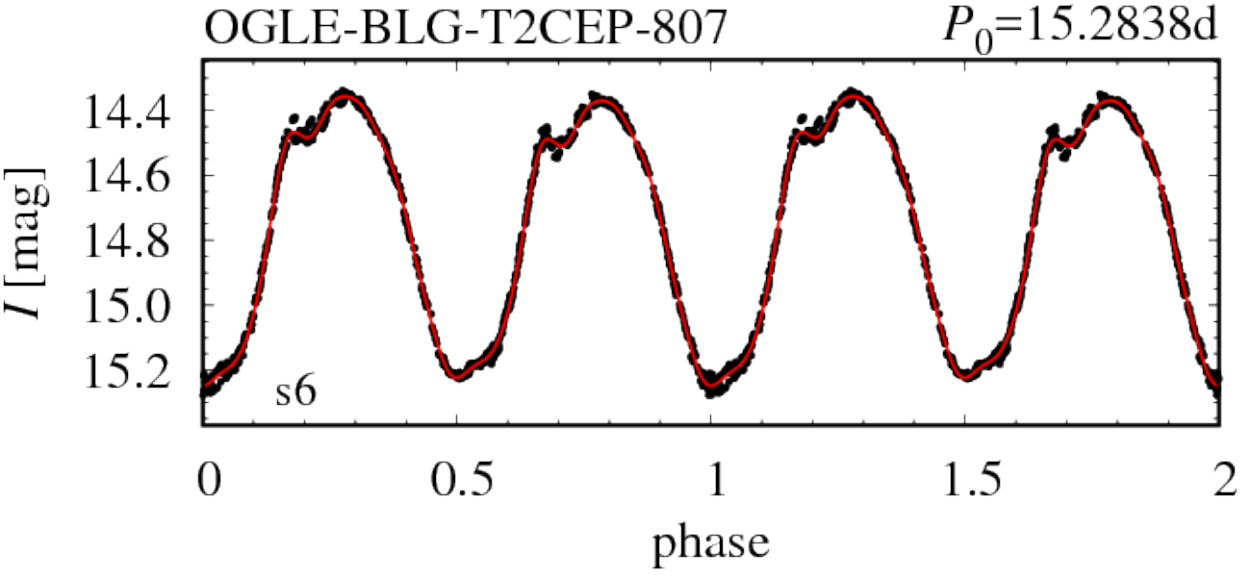}\\ 
\includegraphics[width=.66\columnwidth]{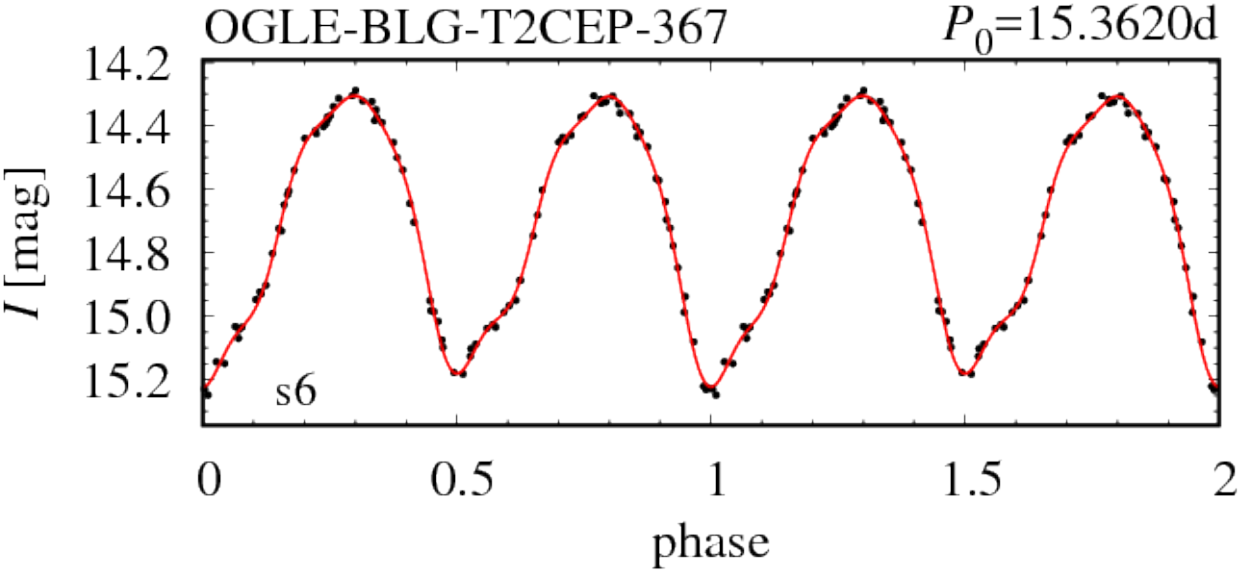}   
\includegraphics[width=.66\columnwidth]{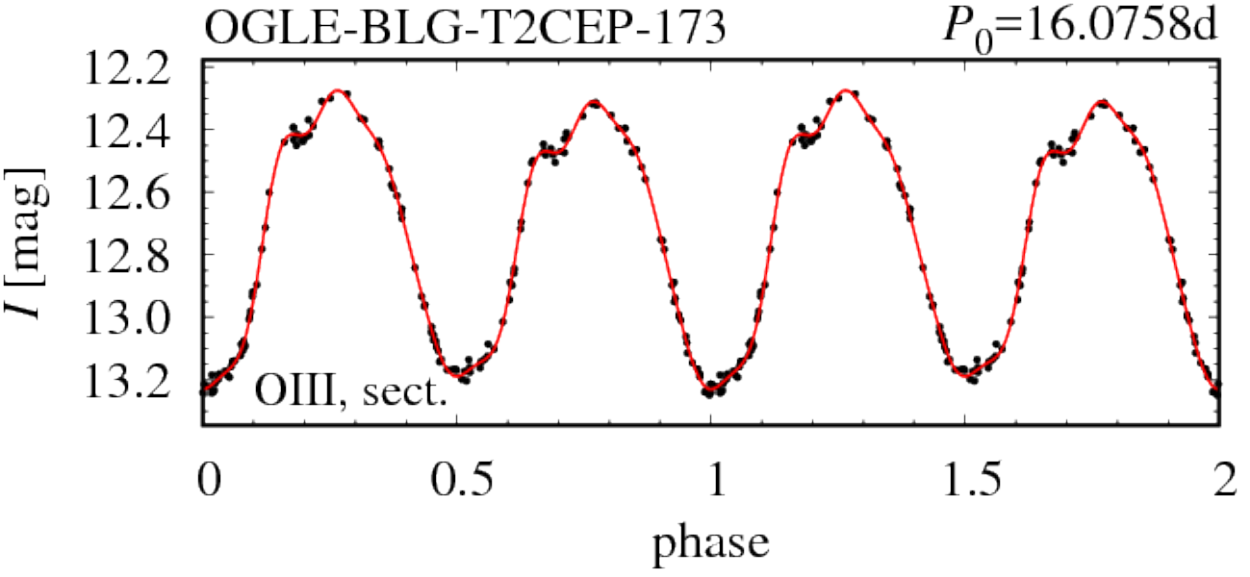} 
\includegraphics[width=.66\columnwidth]{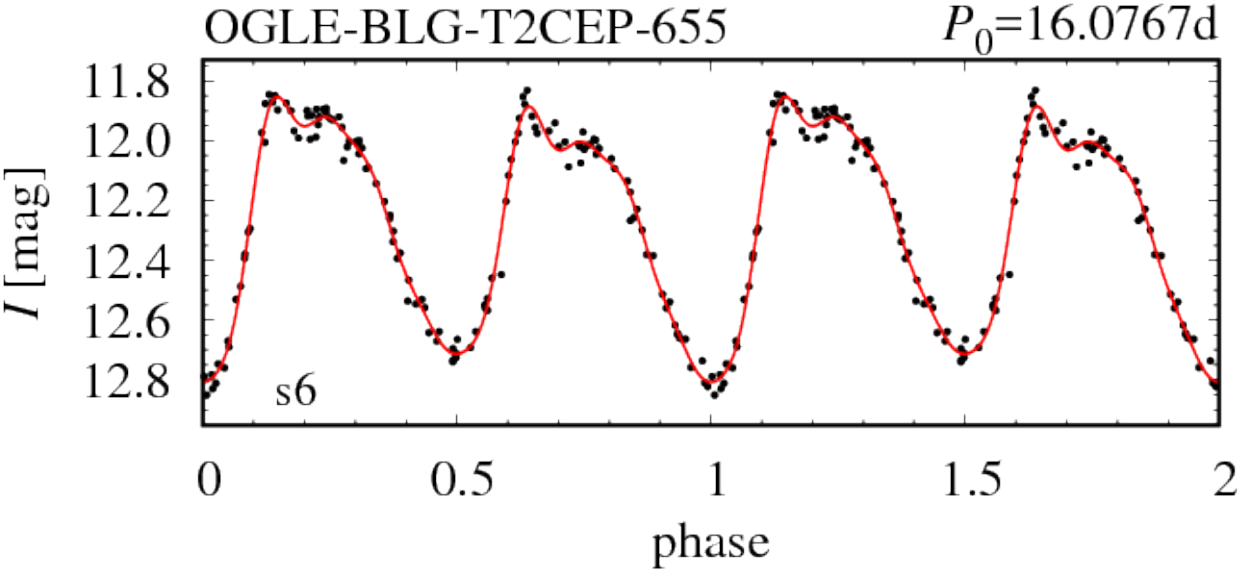}\\ 
\includegraphics[width=.66\columnwidth]{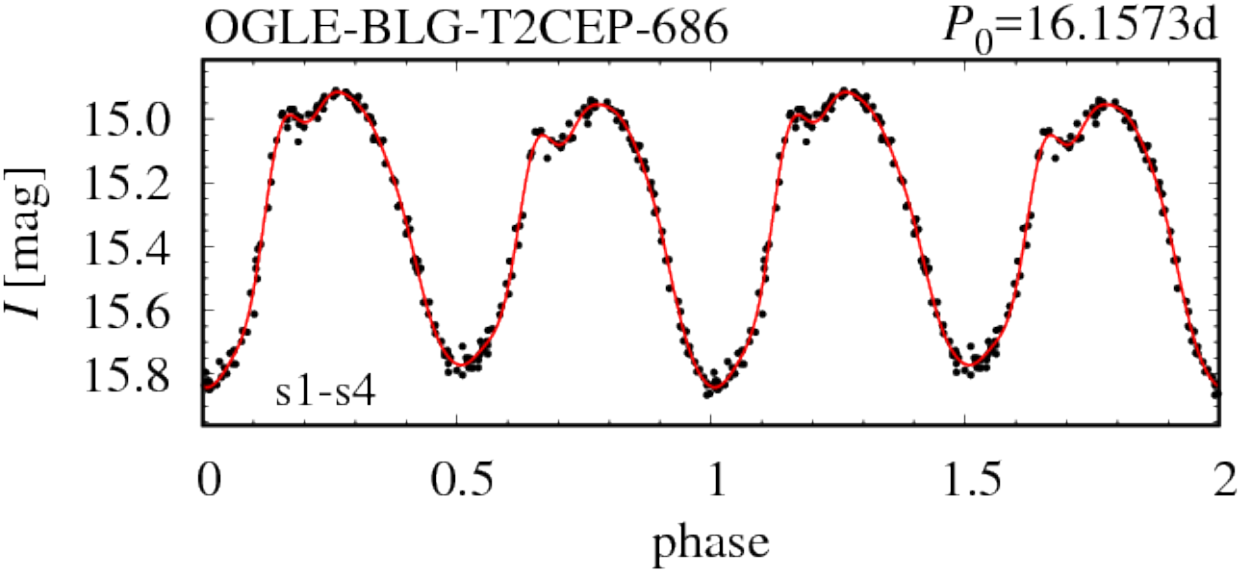}
\includegraphics[width=.66\columnwidth]{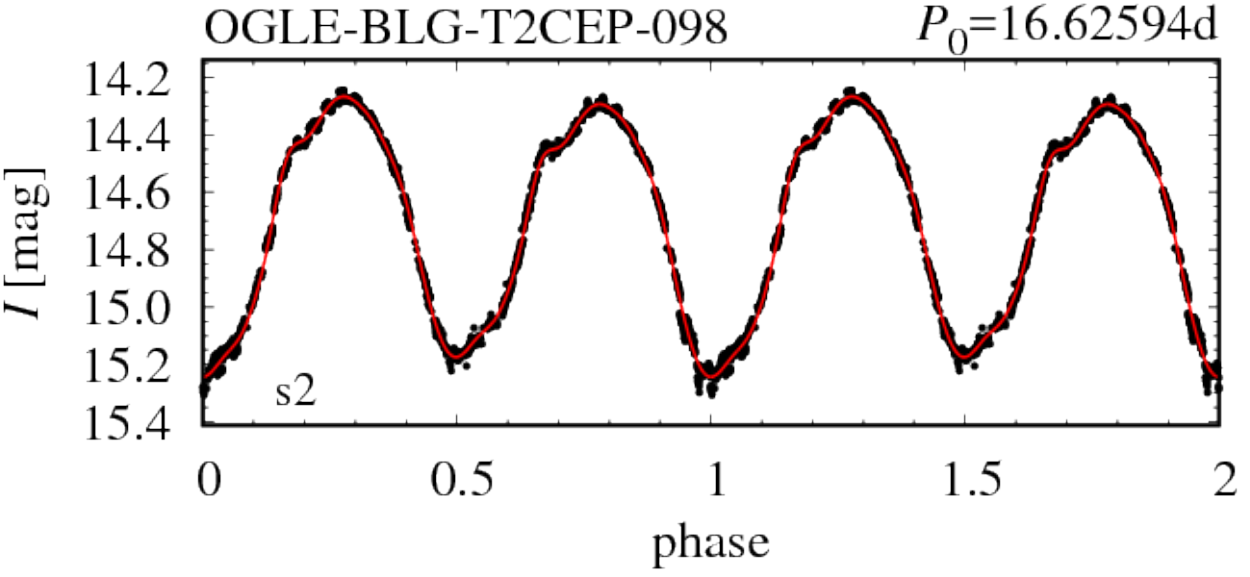}
\includegraphics[width=.66\columnwidth]{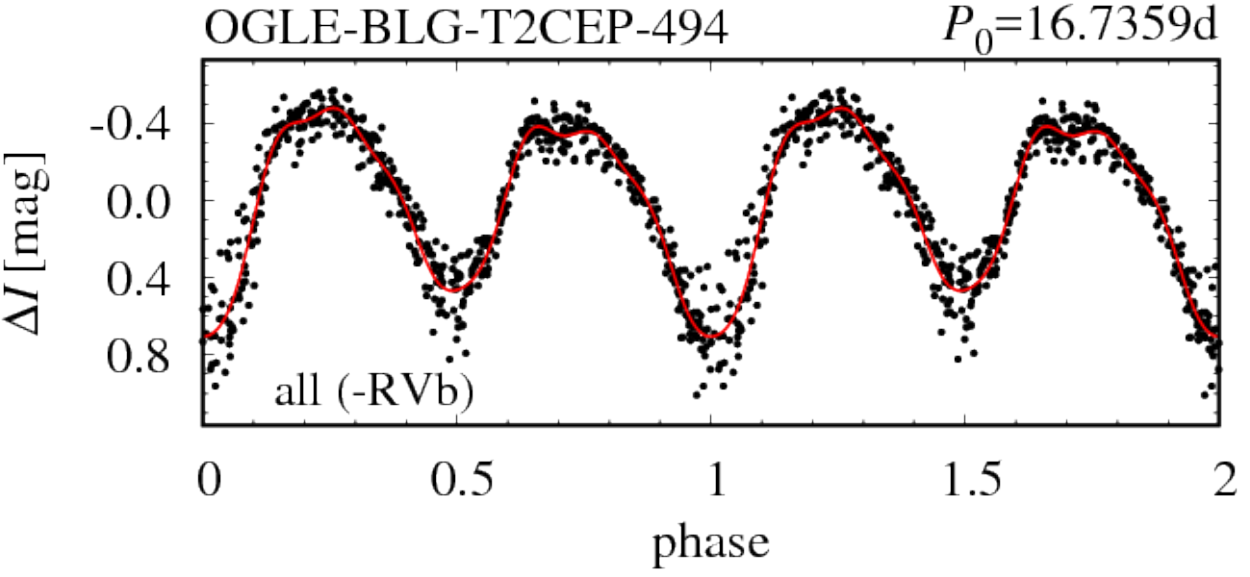}\\ 
\includegraphics[width=.66\columnwidth]{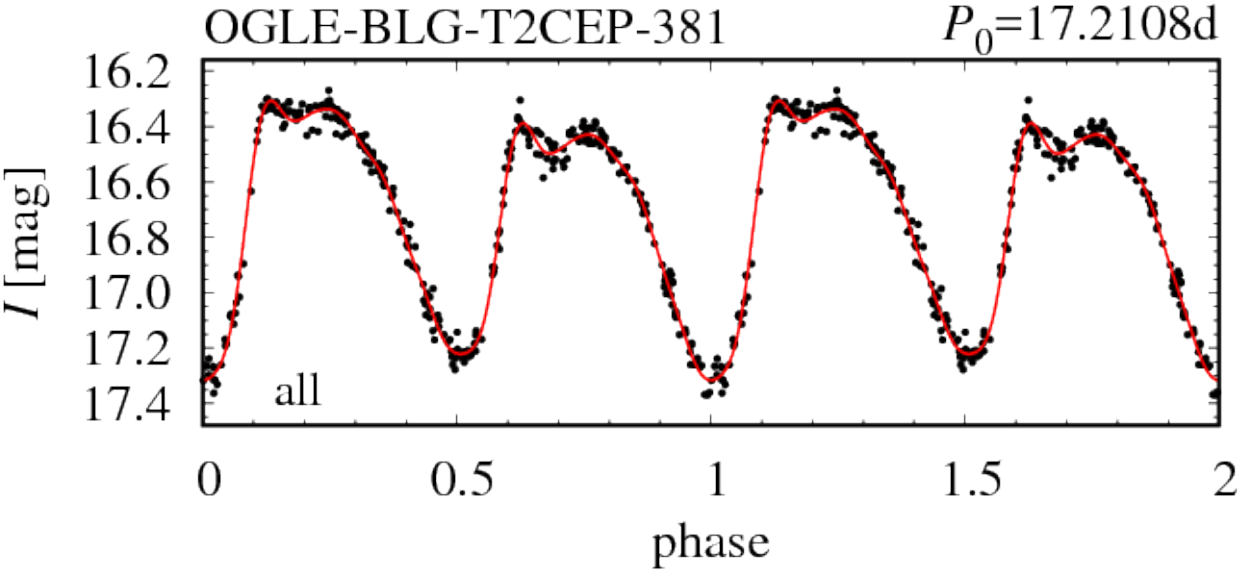} 
\includegraphics[width=.66\columnwidth]{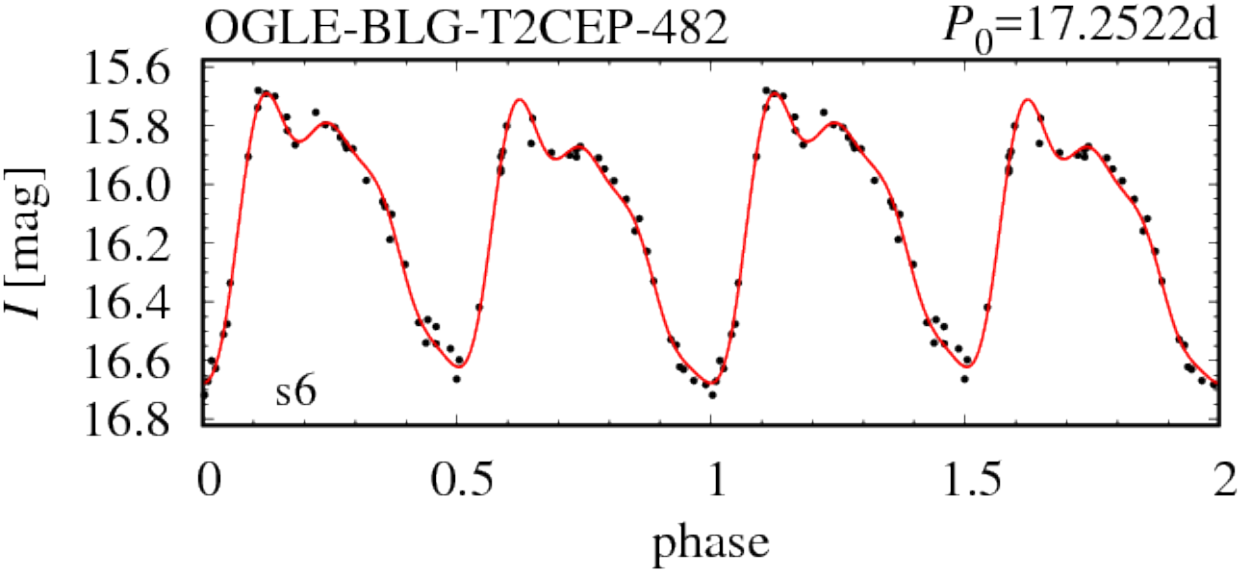}
\includegraphics[width=.66\columnwidth]{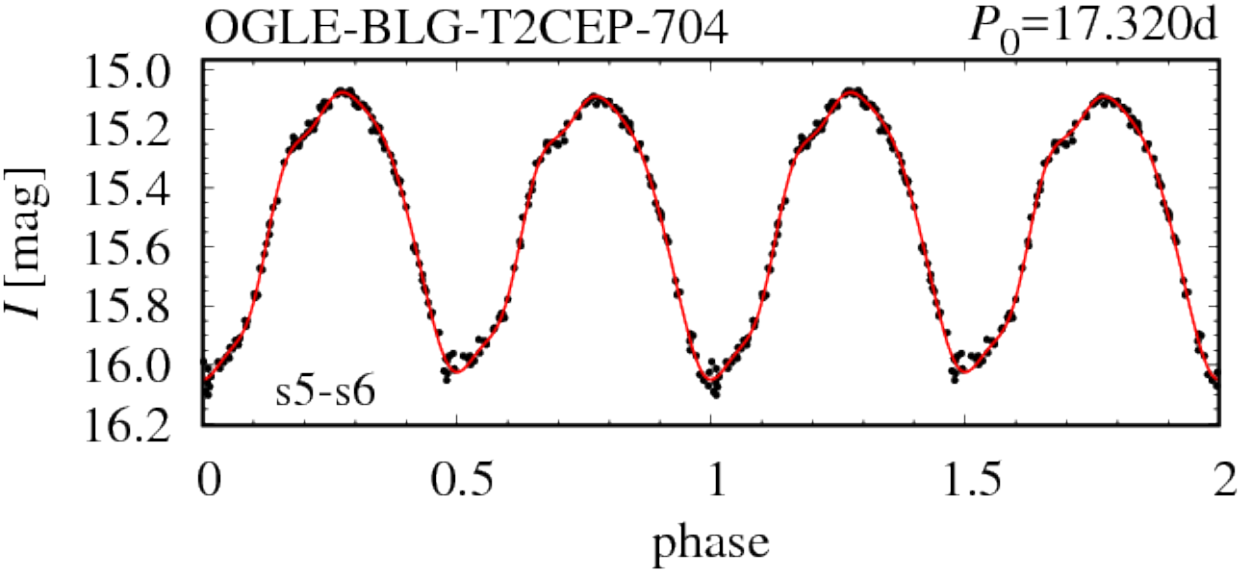} \\
\includegraphics[width=.66\columnwidth]{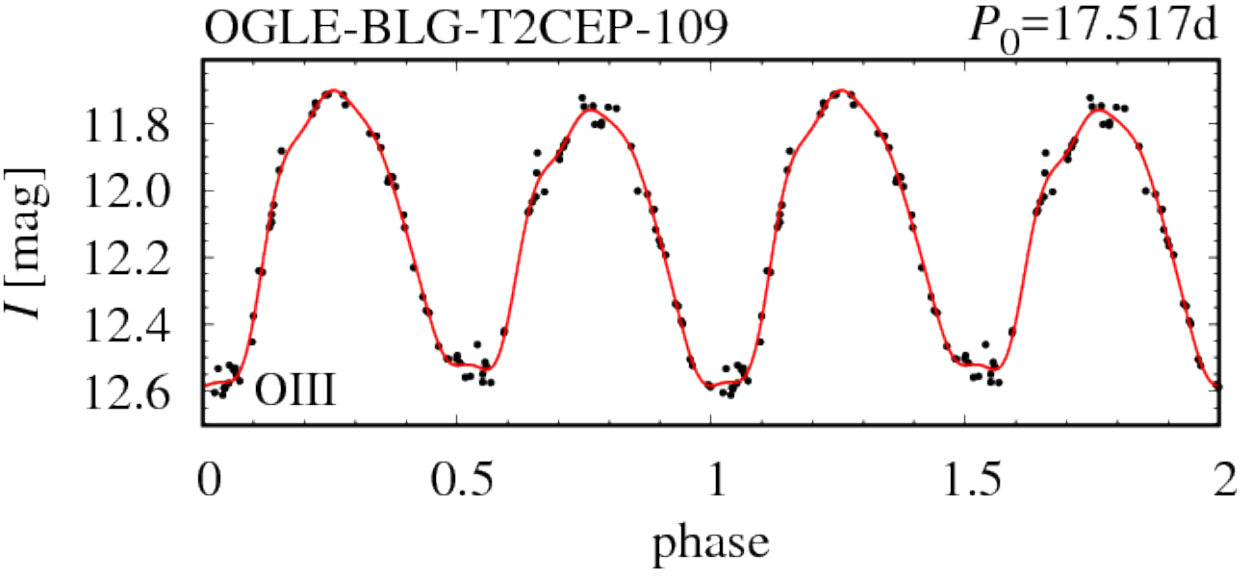} 
\includegraphics[width=.66\columnwidth]{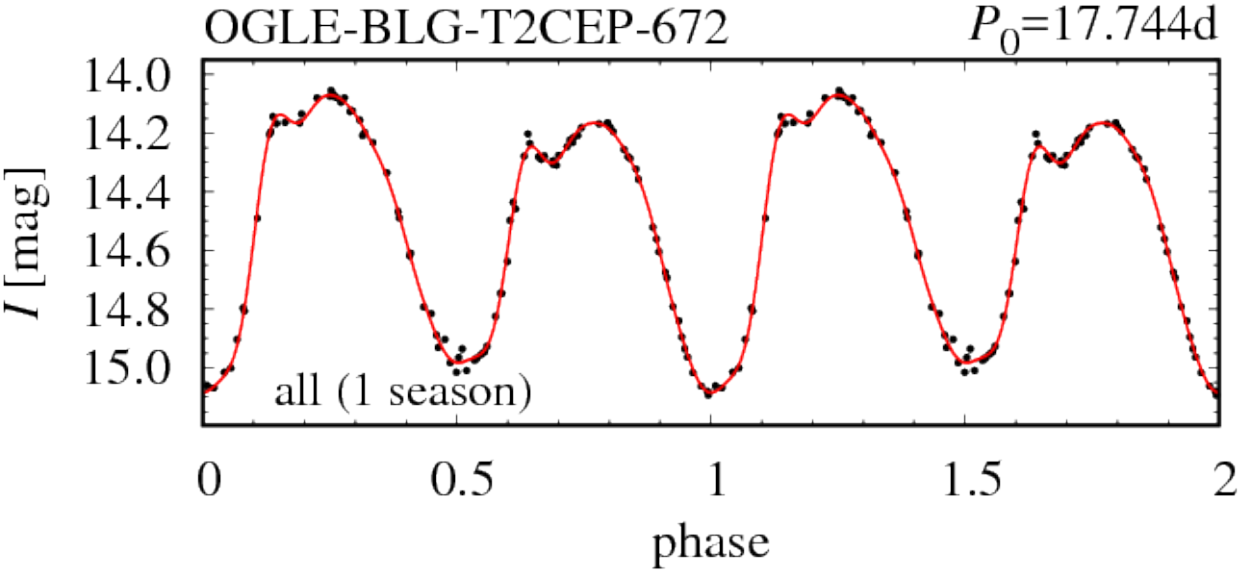}
\includegraphics[width=.66\columnwidth]{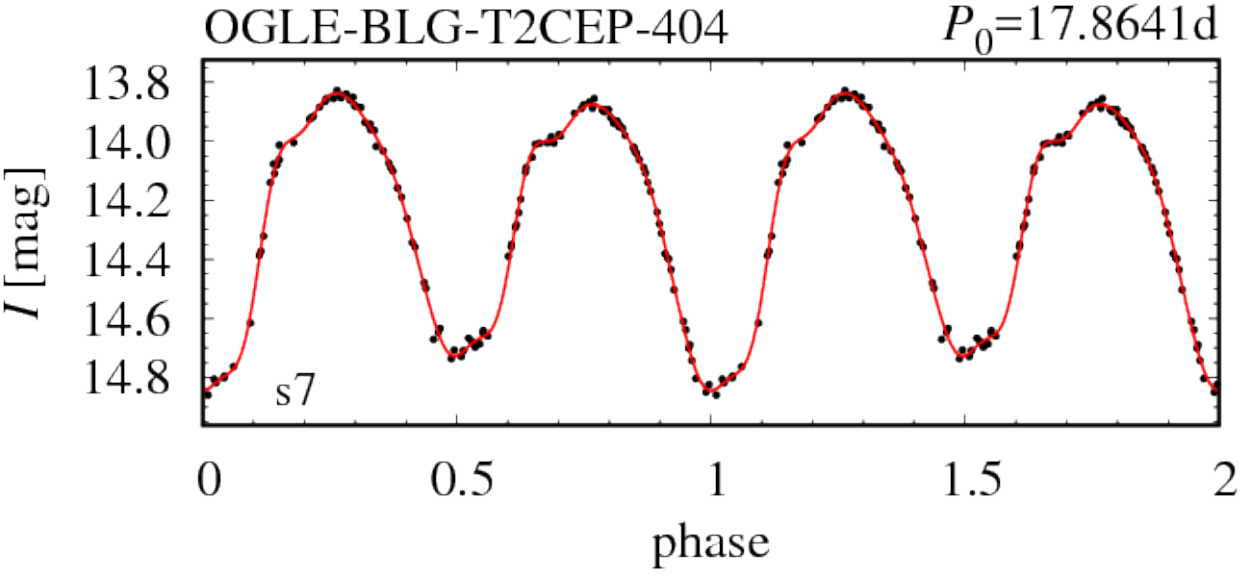} \\
\includegraphics[width=.66\columnwidth]{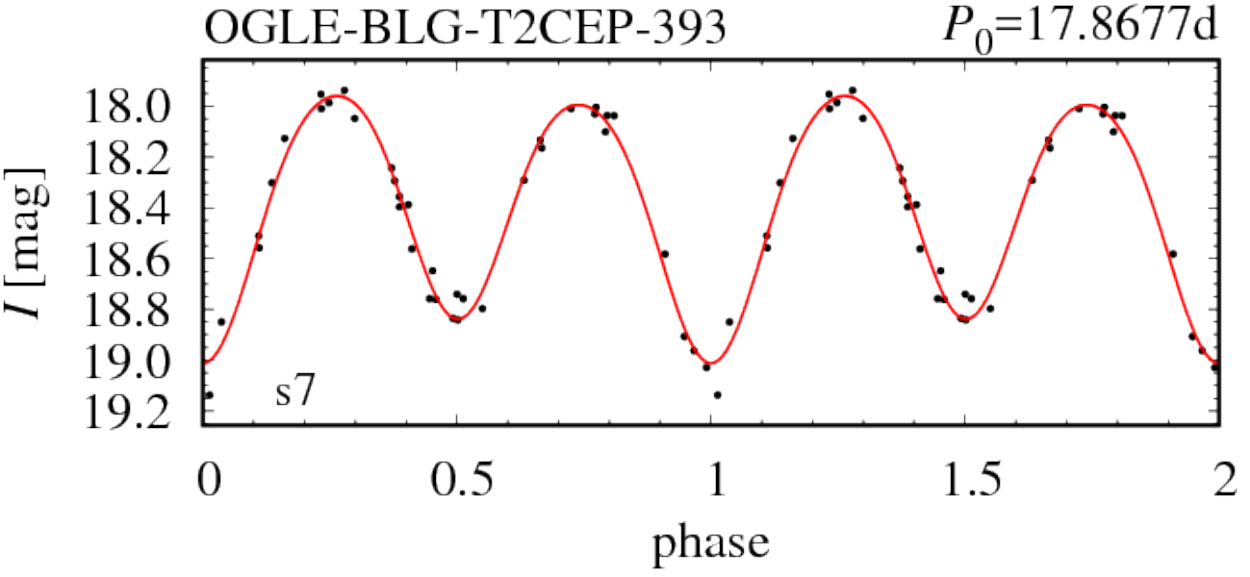}   
\includegraphics[width=.66\columnwidth]{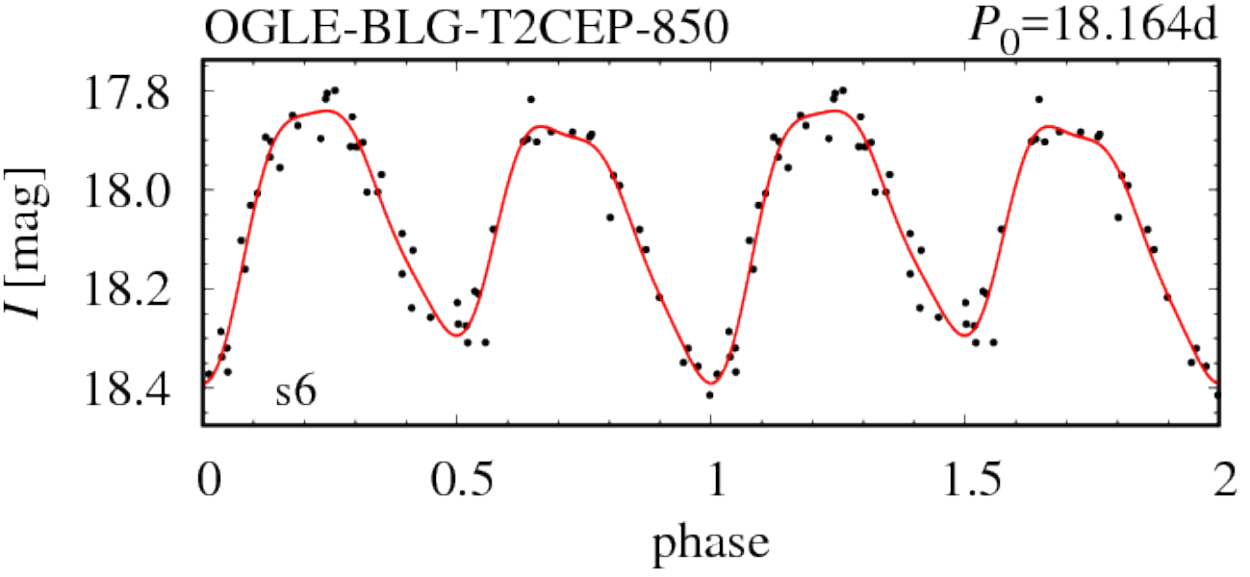}
\includegraphics[width=.66\columnwidth]{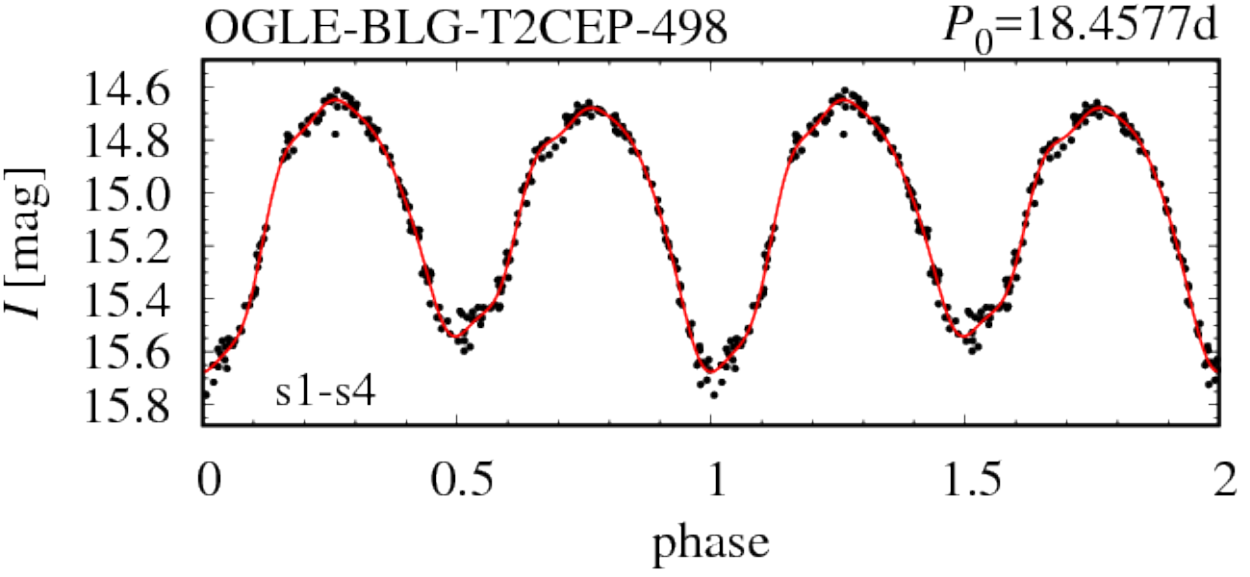}\\   
\includegraphics[width=.66\columnwidth]{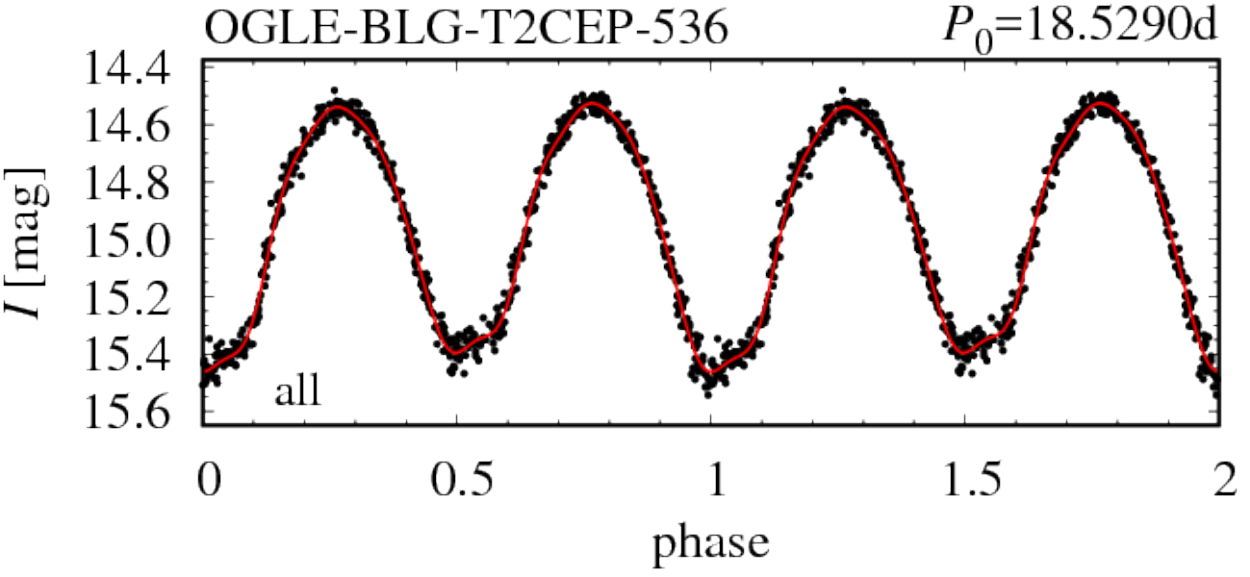}
\includegraphics[width=.66\columnwidth]{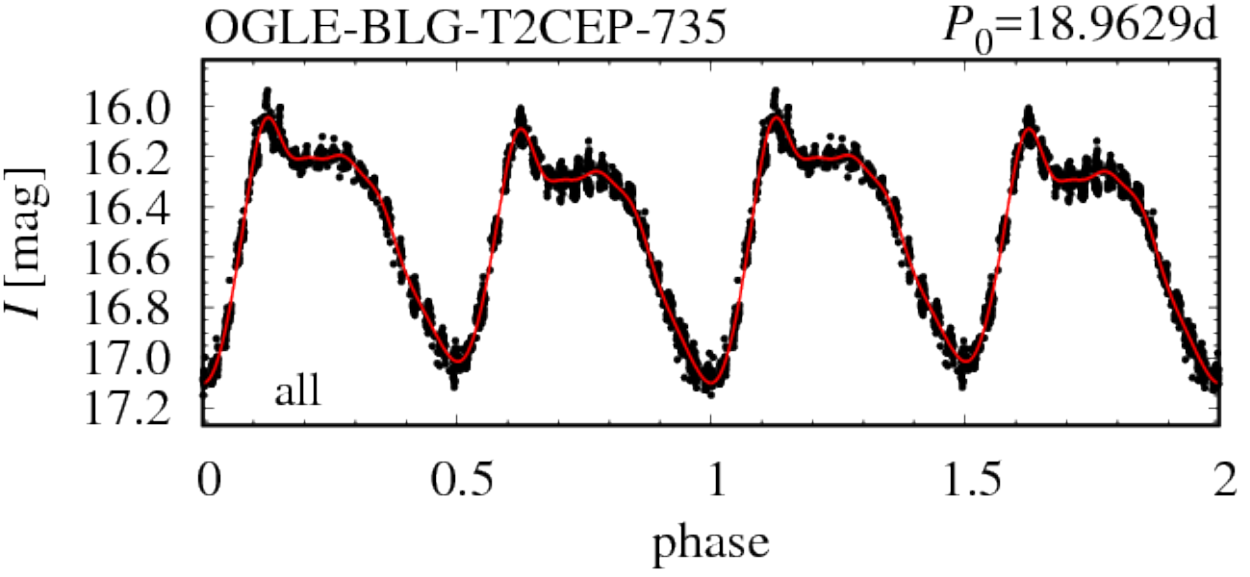}
\includegraphics[width=.66\columnwidth]{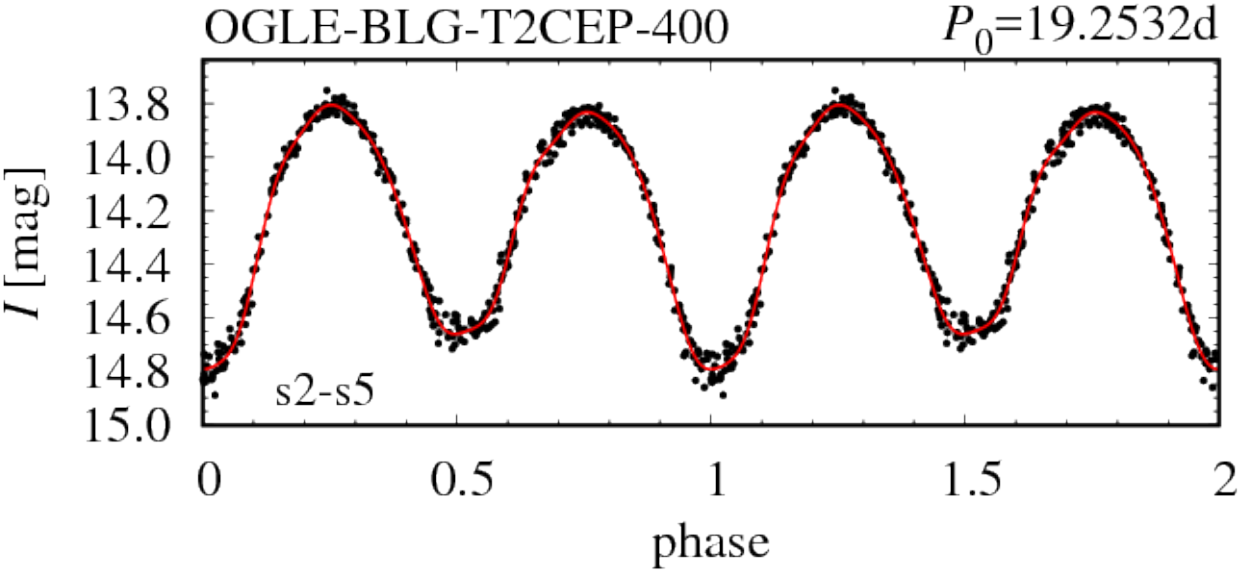}\\   
\includegraphics[width=.66\columnwidth]{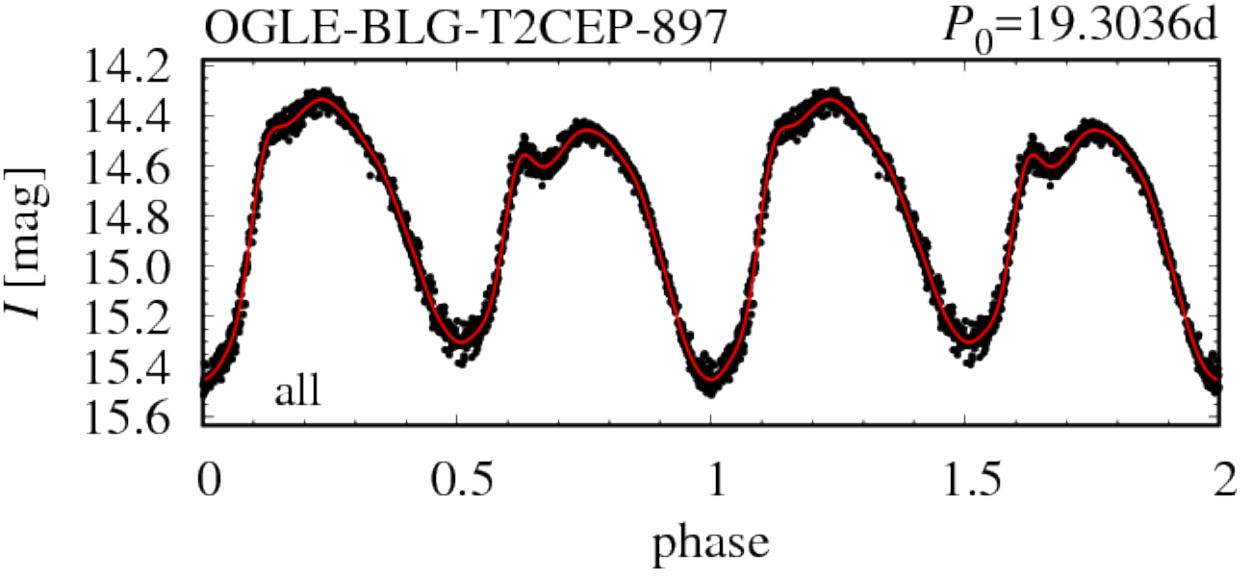}   
\includegraphics[width=.66\columnwidth]{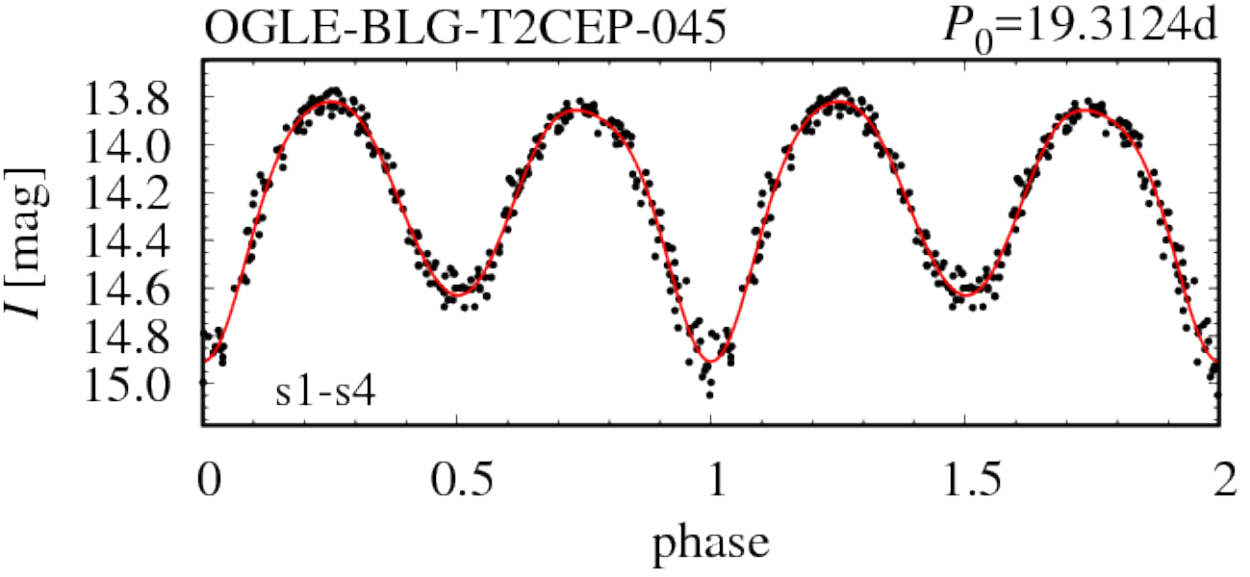}   
\includegraphics[width=.66\columnwidth]{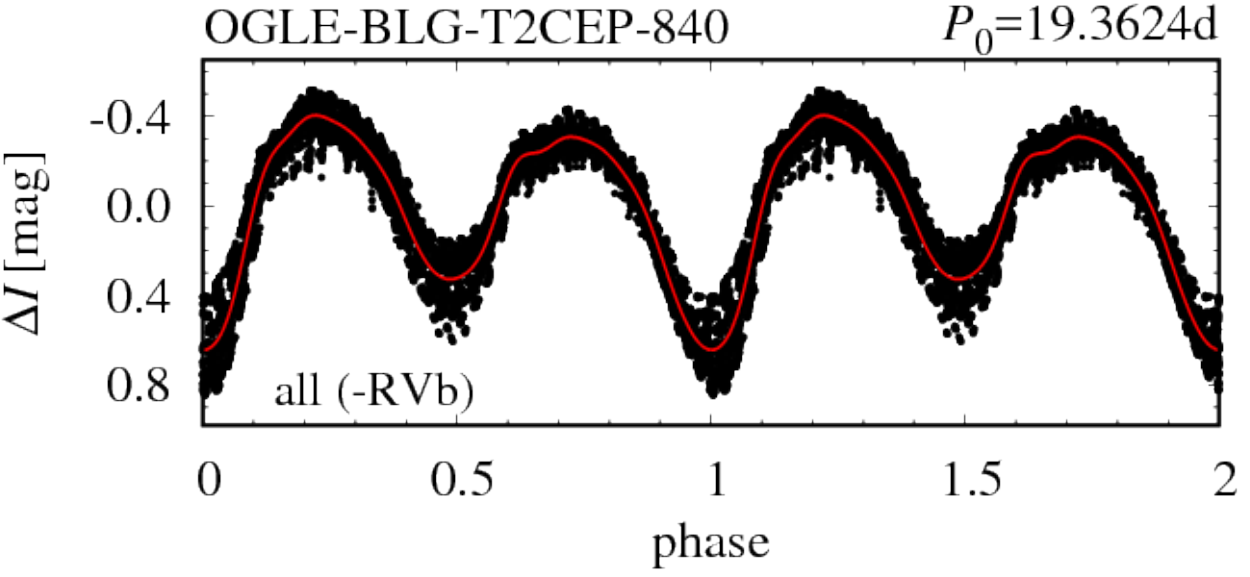}\\
\includegraphics[width=.66\columnwidth]{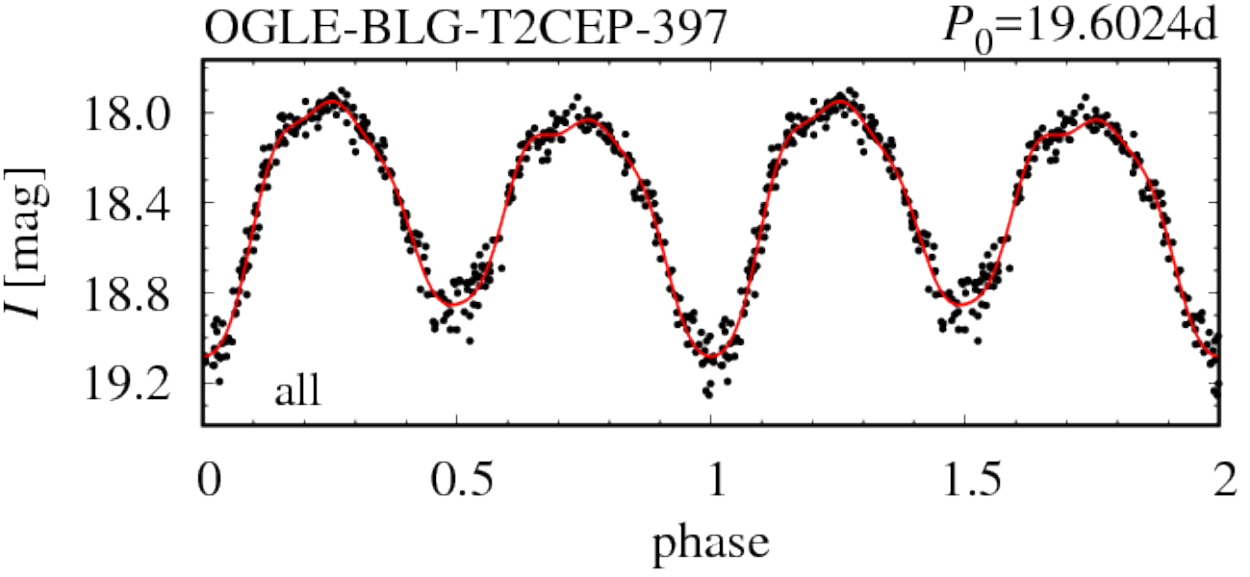}
\caption{A collection of light curves for type II Cepheids with periods in the $15-20$\,d range that show the PD effect. Star's ID, pulsation period and data that were used in the plot are given in each panel. Stars are sorted by the increasing pulsation period.}
\label{fig:wvirpd_lc}
\end{figure*}

\begin{figure*}
\includegraphics[width=.66\columnwidth]{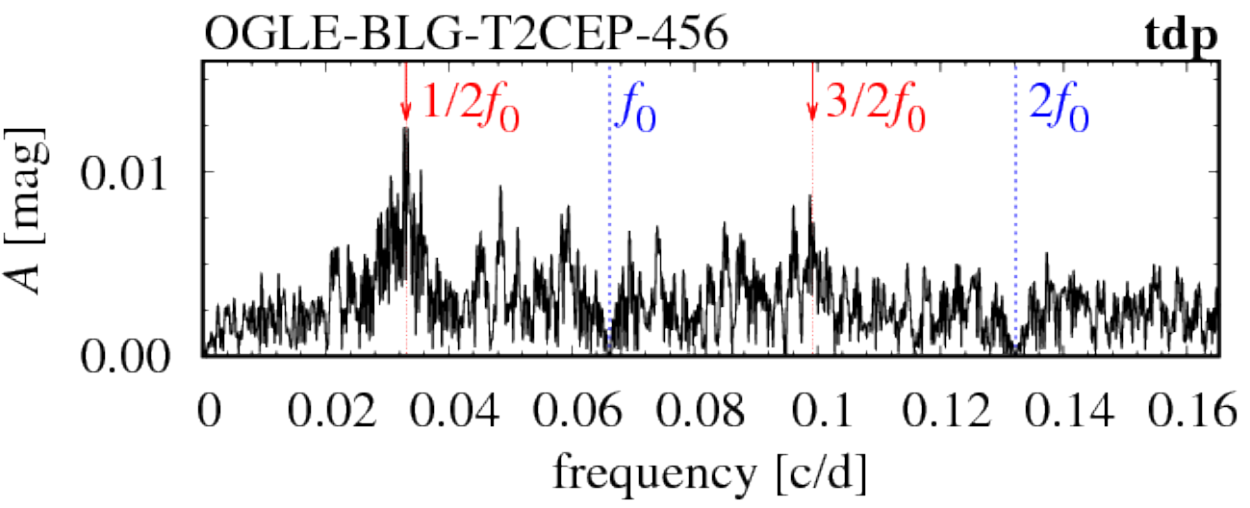} 
\includegraphics[width=.66\columnwidth]{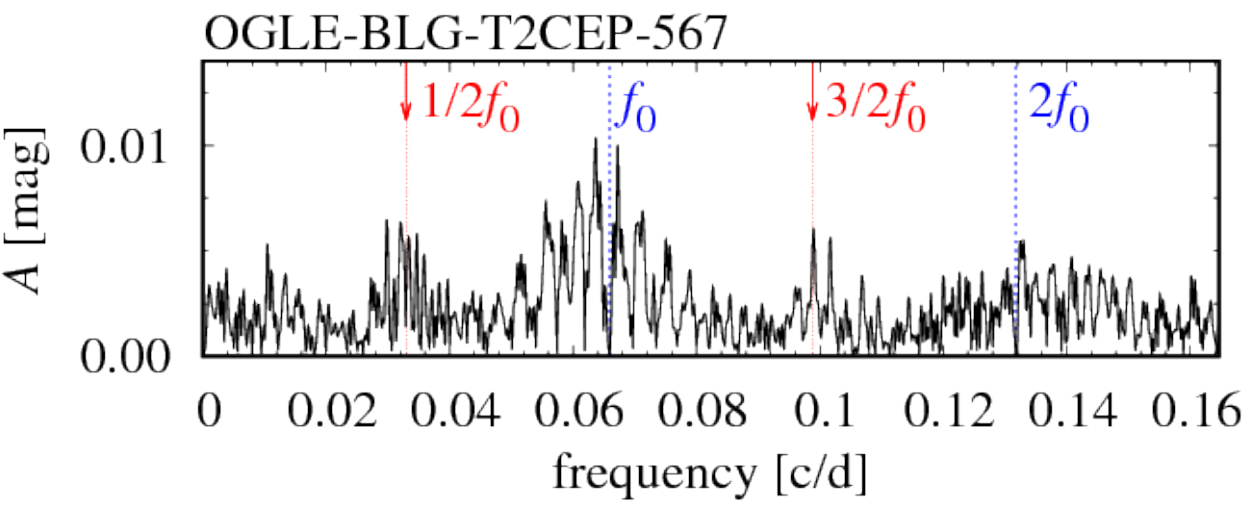}    
\includegraphics[width=.66\columnwidth]{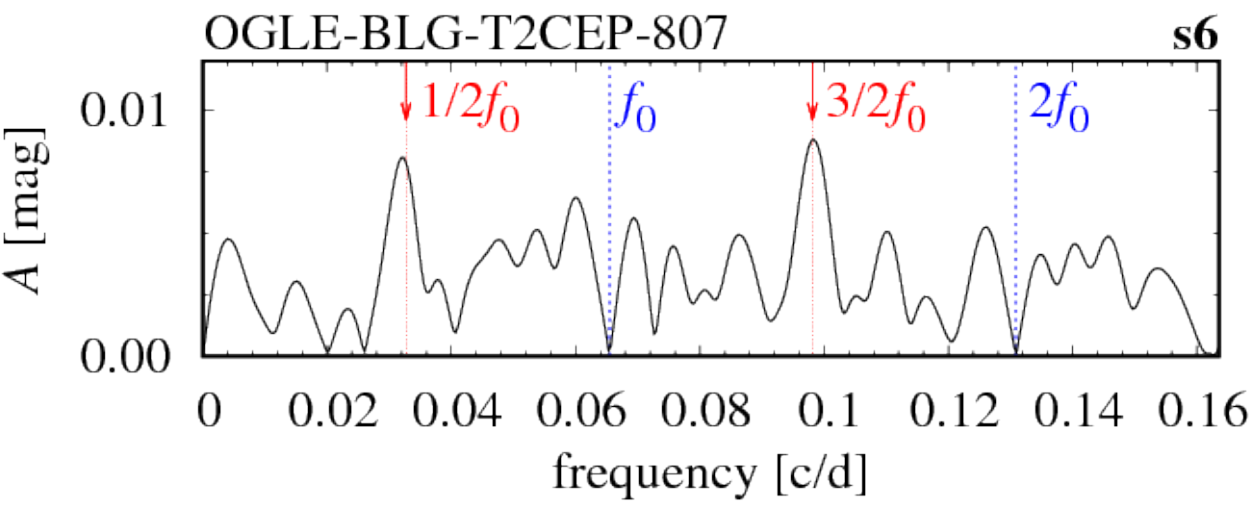}\\ 
\includegraphics[width=.66\columnwidth]{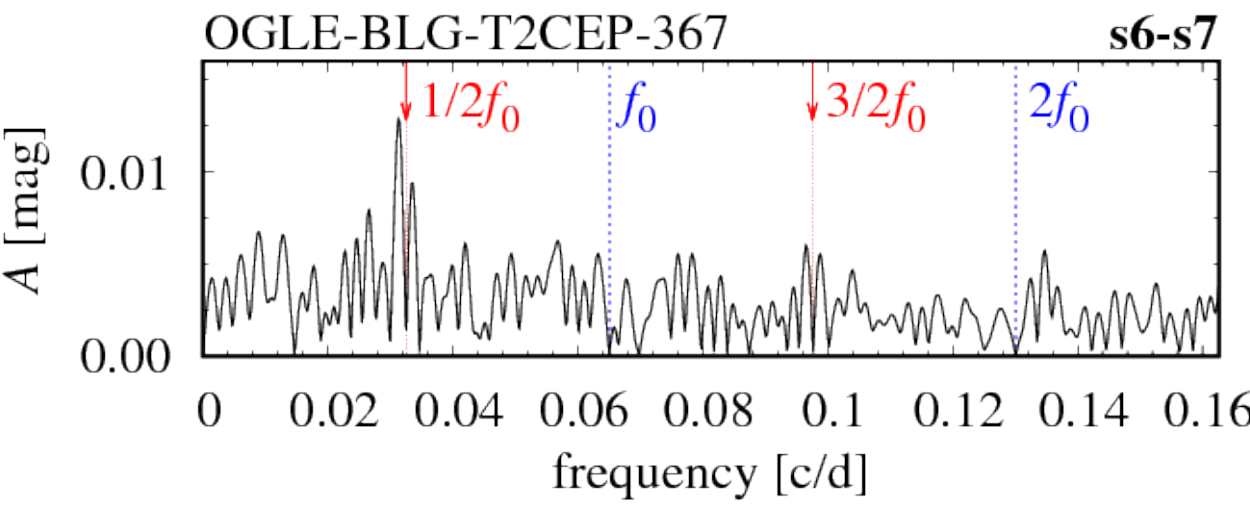}
\includegraphics[width=.66\columnwidth]{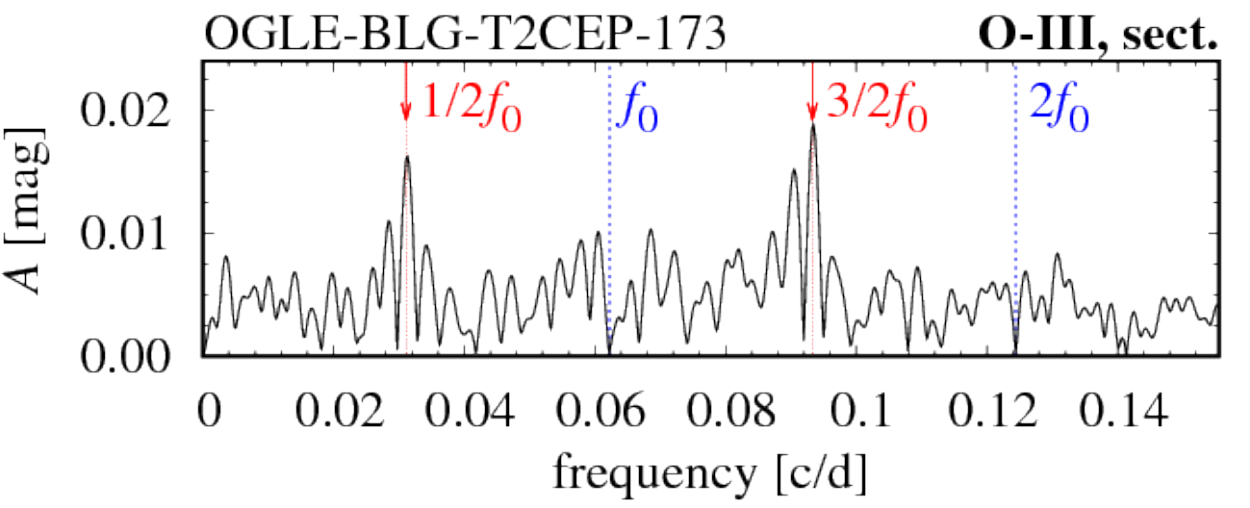} 
\includegraphics[width=.66\columnwidth]{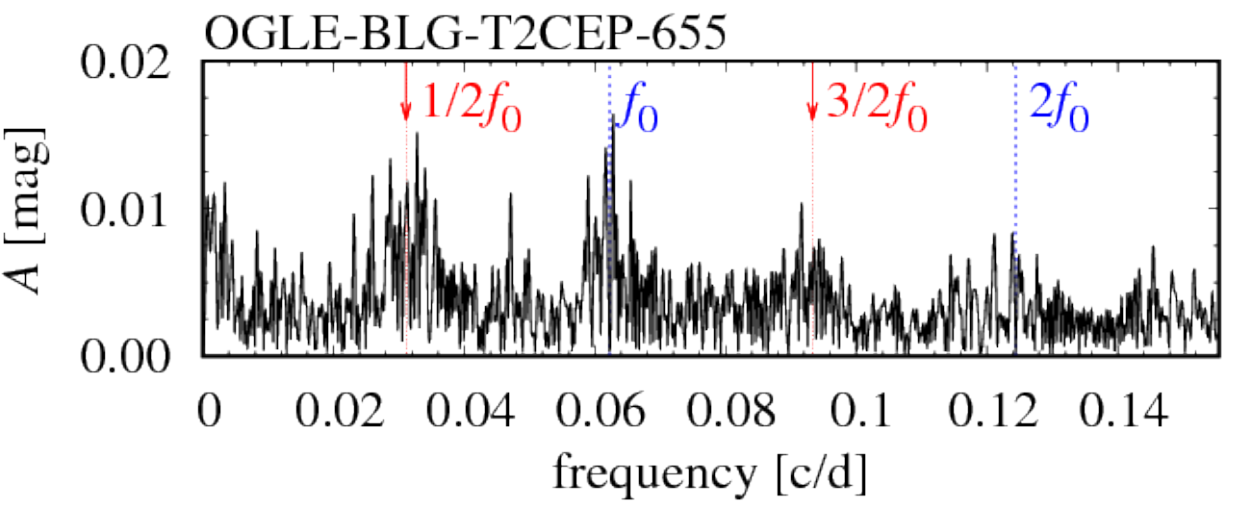}  \\  
\includegraphics[width=.66\columnwidth]{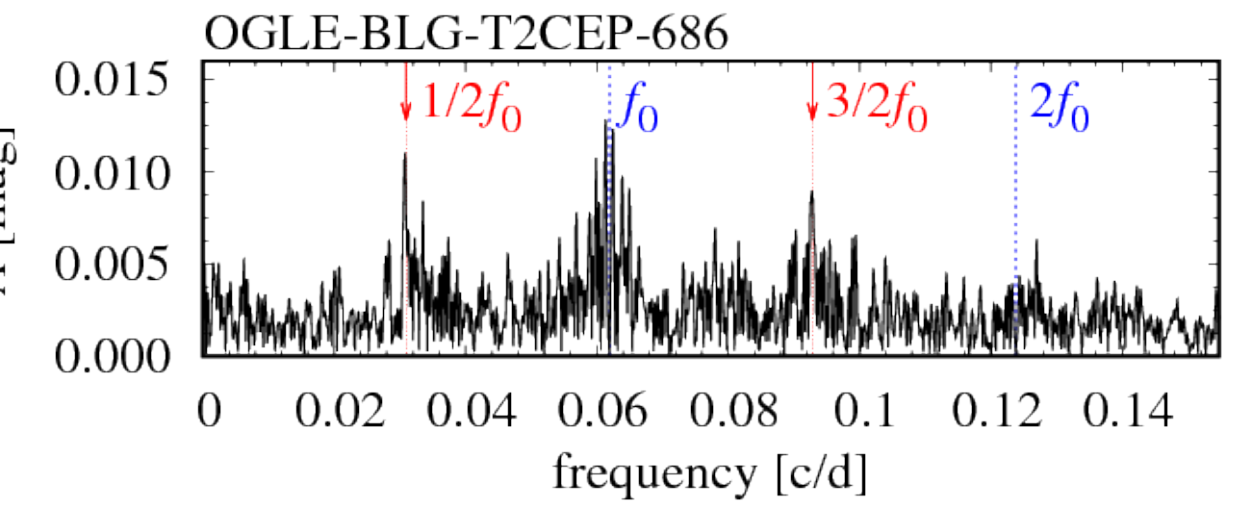}
\includegraphics[width=.66\columnwidth]{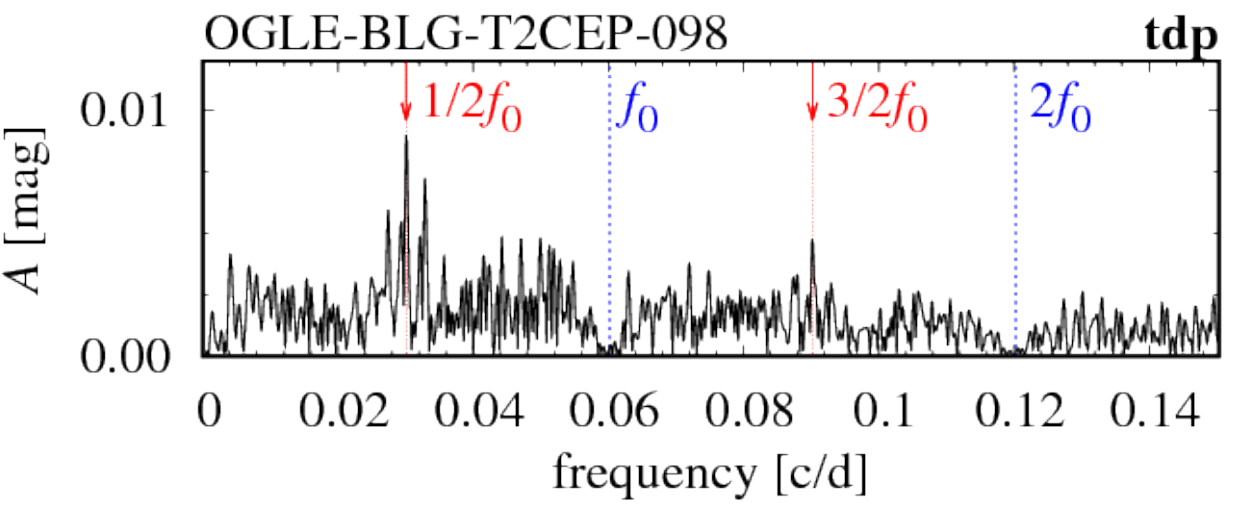}
\includegraphics[width=.66\columnwidth]{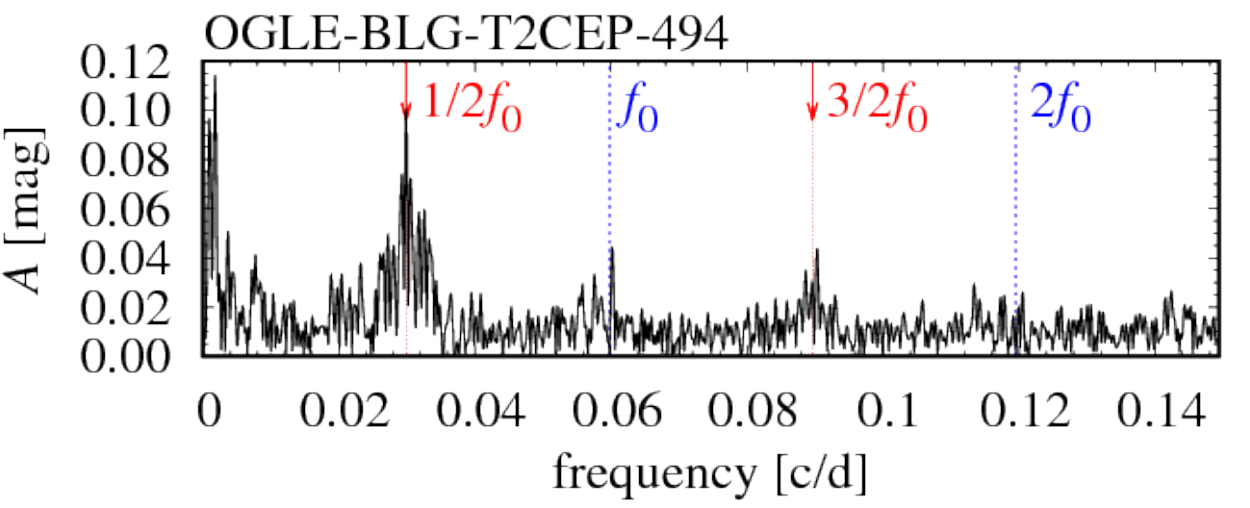}  \\
\includegraphics[width=.66\columnwidth]{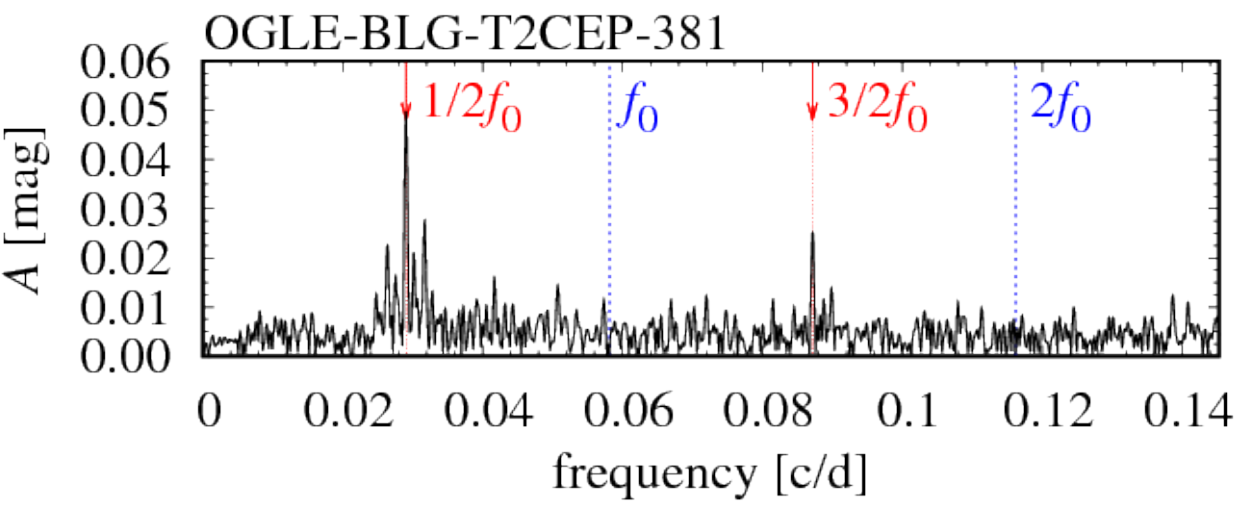}  
\includegraphics[width=.66\columnwidth]{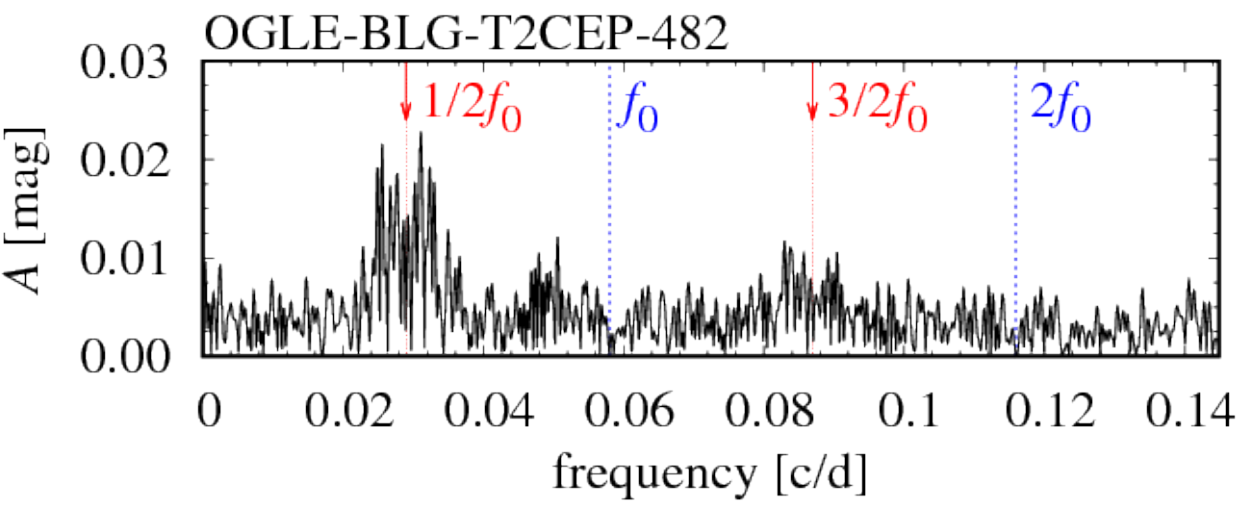}
\includegraphics[width=.66\columnwidth]{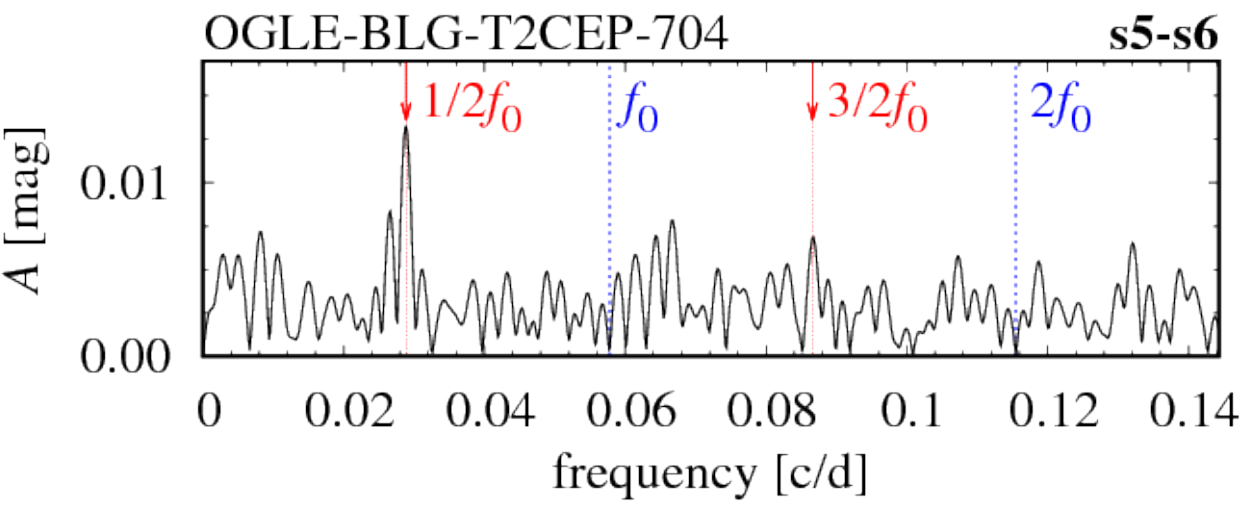}\\
\includegraphics[width=.66\columnwidth]{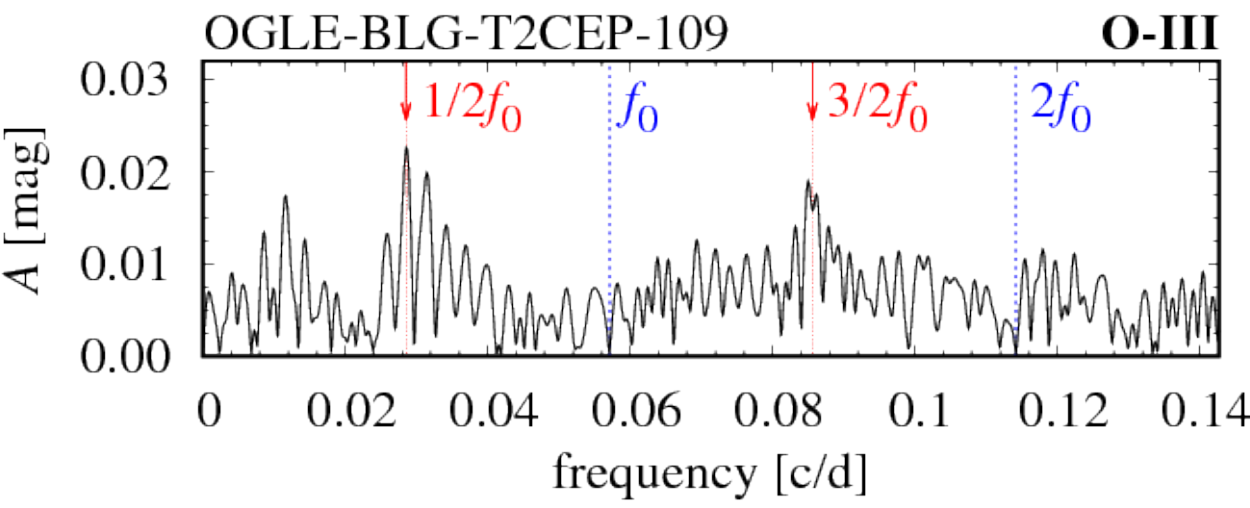} 
\includegraphics[width=.66\columnwidth]{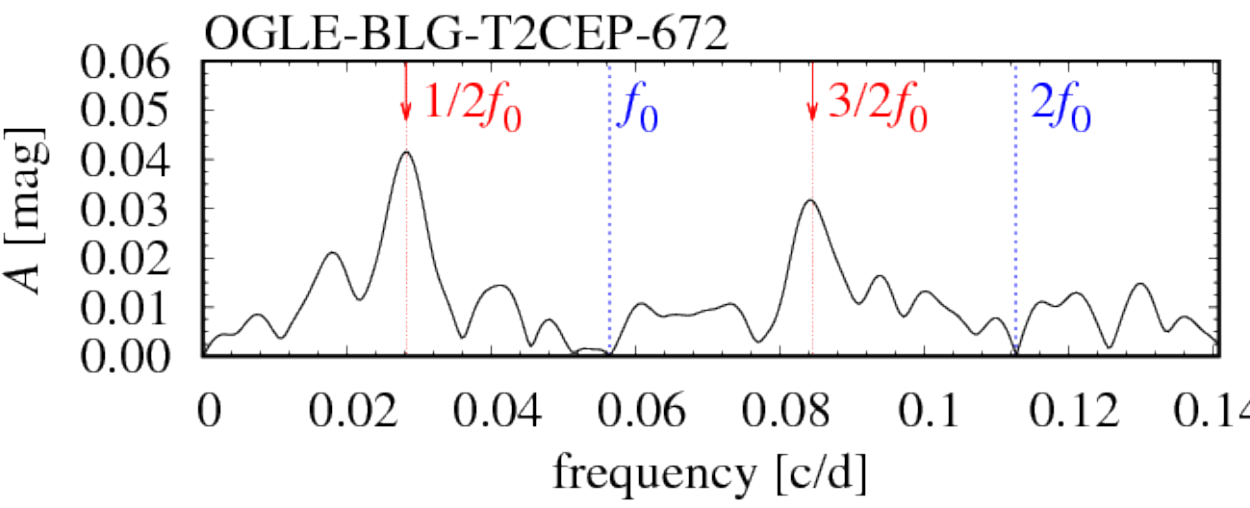} 
\includegraphics[width=.66\columnwidth]{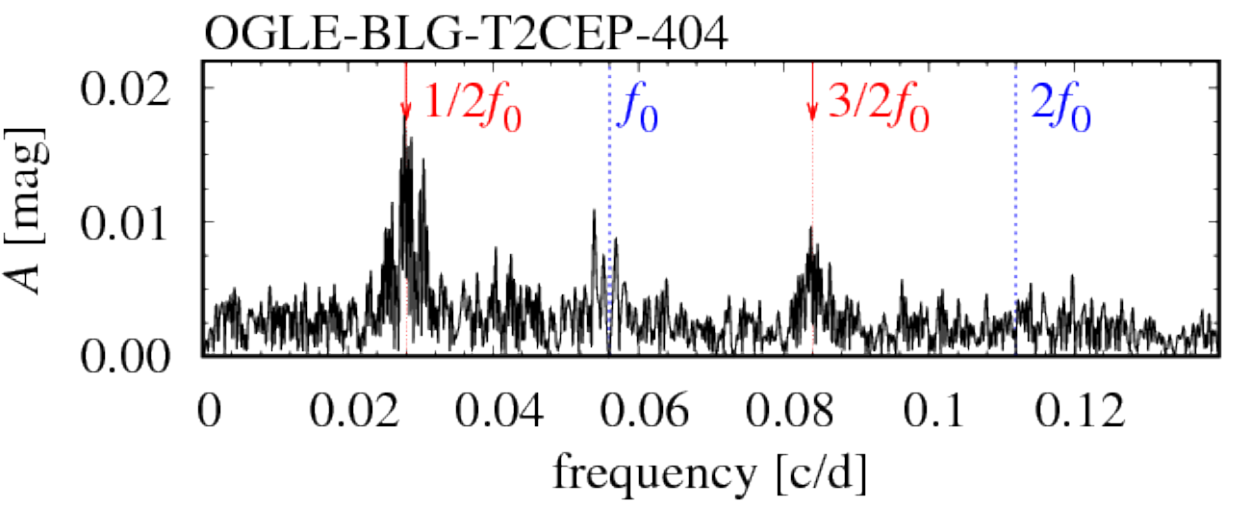}\\ 
\includegraphics[width=.66\columnwidth]{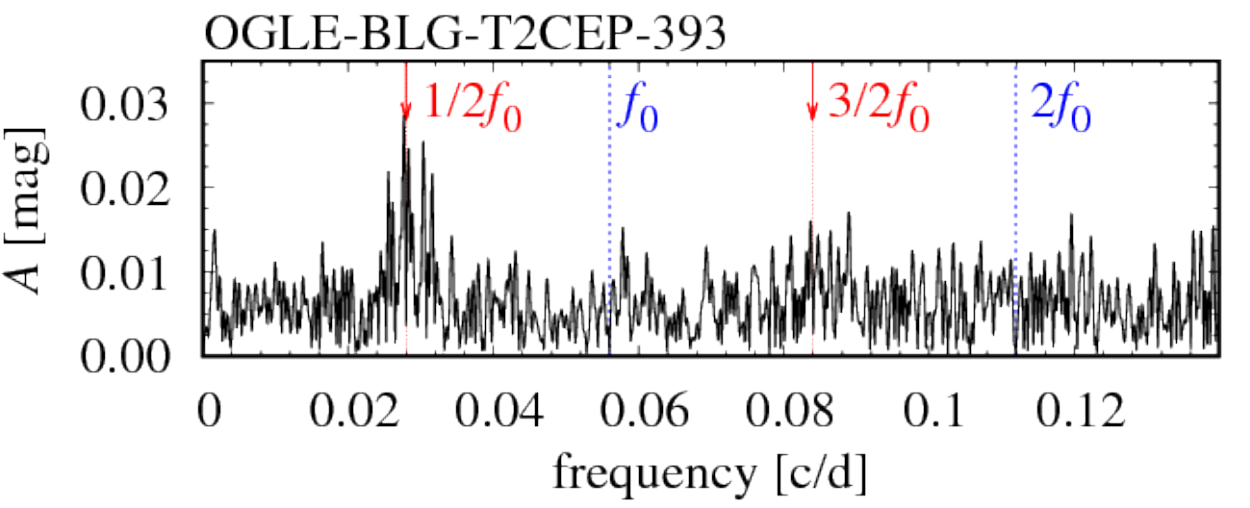}  
\includegraphics[width=.66\columnwidth]{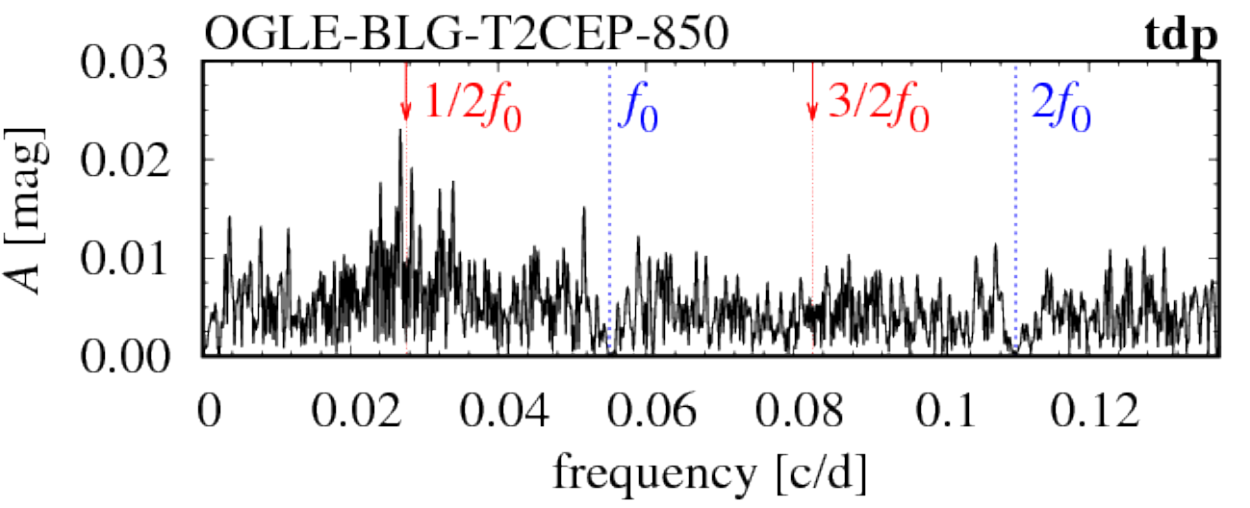}
\includegraphics[width=.66\columnwidth]{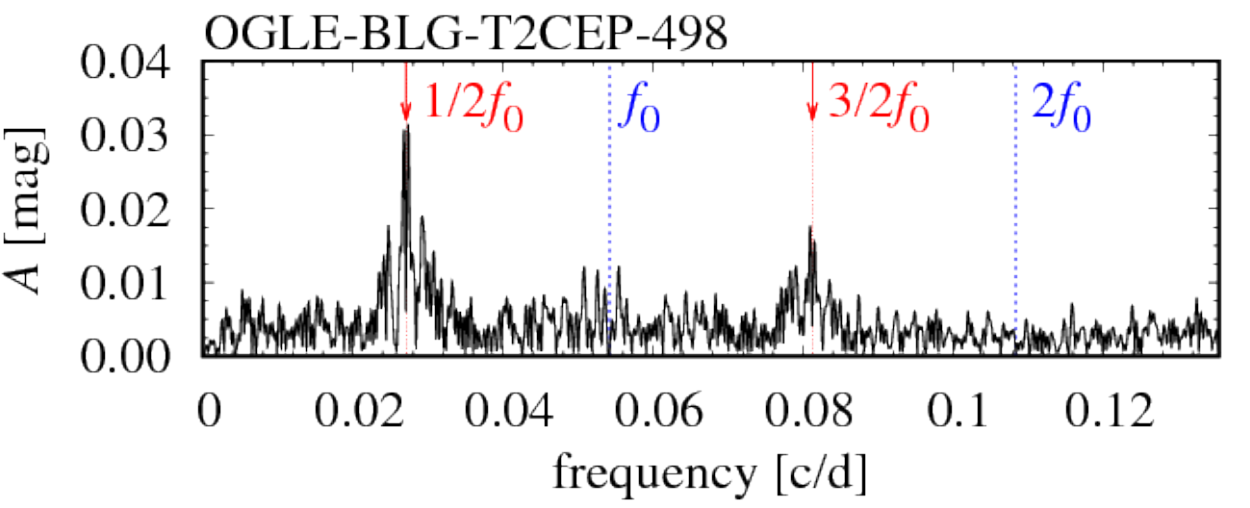}\\  
\includegraphics[width=.66\columnwidth]{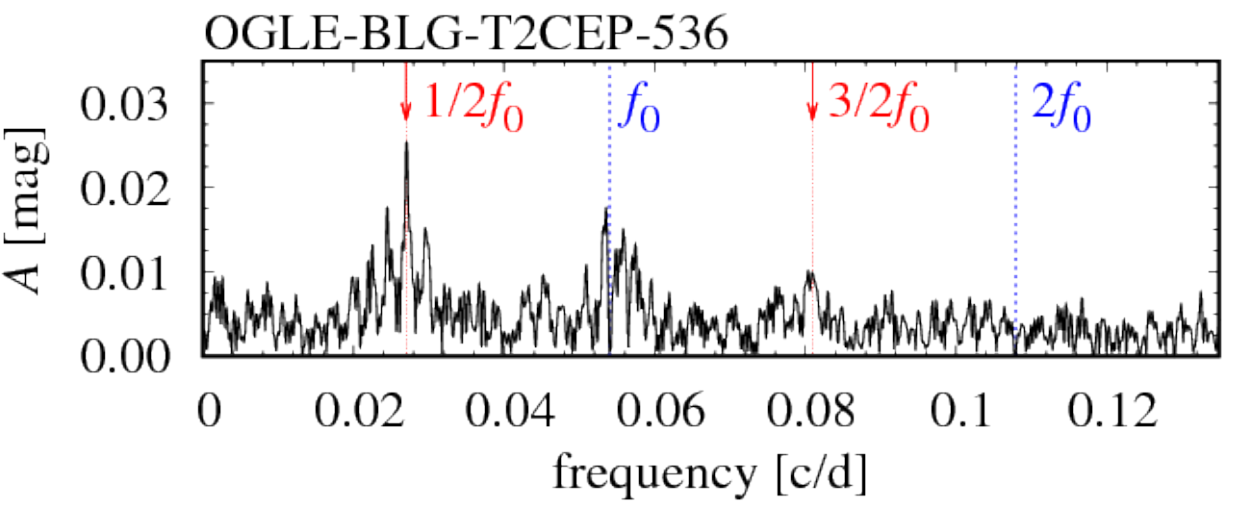}
\includegraphics[width=.66\columnwidth]{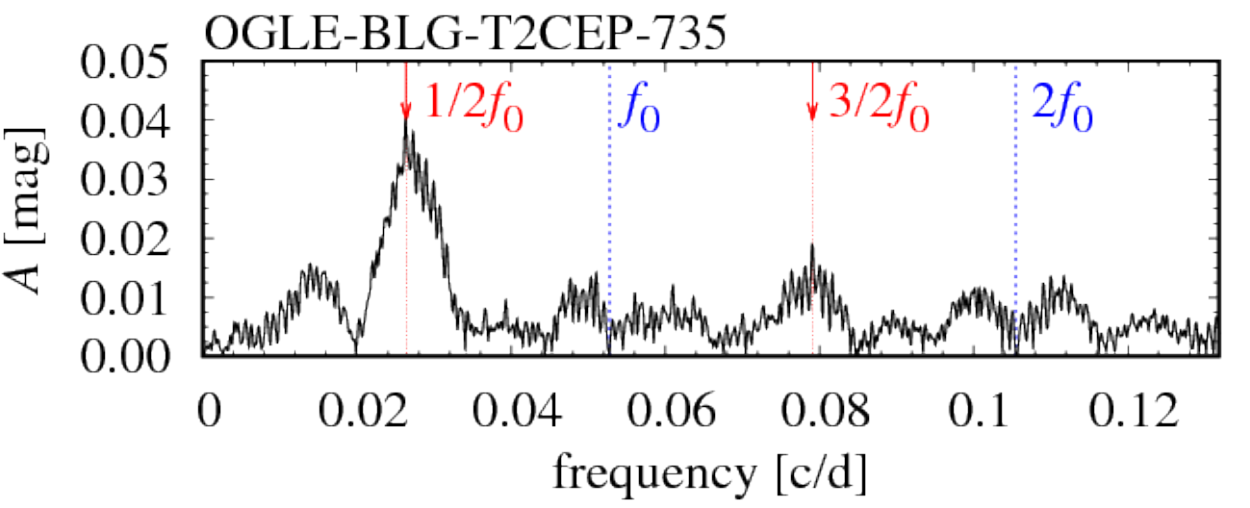}
\includegraphics[width=.66\columnwidth]{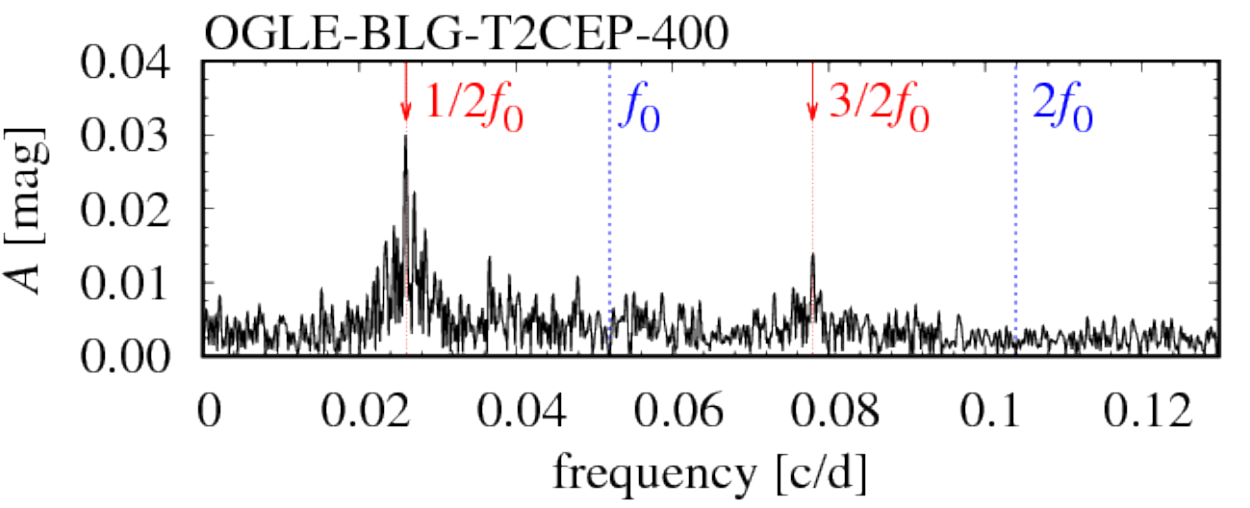}\\  
\includegraphics[width=.66\columnwidth]{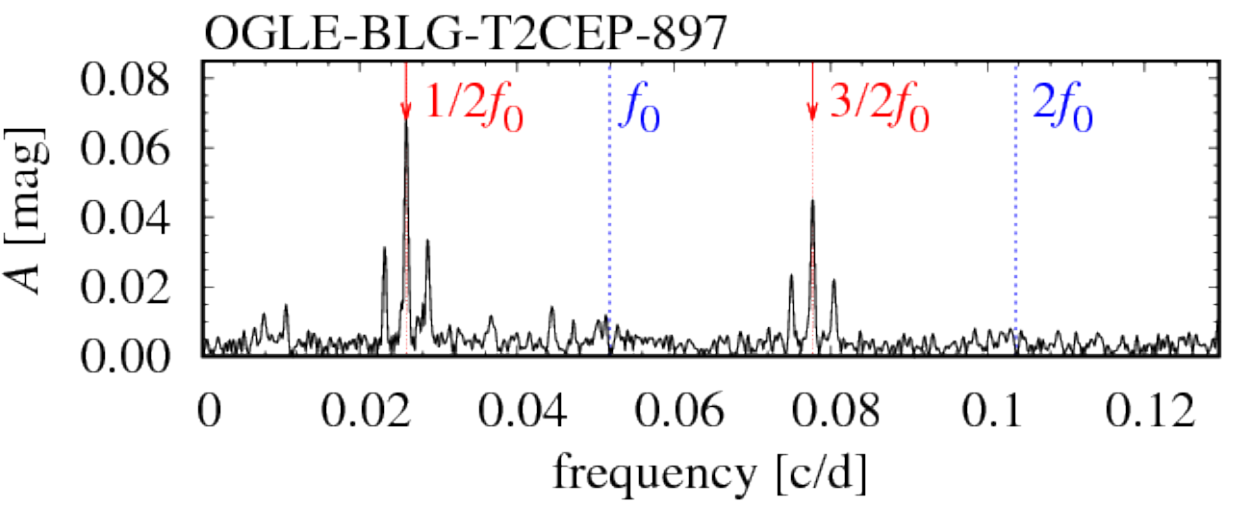}  
\includegraphics[width=.66\columnwidth]{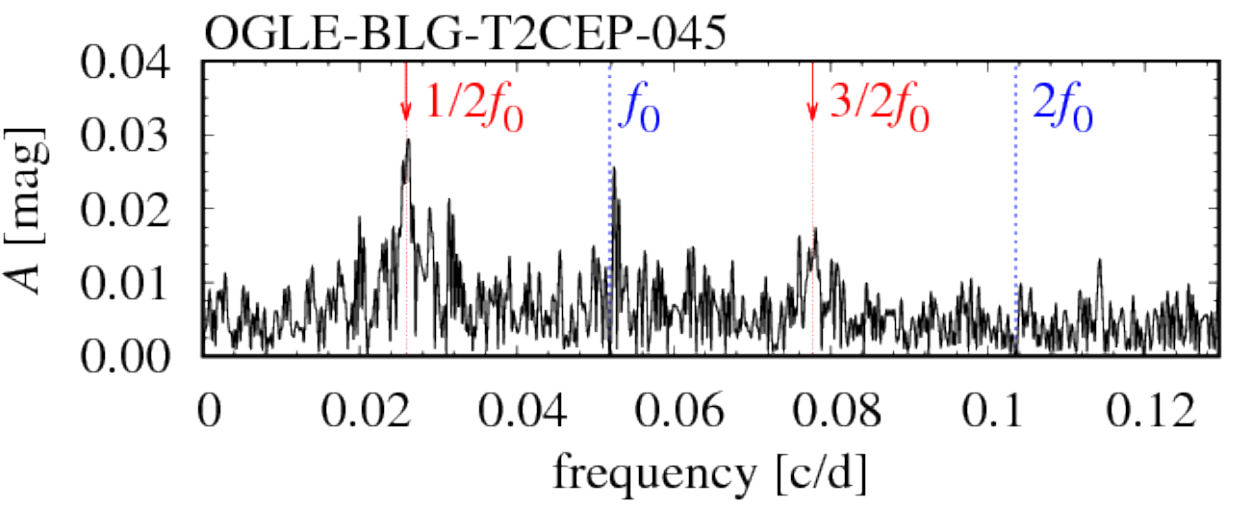}  
\includegraphics[width=.66\columnwidth]{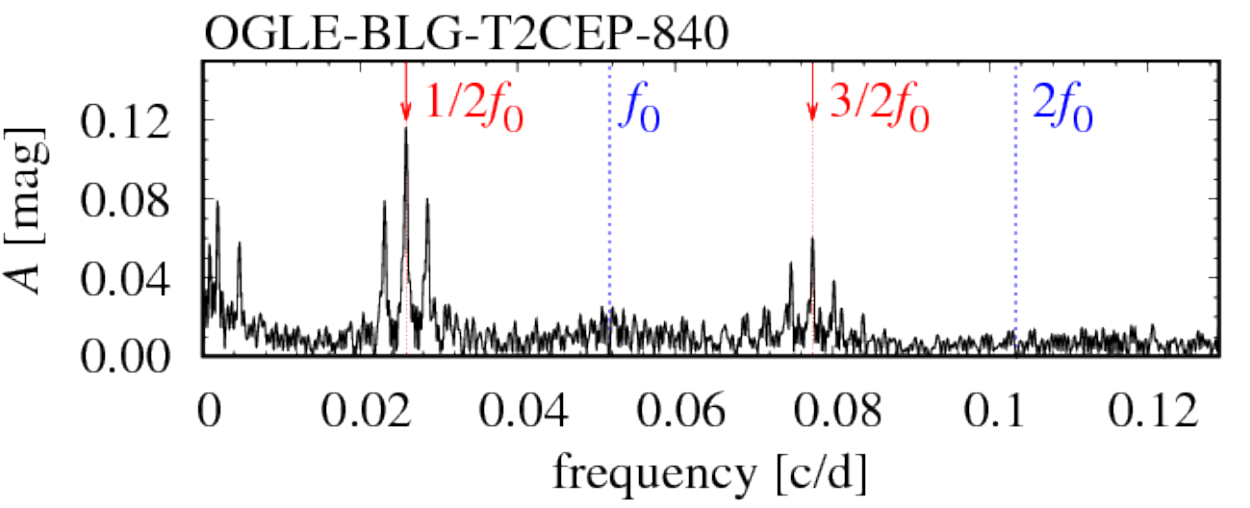}\\ 
\includegraphics[width=.66\columnwidth]{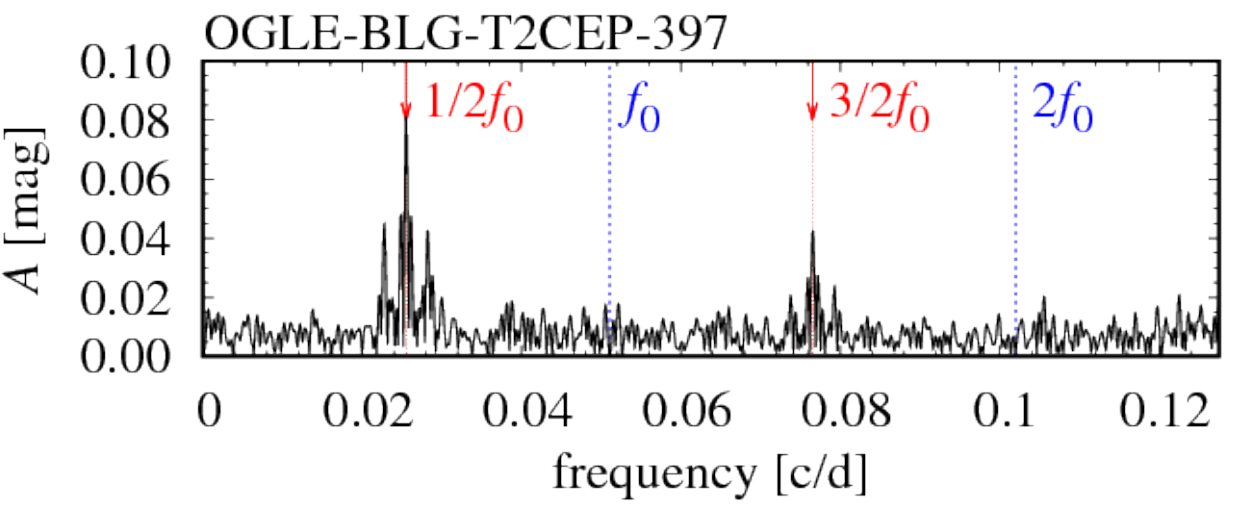}
\caption{A collection of frequency spectra for type II Cepheids with periods in the $15-20$\,d range that show the PD effect. Each panel shows a section of the frequency spectrum in the $(0,\,2.5\fF)$ range after prewhitening with the fundamental mode and its harmonics (blue dashed lines). In some cases, time-dependent prewhitening was used, which is indicated with `tdp' in the top right corner of the plots. Red arrows indicate the location of $\fjd$ and $\ftd$ sub-harmonics. Stars are sorted by the increasing pulsation period.}
\label{fig:wvirpd_fsp}
\end{figure*}

The large number of type II Cepheids in the analysed sample allows us to study the appearance of the PD phenomenon in the transition between W~Vir and RV~Tau classes. In total, we have detected a signature of the period doubling effect in 25 stars with periods shorter than $20$\,d, and thus formally classified as W~Vir stars. Two of them, T2CEP-045 and T2CEP-494, were reported earlier by \cite{blher_ogle, ogle4_gb_t2}. To claim the PD effect we require: (i) the detection of a significant power excess centred at sub-harmonic frequencies in the frequency spectrum of all data or of their section, possibly after time-dependent prewhitening, and (ii) direct detection of the alternations in the light curve for all data or for their section. In Tab.~\ref{tab:wvirpd} we have collected the basic properties of period doubled W~Vir stars. Consecutive columns list: star's ID, period and amplitude of the fundamental mode, $\PF$ and $A_0$, amplitude of the highest peak at $\fjd$, $A_{1/2}$, list of the detected sub-harmonic components (broad structures centred at sub-harmonic frequency are indicated with parenthesis) and remarks. For all the stars light curves folded with $2\PF$ are presented in Fig.~\ref{fig:wvirpd_lc}. In Fig.~\ref{fig:wvirpd_fsp} we show the sections of the corresponding frequency spectra after prewhitening with the fundamental mode and its harmonics (dashed lines). Exact locations of the sub-harmonic frequencies, $\fjd$ and $\ftd$, are marked with arrows and thin dotted lines.
 
\begin{table*}
\centering
\caption{Data for type II Cepheids with periods shorter than $20$\,d, and thus formally classified as W~Vir stars, that show the PD effect. The consecutive columns contain: star's ID, pulsation period, pulsation amplitude, amplitude of the sub-harmonic at $\fjd$, list of all sub-harmonics detected (broad structures centred at sub-harmonic frequency are indicated with parenthesis) and remarks. Amplitudes are derived from the analysis of all OGLE-IV data, unless specific season(s) is(are) indicated in the last column.}
\label{tab:wvirpd}
\begin{tabular}{llllll}
star's ID  & $\PF$\,(d) & $A_0$\,(mag) & $A_{1/2}$\,(mag) & sub-harmonics & remarks\\
\hline
\offi{T2CEP-456}\tech{BLG653.06.2795}  & 15.1251(4) & 0.446(4)  & 0.017(4)  & ($\fjd$)                       & iPD \\ 
\offi{T2CEP-567}\tech{BLG675.13.88344} & 15.1827(1) & 0.467(1)  & 0.007(1)  & ($\fjd$), ($\ftd$)             & iPD \\ 
\offi{T2CEP-807}\tech{BLG500.19.80308} & 15.2838(7) & 0.4451(7) & 0.0081(7) & $\fjd$, $\ftd$                 & iPD, s6 \\ 
\offi{T2CEP-367}\tech{BLG615.05.41624} & 15.3737(7) & 0.439(2)  & 0.013(2)  & ($\fjd$)                       & iPD, s6-s7 \\ 
\offi{T2CEP-173}\tech{}                & 16.076(1)  & 0.458(2)  & 0.017(3)  & $\fjd$, $\ftd$                 & iPD, O-III, s2-s3 \\
\offi{T2CEP-655}\tech{BLG632.23.65428} & 16.0767(2) & 0.418(2)  & 0.015(2)  & ($\fjd$), ($\ftd$)             & iPD, very wide sh \\ 
\offi{T2CEP-686}\tech{BLG648.30.64865} & 16.1573(2) & 0.445(2)  & 0.011(2)  & $\fjd$, $\ftd$                 & sPD \\ 
\offi{T2CEP-098}\tech{BLG501.23.40080} & 16.62594(6)& 0.4679(5) & 0.0088(4) & $\fjd$, $\ftd$                 & iPD \\ 
\offi{T2CEP-494}\tech{BLG653.03.55060} & 16.7359(6) & 0.502(6)  & 0.102(6)  & $\fjd$, $\ftd$                 & sPD, RVb \\ 
\offi{T2CEP-381}\tech{BLG668.06.27652} & 17.2108(4) & 0.436(3)  & 0.050(3)  & $\fjd$, $\ftd$                 &   \\ 
\offi{T2CEP-482}\tech{BLG670.18.34156} & 17.2522(4) & 0.415(3)  & 0.024(3)  & ($\fjd$), ($\ftd$)             & very wide sh    \\ 
\offi{T2CEP-704}\tech{BLG660.20.6883}  & 17.320(1)  & 0.476(2)  & 0.013(2)  & $\fjd$                         & iPD, s5-s6 \\ 
\offi{T2CEP-109}\tech{}                & 17.517(2)  & 0.430(5)  & 0.030(5)  & $\fjd$, $\ftd$                 & O-III (two seasons only) \\ 
\offi{T2CEP-672}\tech{BLG683.06.111544}& 17.744(6)  & 0.463(3)  & 0.048(3)  & $\fjd$, $\ftd$                 & one season only \\ 
\offi{T2CEP-404}\tech{BLG613.20.10348} & 17.8641(2) & 0.468(2)  & 0.018(2)  & ($\fjd$), ($\ftd$)             & sPD \\ 
\offi{T2CEP-393}\tech{BLG668.11.18080} & 17.8677(8) & 0.493(5)  & 0.028(5)  & ($\fjd$)                       & iPD, poor data  \\ 
\offi{T2CEP-850}\tech{BLG643.17.22111} & 18.164(2)  & 0.234(6)  & 0.024(6)  & ($\fjd$)                       & iPD, poor data  \\ 
\offi{T2CEP-498}\tech{BLG654.28.40781} & 18.4577(3) & 0.471(2)  & 0.032(2)  & ($\fjd$), ($\ftd$)             & sPD \\ 
\offi{T2CEP-536}\tech{BLG653.17.5511}  & 18.5290(3) & 0.462(2)  & 0.027(2)  & $\fjd$, $\ftd$                 & iPD \\ 
\offi{T2CEP-735}\tech{BLG501.16.66184} & 18.9629(2) & 0.411(1)  & 0.042(1)  & $\fjd$, $\ftd$                 & poor data, pec. window fun.\\ 
\offi{T2CEP-400}\tech{BLG613.28.36418} & 19.2532(3) & 0.469(2)  & 0.030(2)  & $\fjd$, $\ftd$                 & sPD \\ 
\offi{T2CEP-897}\tech{BLG520.17.89531} & 19.3036(2) & 0.483(1)  & 0.069(1)  & $\fjd$, $\ftd$, $\fpd$, $\fsd$ &     \\ 
\offi{T2CEP-045}\tech{BLG633.29.90299} & 19.3124(7) & 0.463(4)  & 0.030(4)  & $\fjd$, $\ftd$                 & sPD \\ 
\offi{T2CEP-840}\tech{BLG505.21.205063}& 19.3624(2) & 0.4187(9) & 0.1136(9) & $\fjd$, $\ftd$, $\fpd$         & RVb \\ 
\offi{T2CEP-397}\tech{BLG672.15.18746} & 19.6024(9) & 0.492(5)  & 0.084(5)  & $\fjd$, $\ftd$                 &     \\ 
 \hline  
\end{tabular}
\end{table*}

We note that T2CEP-672 was observed for only one season, hence the peaks in the frequency spectrum in Fig.~\ref{fig:wvirpd_fsp} are particularly wide. For T2CEP-735 scarce data are available for all observing seasons, but only two seasons are densely sampled, which gives rise to peculiar appearance of peaks in its frequency spectrum. Nevertheless, for both stars the PD effect is clearly visible in the light curves as Fig.~\ref{fig:wvirpd_lc} shows (for T2CEP-735 alternations are best visible for brightness maxima).

In T2CEP-494 and T2CEP-840 additional long-period variation is superimposed on the pulsation light curves -- a phenomenon observed in RV~Tau stars of the `b'-type (RVb). It was subtracted from the light curves displayed in Fig.~\ref{fig:wvirpd_lc}.

The PD effect becomes more regular as period increases. This is well visible in the frequency spectra shown in Fig.~\ref{fig:wvirpd_fsp}. For stars with shorter periods, below $\approx$18\,d, the peaks located at sub-harmonic frequencies are typically non-coherent (remnant power remains after the prewhitening), or we observe more complex structures centred at sub-harmonic frequencies (e.g. wide structures in T2CEP-655, T2CEP-494, or in T2CEP-482). In the time domain, such structures correspond either to the transient (PD is present in only some of the observing seasons) or irregular nature of the effect (e.g. varying depth of alternations; `iPD' in the remarks column of Tab.~\ref{tab:wvirpd}), or to the switching of the deep and shallow pulsation cycles (`sPD' in the remarks column of Tab.~\ref{tab:wvirpd}). For this reason, for some stars displayed in Fig.~\ref{fig:wvirpd_lc}, we used data for specific observing seasons only (indicated in the bottom left corners of the panels) to better visualise the PD effect. Switching of the deep and shallow cycles is detected in T2CEP-686, T2CEP-494, T2CEP-404, T2CEP-498, T2CEP-400 and T2CEP-045. This phenomenon is frequent in longer-period RV~Tau stars and will be discussed in more detail in Sect.~\ref{ssec:rvt-switch}.

This numerous sample of period-doubled W~Vir stars allows us to demonstrate that the emergence of the PD effect is a gradual process that starts at pulsation periods slightly above 15\,d. This is illustrated in Tab.~\ref{tab:transition}. We have divided the sample into five bins with pulsation periods in the range $15-16$\,d, $16-17$\,d, till $19-20$\,d, as given in the first column of Tab.~\ref{tab:transition}. The second column shows the number of period doubled stars within a bin, the third column shows the total number of stars in a bin and the fourth column gives the corresponding incidence rate. In all bins  a few (4--7) period doubled stars are present. But the overall number of stars rapidly decreases as pulsation period gets longer. Consequently, the incidence rate of the period doubling effect increases from 12 per cent ($15-16$\,d) to 70 per cent ($19-20$\,d). For periods above 20\,d the period doubling effect is common and only in a few stars with periods longer than 20\,d the effect is not present.

We also note that the amplitude of the PD increases with the increasing pulsation period. This is quantified with the average amplitude ratios, $\langle A_{1/2}/A_0\rangle$, computed in the just discussed period bins, given in the last column of Tab.~\ref{tab:transition}. The effect is clear, but not very pronounced.

\begin{table}
\centering
\caption{The occurrence of period doubled W~Vir stars in the 1-d wide period bins defined in the first column. The second and third columns give a number of period doubled ($N_{\rm PD}$) and all ($N$) W~Vir stars, respectively. The fourth column gives the incidence rate and the last column lists the average amplitude ratio, $\langle A_{1/2}/A_0\rangle$.}
\label{tab:transition}
\begin{tabular}{rrrrr}
$P$   & $N_{\rm PD}$ & $N$ &   $N_{\rm PD}/N$ & $\langle A_{1/2}/A_0\rangle$\\
\hline
15\ldots 16\,d & 4 & 33 & 0.12$\pm$0.06 & 0.025$\pm$0.005\\
16\ldots 17\,d & 5 & 23 & 0.22$\pm$0.09 & 0.06$\pm$0.03\\
17\ldots 18\,d & 7 & 14 & 0.50$\pm$0.13 & 0.07$\pm$0.01\\
18\ldots 19\,d & 4 &  9 & 0.44$\pm$0.17 & 0.08$\pm$0.01\\
19\ldots 20\,d & 5 &  7 & 0.71$\pm$0.17 & 0.14$\pm$0.04\\
\hline
\end{tabular}
\end{table}

\subsection{Period doubling in RV~Tau stars}\label{ssec:rvt-switch}

Period doubling is a characteristic feature of RV~Tau variables. Stars classified as RV~Tau, but without noticeable alternating deep and shallow minima are rare, see \cite{ogle3_lmc_t2,ogle4_gb_t2} for details about the OGLE classification. Period doubling in RV~Tau stars manifests as pronounced alternations of the deep and shallow brightness minima/maxima. The excellent coverage of OGLE data allows us to study the effect in detail. There are only a few RV~Tau stars in which the effect is very regular, i.e. light curve shape repeats after two pulsation cycles with a roughly constant minima and maxima (a very nice example is, e.g. T2CEP-193). In such a case, in the frequency spectrum we observe coherent signals located at sub-harmonic frequencies.

\begin{figure}
\includegraphics[width=\columnwidth]{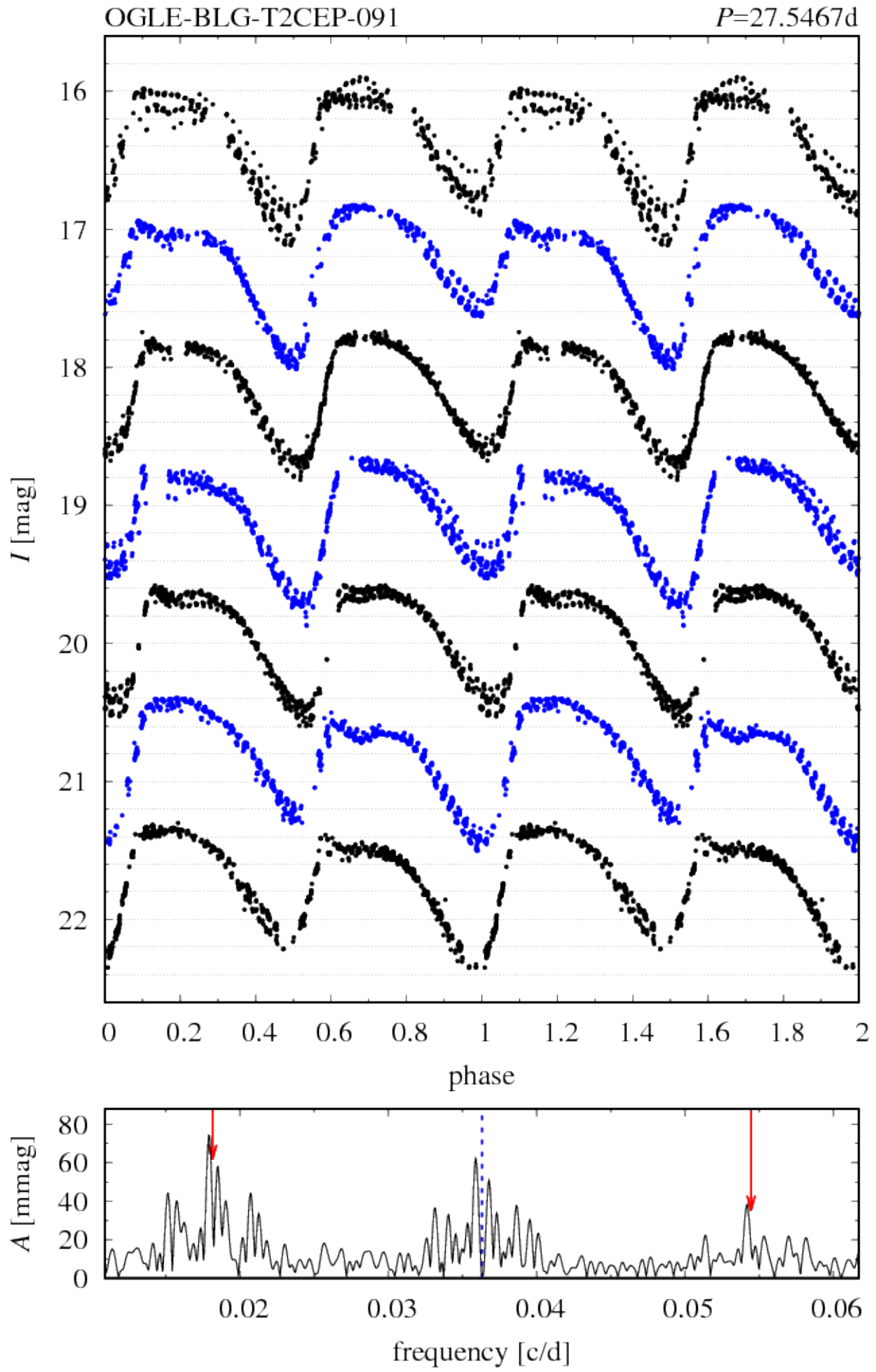}
\includegraphics[width=\columnwidth]{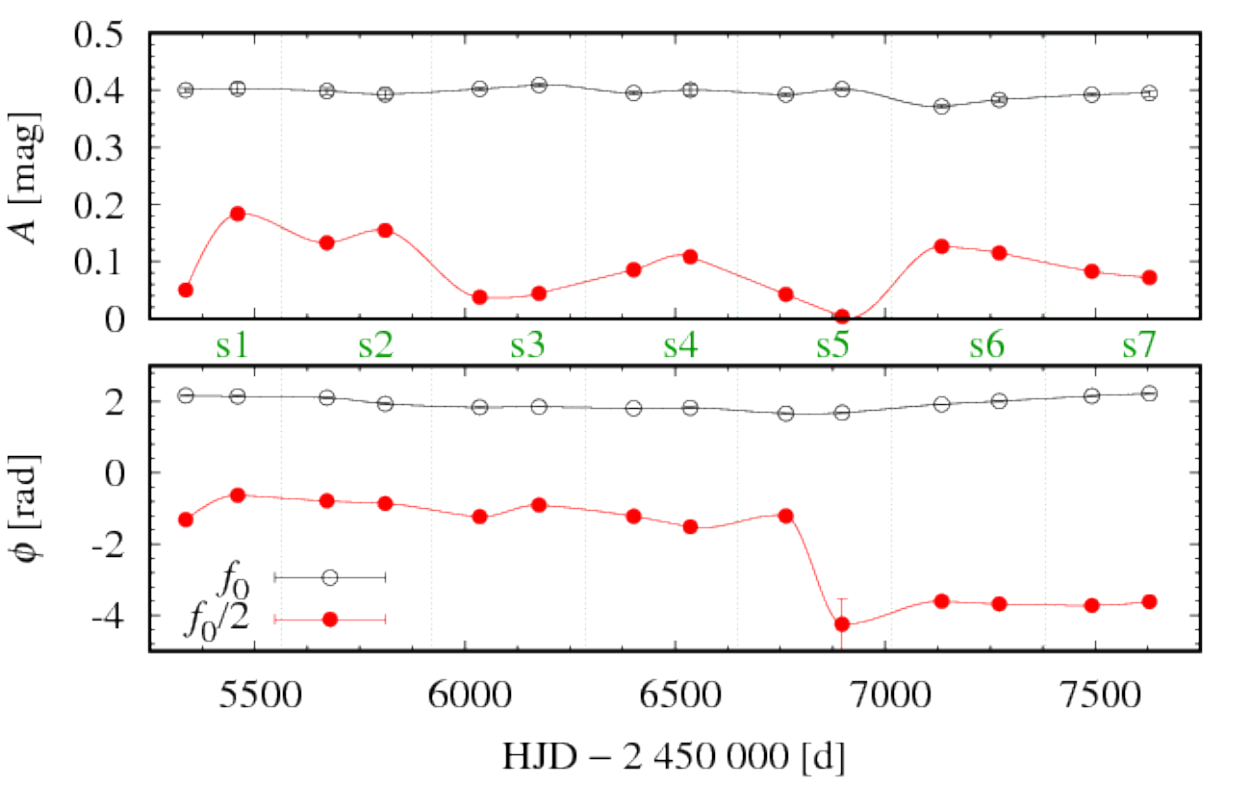}
\caption{Illustration of switching of the deep and shallow pulsation cycles in T2CEP-091. The top panel shows light curves folded with $2\PF$ separately for seven observing seasons: from s1 (top light curves) to s7 (bottom light curve) artificially shifted in vertical direction to allow comparison. The middle panel shows frequency spectrum after prewhitening with the fundamental mode and its harmonics (dashed line). Arrows indicate the exact location of $\fjd$ and $\ftd$. The two bottom panels show the results of time-dependent Fourier analysis -- amplitude and phase changes -- for signals at $\fF$ and $\fjd$ (followed with a time resolution of half of the observing season; open and filled symbols, respectively; formal errors are typically smaller than symbol's size). Data were fitted with spline function just to guide the eye.}
\label{fig:rvts091}
\end{figure}

In the majority of stars, the effect is irregular on a short time scale, as already noted, e.g. by \cite{ogle3_lmc_t2,ogle4_gb_t2}. There are stars in which the effect is transient: alternations are present in some observing seasons only. In other stars we observe varying depth of the minima/maxima, but the deep and shallow cycles do not switch. In the frequency spectra, we observe non-coherent signals centred at sub-harmonic frequencies then. In several stars we observe switching of the deep and shallow cycles. In the frequency spectra of such stars we observe signals centred at sub-harmonic frequencies with a local minimum in the middle. It was already illustrated in Sect.~\ref{sec:data}, in the bottom panel of Fig.~\ref{fig:pd034} for T2CEP-034. Here we show the effect in more detail with the help of T2CEP-091 ($\PF=27.5467$\,d). The top panel of Fig.~\ref{fig:rvts091} shows the data folded with $2\PF$ separately for the seven observing seasons. The shallow minimum visible in s1--s4 at phase 1, becomes the deep minimum for s6 and s7. Middle panel of Fig.~\ref{fig:rvts091} shows the frequency spectrum after prewhitening with the fundamental mode and its harmonics. Signals centred at $\fjd$ and $\ftd$ (weak), with minimum in the middle, are well visible. Further insight into the switching phenomenon is possible thanks to the time-dependent Fourier analysis illustrated in the bottom panel of Fig.~\ref{fig:rvts091}. To increase the time resolution, each observing season was divided into two equal parts and amplitude and phase change of signals at $\fF$ and $\fjd$ was followed (open and filled symbols, respectively). The amplitude and phase of the fundamental mode slowly vary which gives rise to unresolved remnant power at $\fF$ well visible in the middle panel of Fig.~\ref{fig:rvts091}. The changes are much more pronounced for the signal at $\fjd$. In s1 the amplitude significantly varies, which is also well visible in the top panel of Fig.~\ref{fig:rvts091}. Light curve for s1 is characterised with relatively large scatter. For s2--s4 the depths of the minima change, but otherwise light curve phases well with $2\PF$. The switching occurs within s5, in which amplitude of the signal at $\fjd$ drops to $\approx$0. Indeed, no alternations are visible in the time series at the end of s5. The phase (bottom panel of Fig.~\ref{fig:rvts091}) changes by $\pi$, and switch of the deep and shallow cycles occurs. In s6 and s7 the minimum at phase 1 is the deeper one.

Switching of the deep and shallow cycles clearly occurs in at least 23 stars of our sample. They cover a wide period range, from slightly above 20\,d to more than 40\,d. For these stars, the effect is illustrated in a series of figures analogous to that of Fig.~\ref{fig:rvts091} (folded light curves for s1--s7 and frequency spectra) attached to this paper as Supplementary Online Material (Appendix~A).

For some long-period stars the PD effect becomes strongly irregular. However, there is no rule that the longer the pulsation period, the more irregular the PD is. There are very nice examples of pretty regular long-period RV~Tau stars (e.g. T2CEP-782, T2CEP-761, T2CEP-095 and T2CEP-542, all with $\PF>45$\,d).

\subsection{Period-4 RV~Tau candidate}\label{ssec:rvt-p4}

\begin{figure}
\includegraphics[width=\columnwidth]{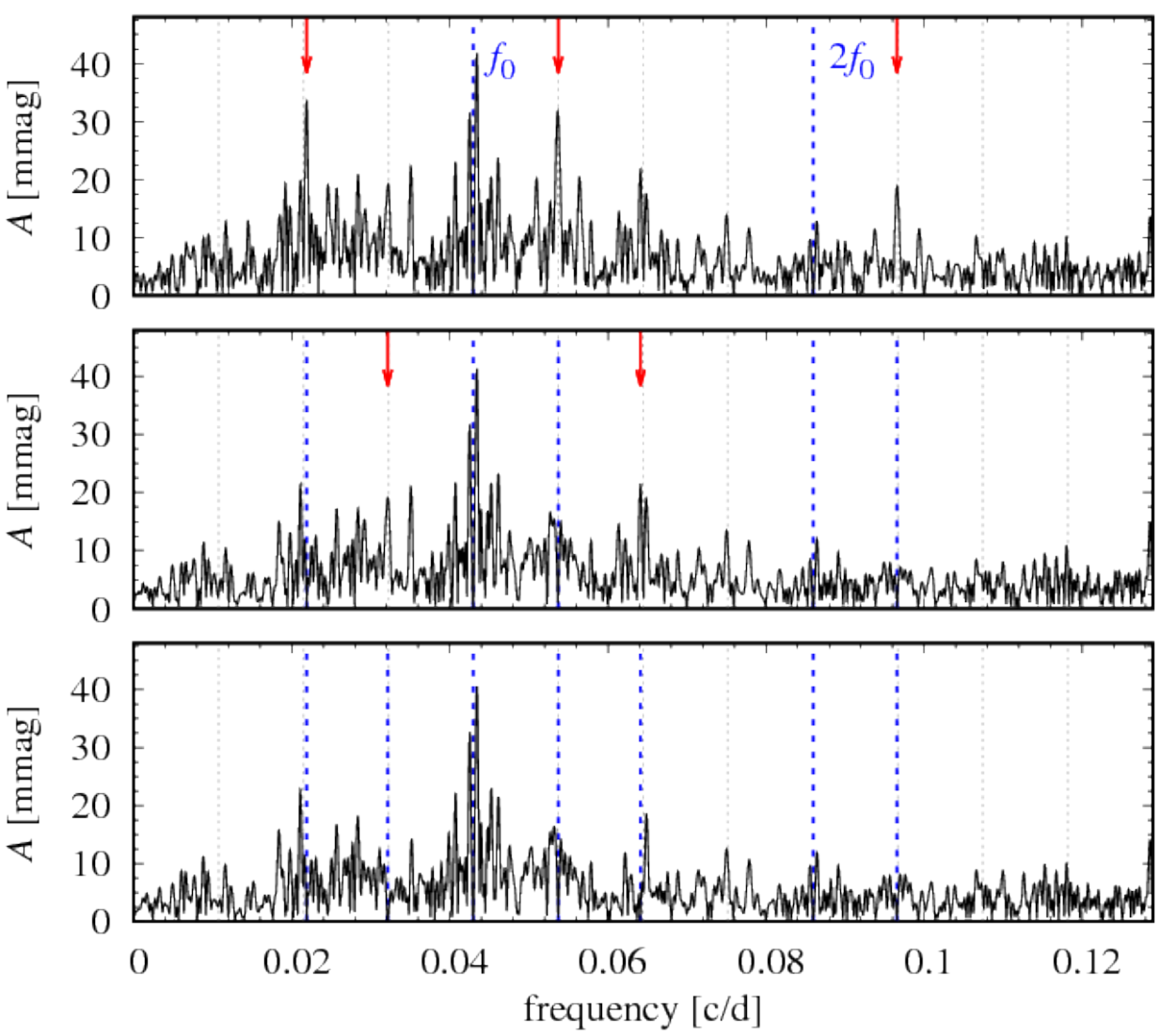}
\caption{Frequency spectra during consecutive prewhitening steps for T2CEP-135, a period-4 candidate. Thin grey vertical lines form a grid with a separation of $\fjc$. Prewhitened frequencies are marked with thick dashed blue lines, new identified frequencies are marked with red arrows. Fundamental mode and its harmonics are prewhitened in all panels.}
\label{fig:135_fsp}
\end{figure}

In the frequency spectrum of \offi{T2CEP-135}\tech{BLG534.03.40329-PM}, illustrated in the top panel of Fig.~\ref{fig:135_fsp}, we detect a signature of \mbox{period-4} pulsation. After prewhitening with the fundamental mode ($\PF=23.2681(4)$\,d) and its harmonics, peaks centred at $\fpc$ ($f/\fF=1.2500$) and $\fdc$ ($f/\fF=2.2471$) are well visible and significant. In addition we see unresolved remnant power at $\fF$ and a peak located close to $\fjd$ ($f/\fF=0.5094$). After prewhitening with the above signals, middle panel of Fig.~\ref{fig:135_fsp}, a structure centred at $\ftd$ becomes visible. In addition a signal centred at $\ftc$ ($f/\fF=0.7483$) can be noticed, but it is very weak. The bottom panel of Fig.~\ref{fig:135_fsp} shows the frequency spectrum prewhitened with these two additional peaks. The most pronounced peaks that remain in the spectrum correspond to remnant unresolved power at $\fF$, $\fjd$ and $\ftd$.

\begin{figure}
\includegraphics[width=\columnwidth]{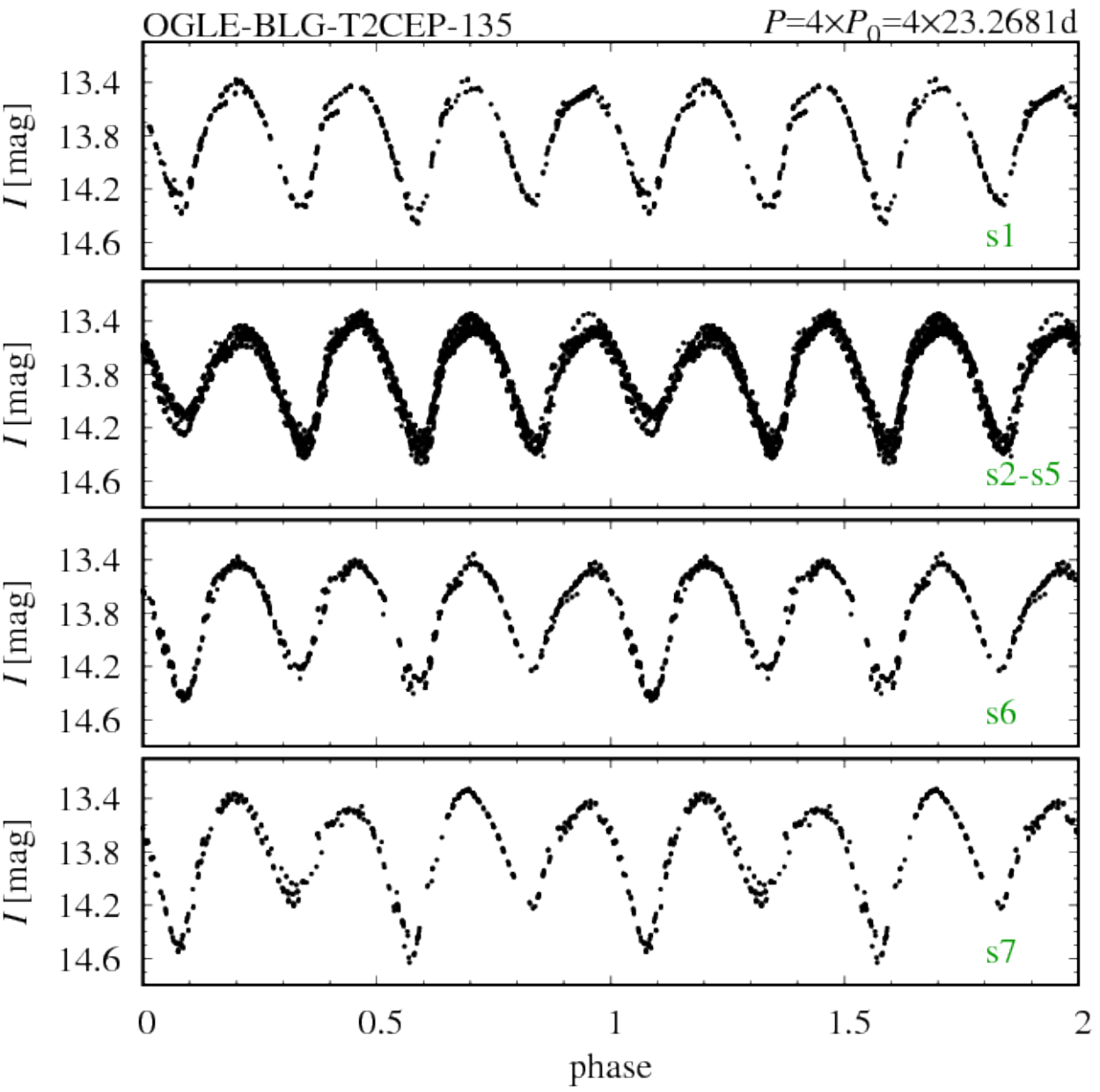}\\
\includegraphics[width=\columnwidth]{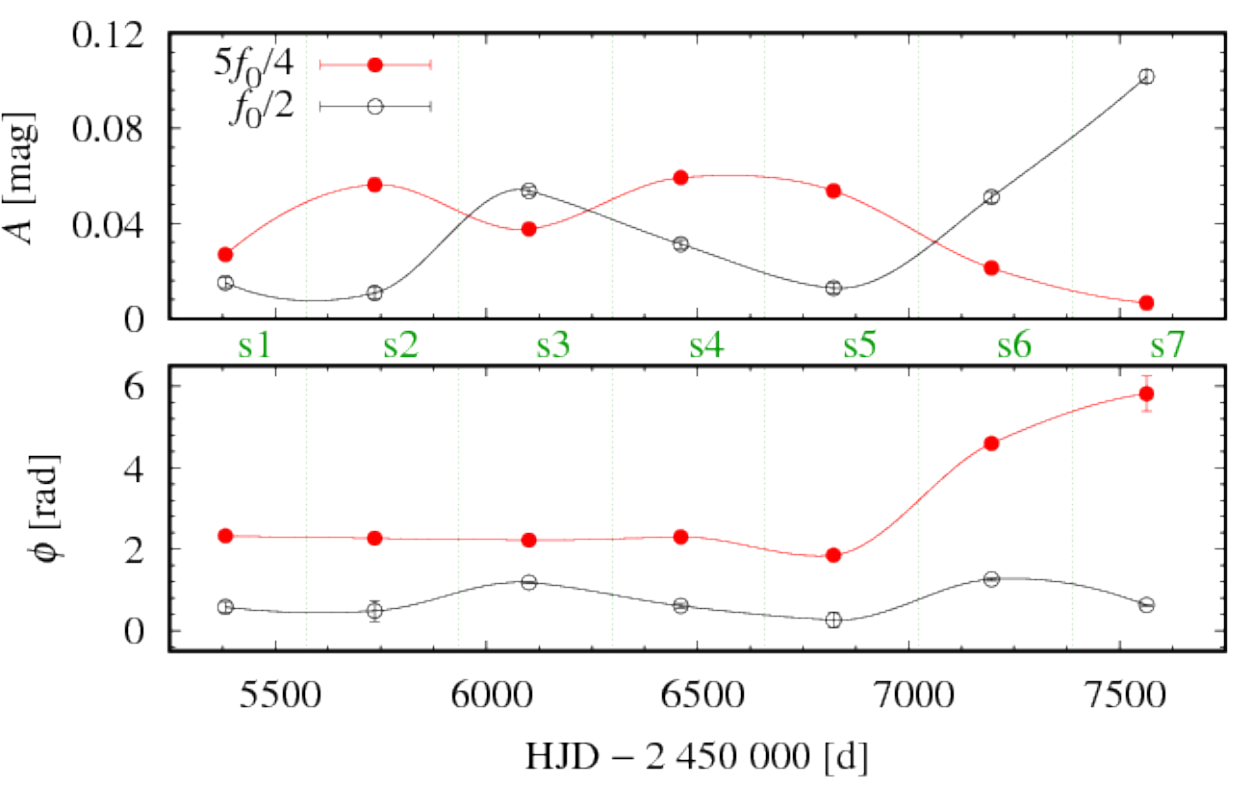}
\caption{Top four panels: data for T2CEP-135 folded with $4\PF$ for seasons s1, s2--s5, s6 and s7. Bottom panels: amplitudes and phases of signals centred at $\fpc$ and $\fjd$ (filled and open circles, respectively; formal errors are usually smaller than point's size) followed on a season-to-season basis using the time-dependent Fourier analysis. Data were fitted with spline function just to guide the eye.}
\label{fig:135_p4}
\end{figure}

Period-4 pulsation is also visible in the time domain, but it is clearly an irregular effect, not present in all observing seasons. It is best visible in s2--s5. In the consecutive panels of Fig.~\ref{fig:135_p4}, we plot the data folded with $4\PF$ for s1, jointly for s2--s5, s6 and s7. While data for s1, s6 and s7 phase well also with $2\PF$, with relatively little scatter around light curve minima/maxima, it is not the case for s2--s5. Data in these seasons follow a very similar pattern and phase much better with $4\PF$. These observations are nicely confirmed by the time-dependent analysis in which we followed the amplitudes and phases of signals at $\fjd$ and $\fpc$ on a season-to-season basis -- bottom panels of Fig.~\ref{fig:135_p4}. We recall that these are the highest peaks corresponding to, respectively, the period doubling and the period-4 behaviour. Amplitude of the $\fpc$ signal is the highest during s2--s5 and is noticeably lower during s1, s6 and s7. In s7 it is very small, while amplitude of the signal at $\fjd$ is the highest. Indeed, nice PD effect is visible in s7. Interestingly, amplitudes of the signals at  $\fjd$ and at $\fpc$ are anticorrelated. Phase variation for both signals is rather moderate. For the last two seasons, we observe strong increase of phase of the signal at $\fpc$. Since its amplitude nearly vanishes in s7, its phase is not well determined then. Clearly, it will be interesting to revisit T2CEP-135 in the future, to check whether period-4 behaviour will appear with larger amplitude again.

\begin{figure}
\includegraphics[width=\columnwidth]{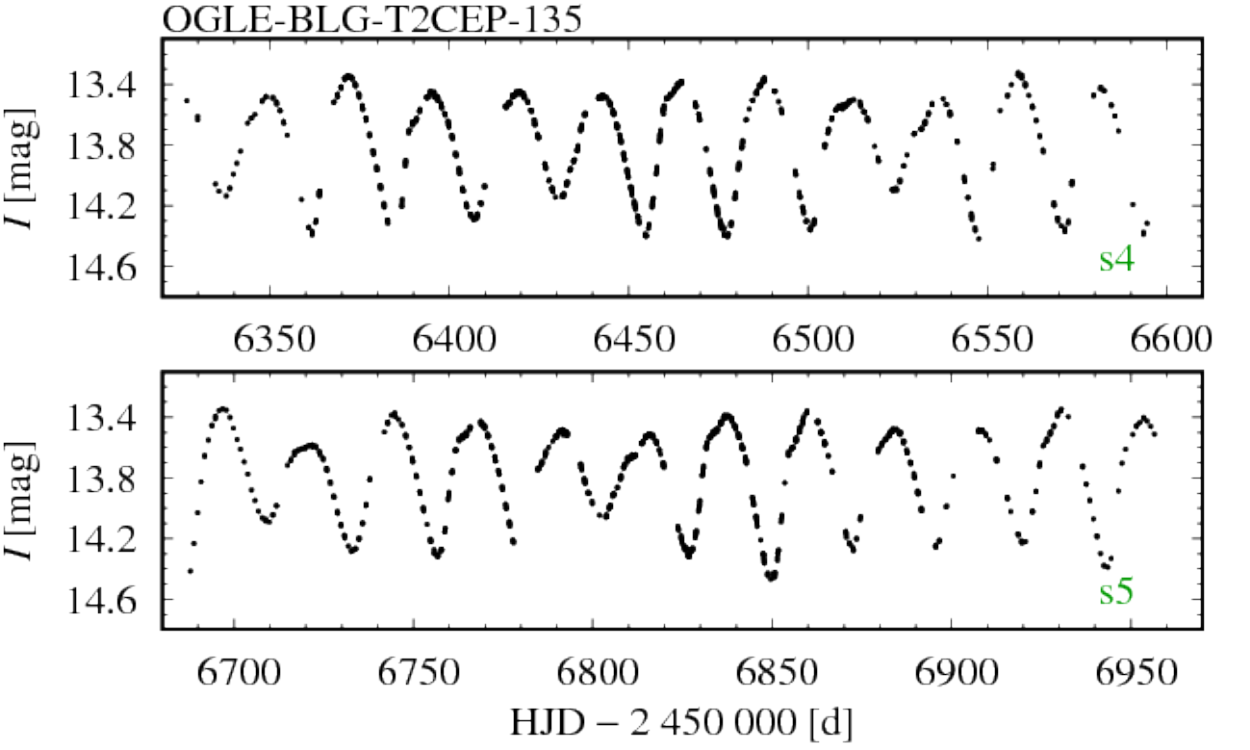}
\caption{Photometric data for T2CEP-135 during s4 and s5.}
\label{fig:135_data}
\end{figure}

For seasons s2--s5, some scatter around light curve minima/maxima is still noticeable. Period-4 pulsation is only a {\it mean} state around which some variation is present. It is best visible in the data alone, which have good enough quality to trace the individual pulsation cycles. They are displayed in Fig.~\ref{fig:135_data} for seasons s4 and s5.

OGLE-III data for T2CEP-135 are also available, but the sampling is much more scarce. The star appears as ordinary, period-doubled RV~Tau star. No trace of period-4 pulsation is apparent in the time domain. In the frequency spectrum, after prewhitening with the fundamental mode and its harmonics, signals centred at $\fjd$ and $\ftd$ dominate the spectrum. In addition, there is a signal near $\fpc$ but it is weak and significantly offsets the exact $\fpc$ position ($f/\fF=1.2385$).

In a few other stars we have found rather weak hints for period-4 pulsation; definitely more data with better sampling are needed for verification of these cases. In T2CEP-224 ($\PF=37.746$\,d), for which OGLE-III data are only available, RV~Tau alternations are pronounced, but rather large scatter is visible in the light curve. In the frequency spectrum, after prewhitening with $\fF$, its harmonics, and sub-harmonics, two peaks symmetrically placed around $3\fF/4$ are detected, which suggests period-4 pulsation. Inspection of seasonal light curves folded with $4\PF$ is inconclusive however, due to poor data sampling. T2CEP-853 ($\PF=28.268$\,d) is a similar case; here we detect a weak signal centered at $5\fF/4$, but again, data sampling is too poor to reveal possible period-4 pulsation in the seasonal light curves folded with $4\PF$. Finally, in the frequency spectrum of T2CEP-810  ($\PF=28.6085$\,d) we detect, besides $\fF$, its harmonics and sub-harmonics ($\fjd$, $\ftd$, and $\fpd$), a significant power excess centred on $2\fF/3$. All data nicely fold with $6\PF$. However we do not detect any other signals that would support period-6 pulsation. Excitation of additional non-radial mode or contamination are likely explanations too.

\subsection{Periodic modulation of pulsation: BL~Her variables}\label{ssec:mod}

\begin{figure*}
\includegraphics[width=.66\columnwidth]{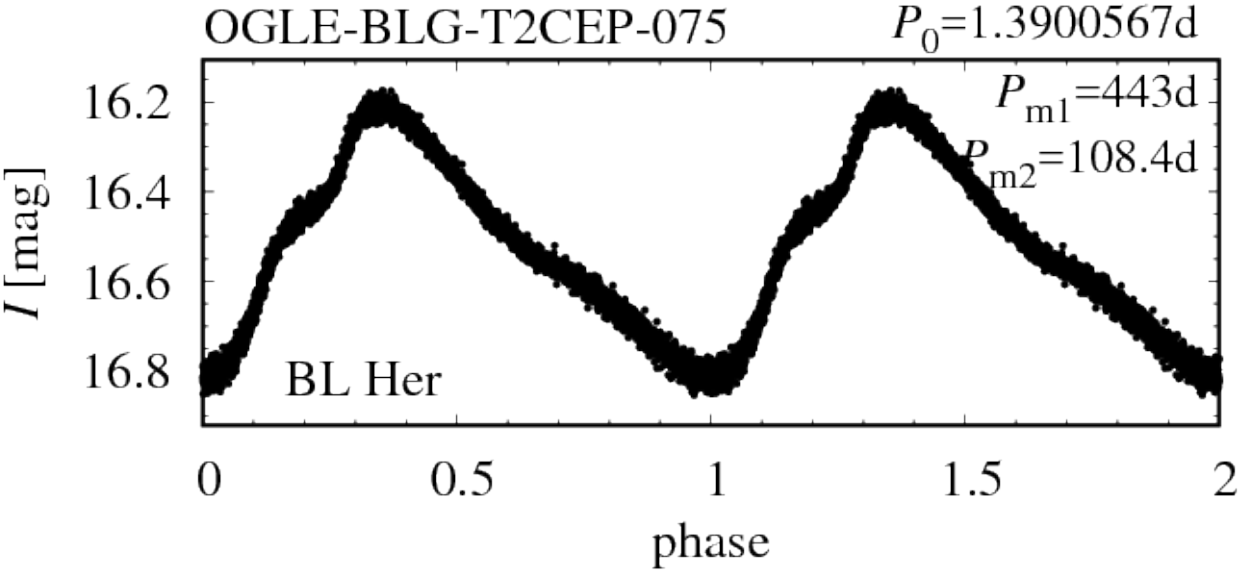}
\includegraphics[width=.66\columnwidth]{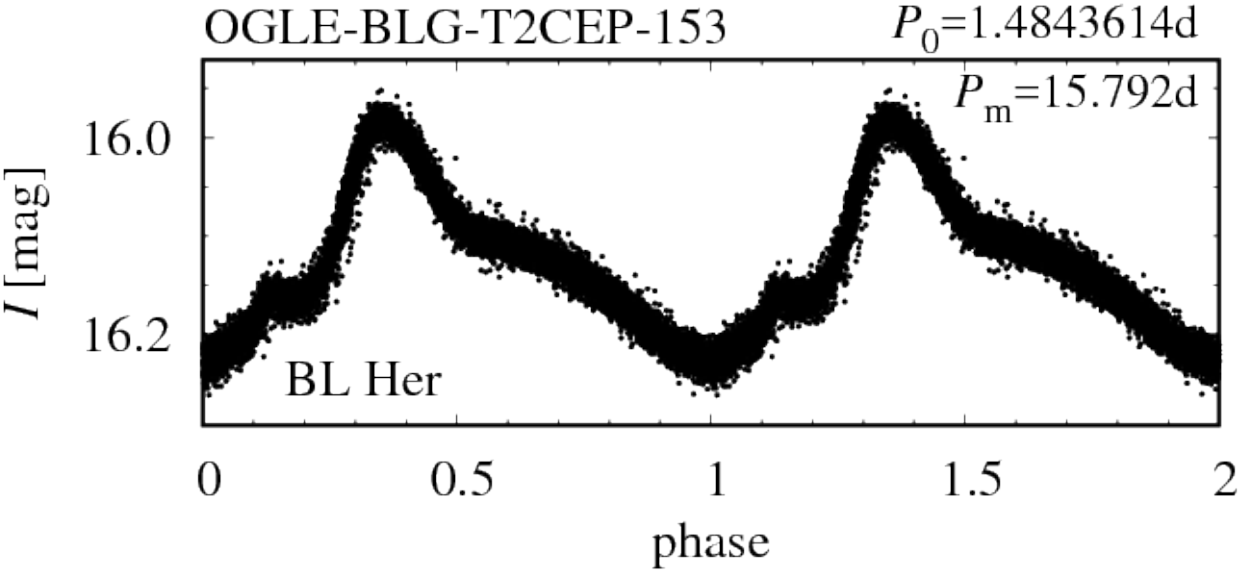}
\includegraphics[width=.66\columnwidth]{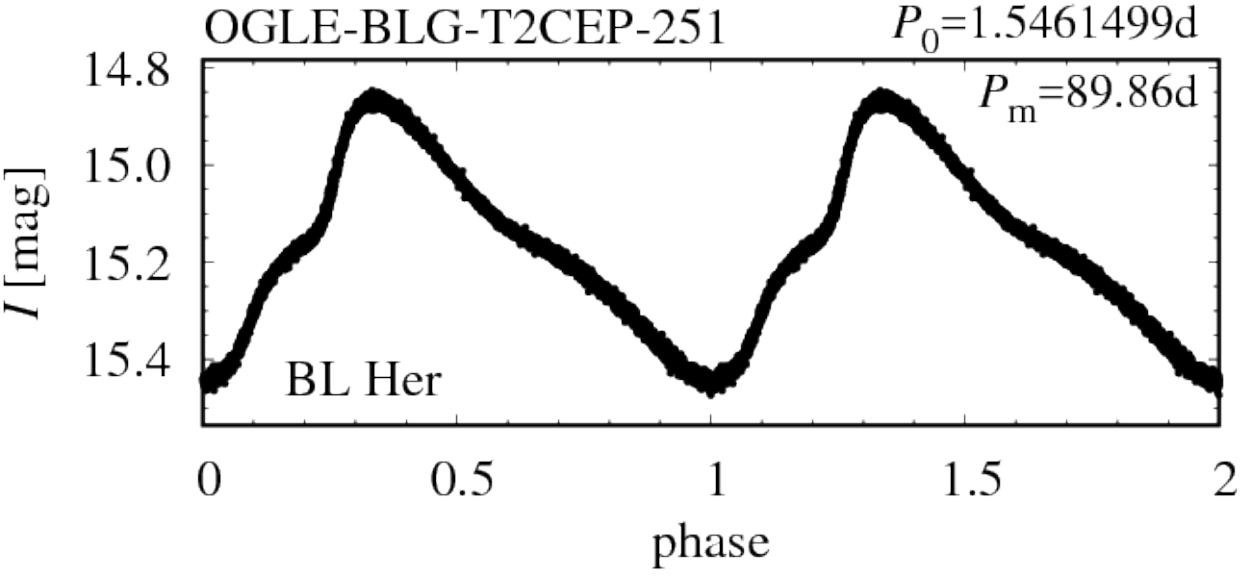}
\includegraphics[width=.66\columnwidth]{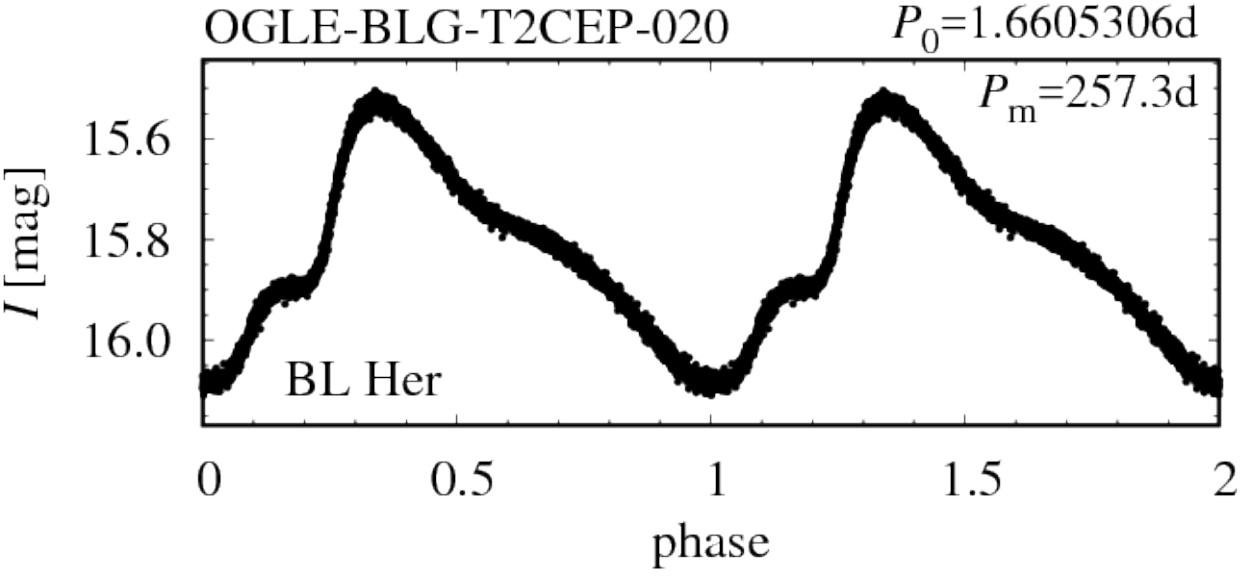}
\includegraphics[width=.66\columnwidth]{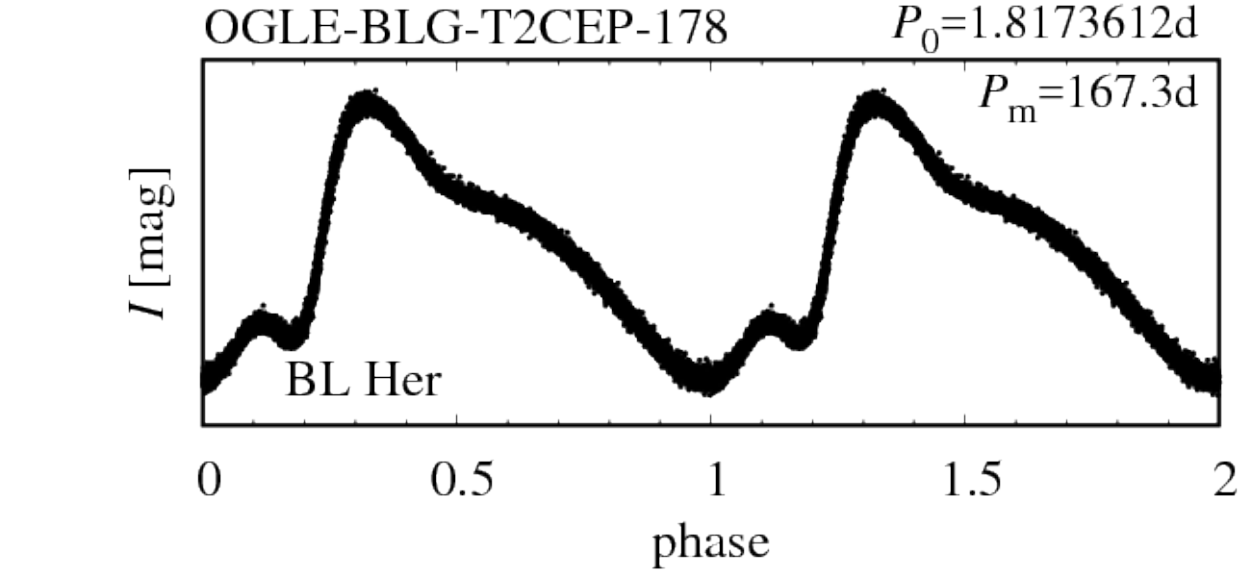}
\includegraphics[width=.66\columnwidth]{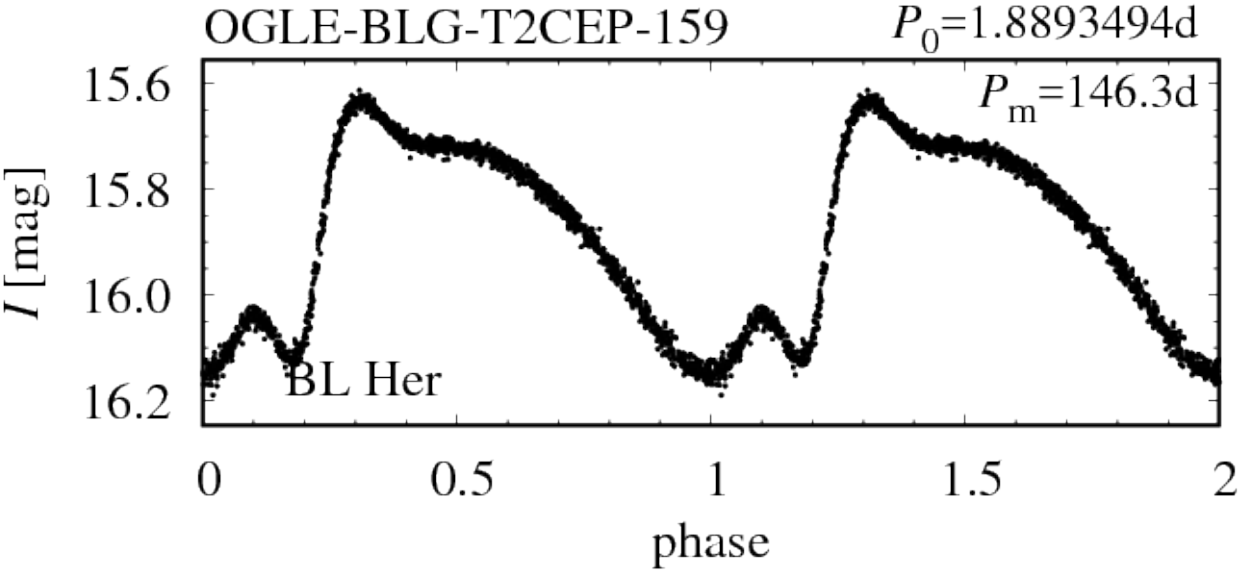}
\includegraphics[width=.66\columnwidth]{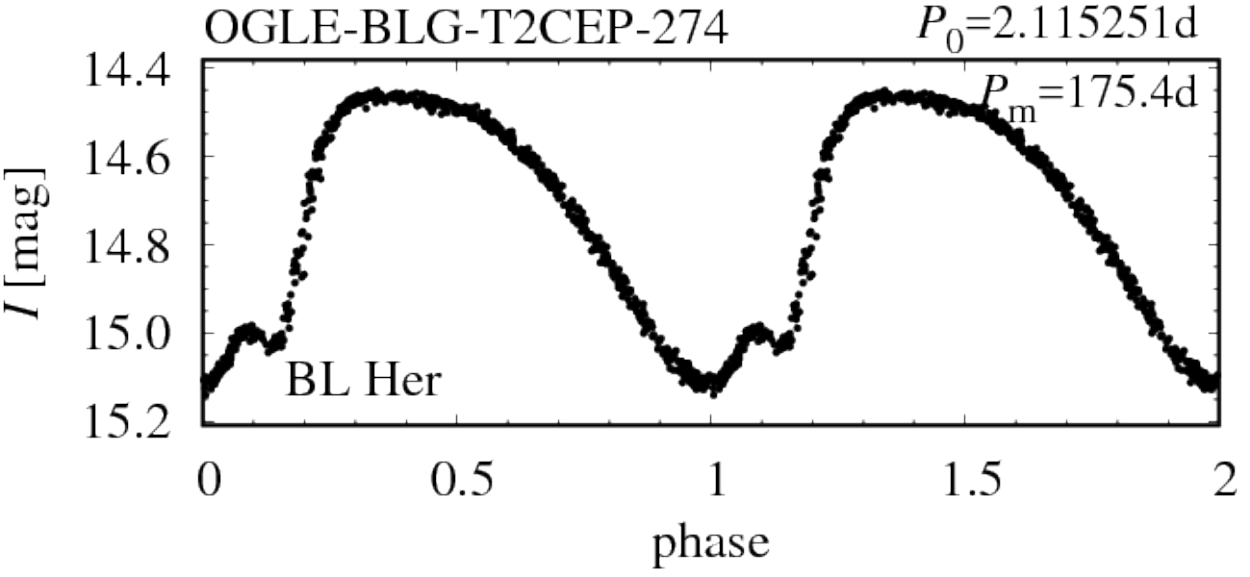}
\includegraphics[width=.66\columnwidth]{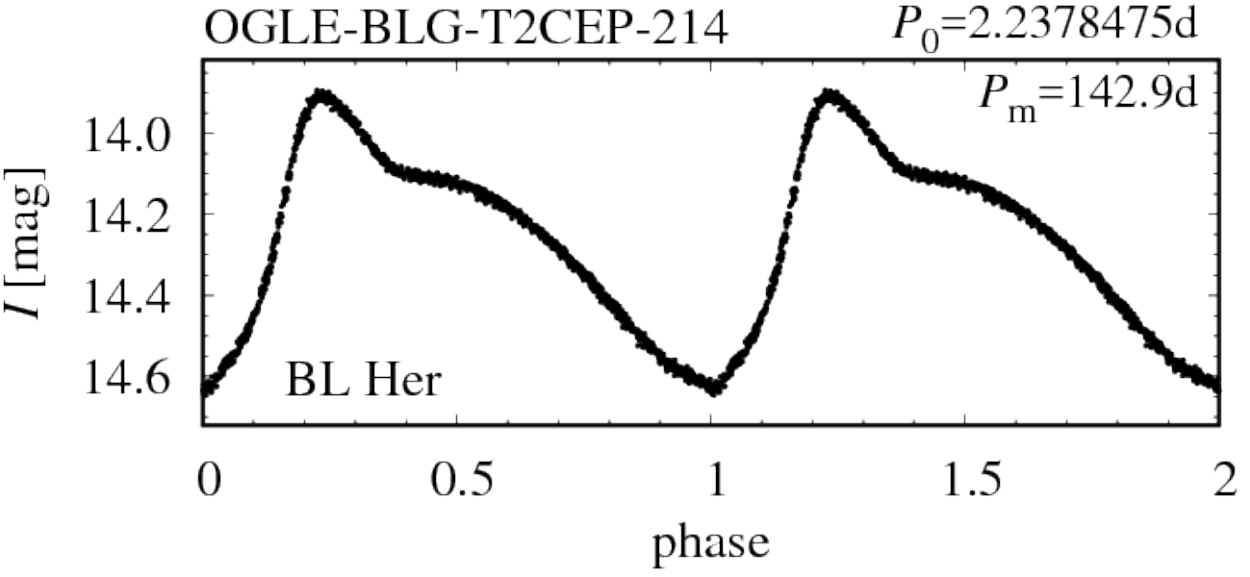}
\includegraphics[width=.66\columnwidth]{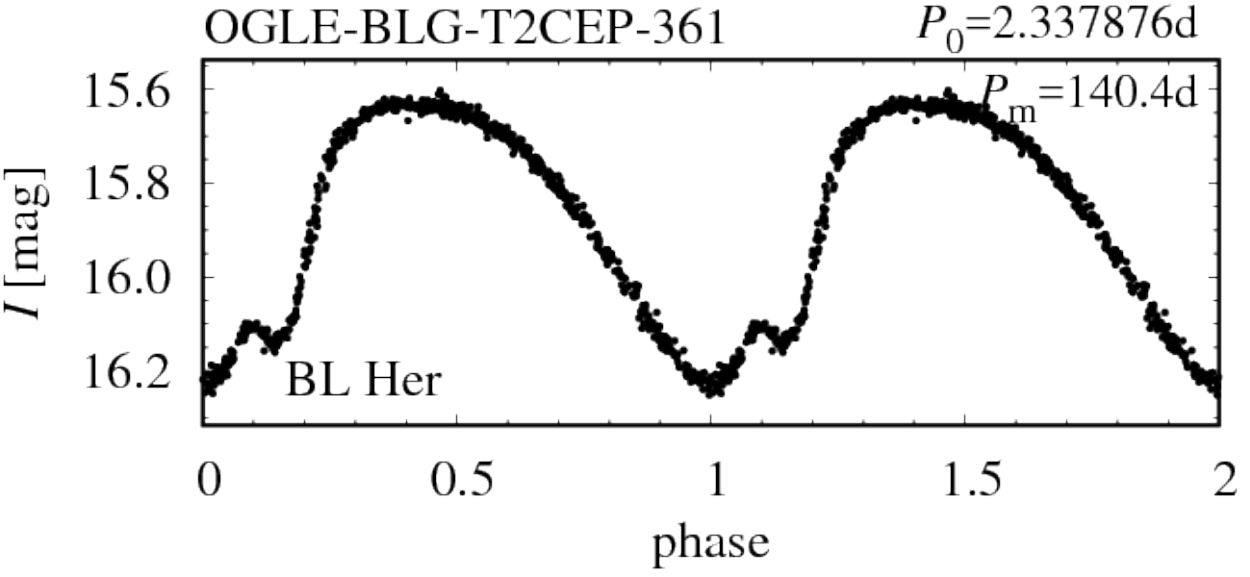}
\includegraphics[width=.66\columnwidth]{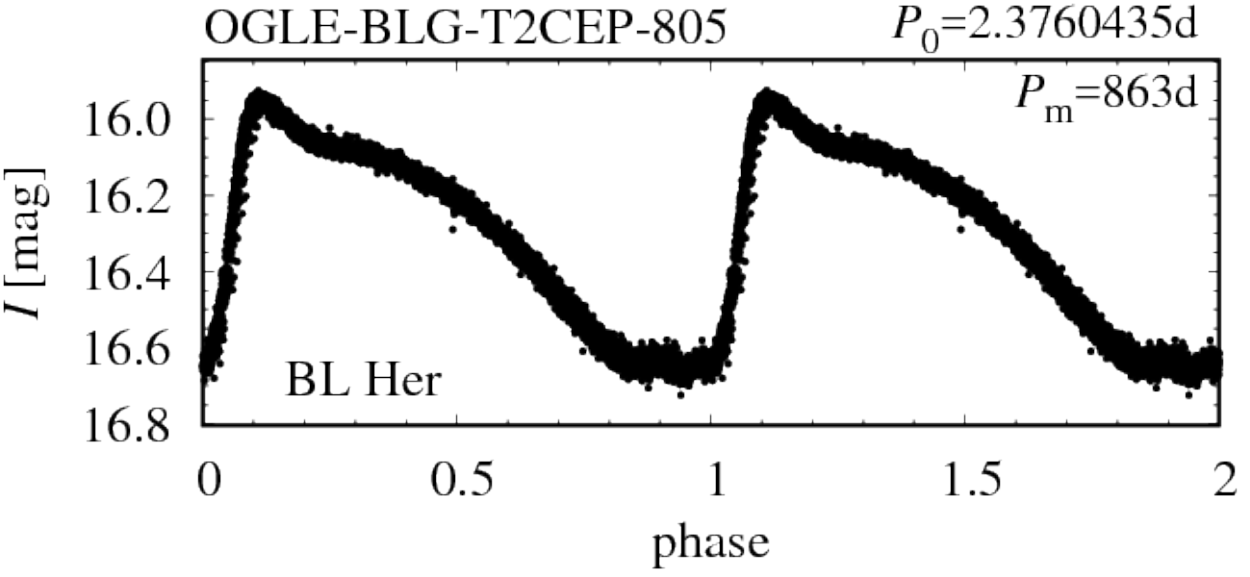}
\includegraphics[width=.66\columnwidth]{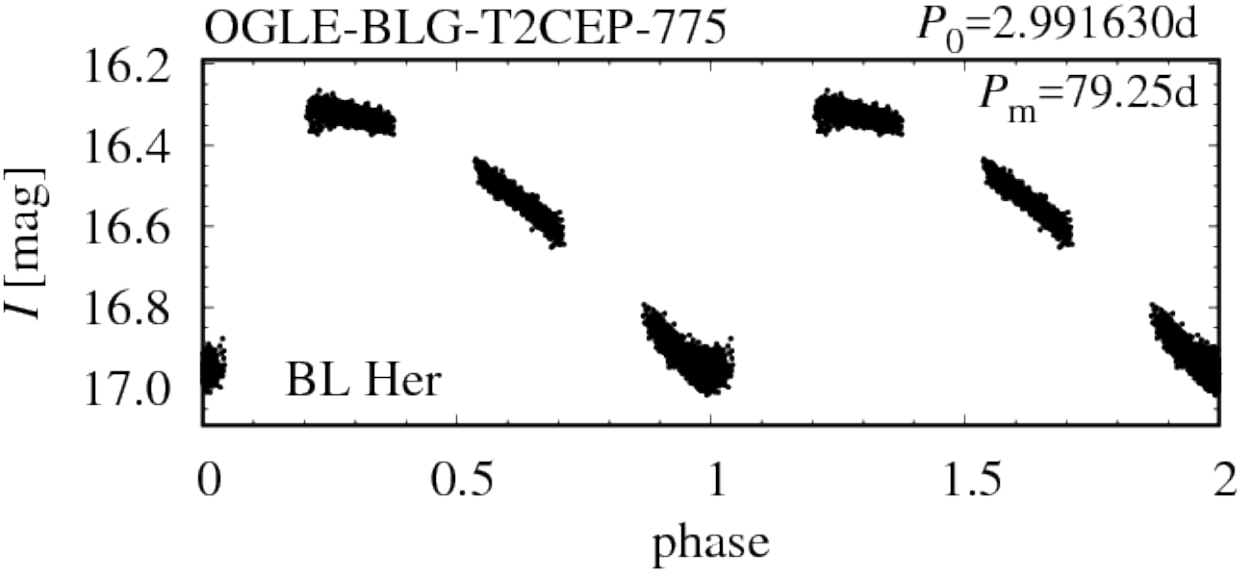}
\includegraphics[width=.66\columnwidth]{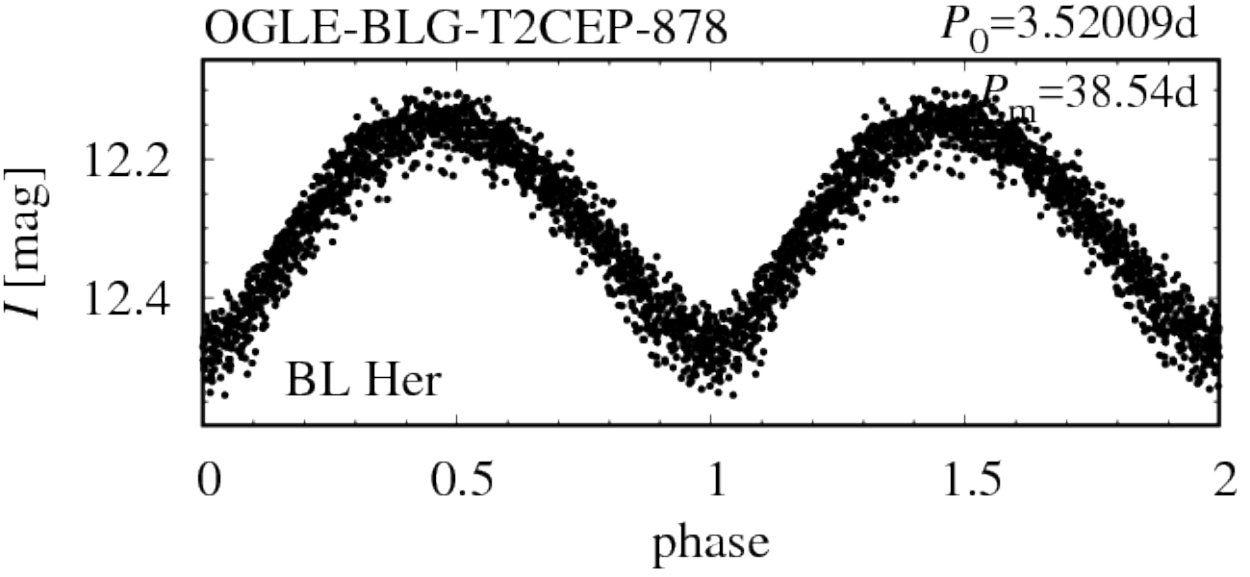}
\includegraphics[width=.66\columnwidth]{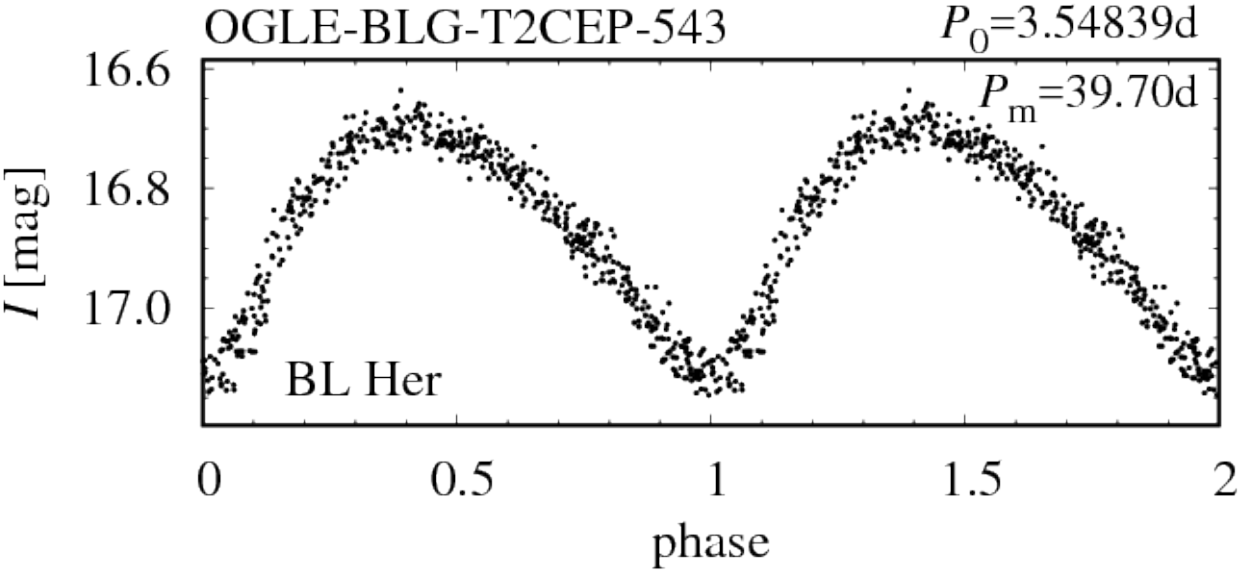}
\includegraphics[width=.66\columnwidth]{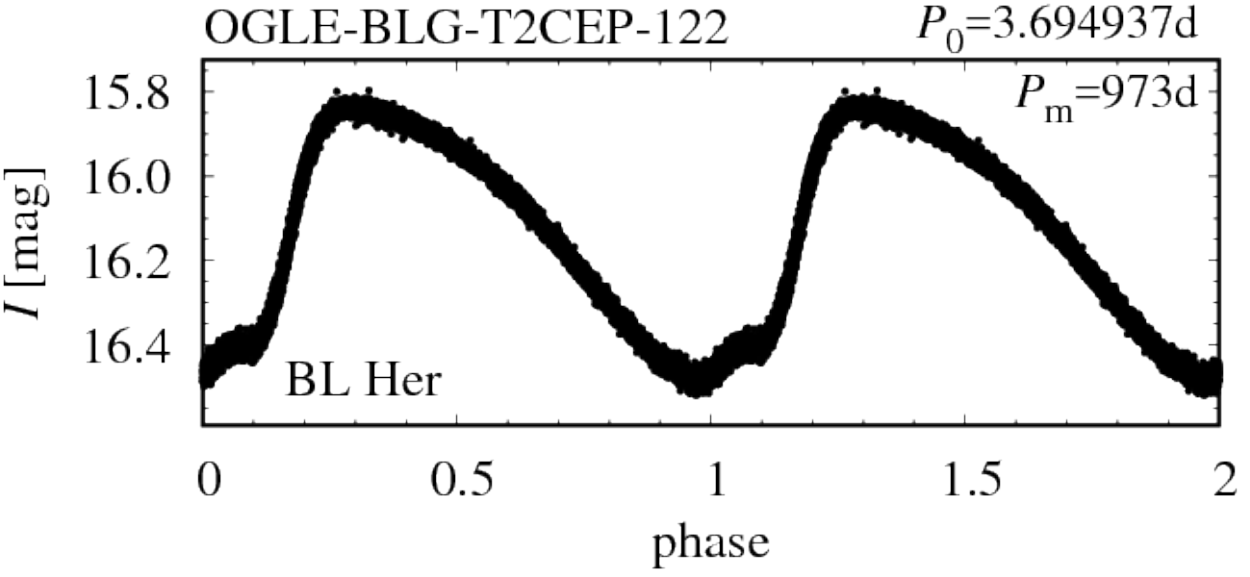}
\includegraphics[width=.66\columnwidth]{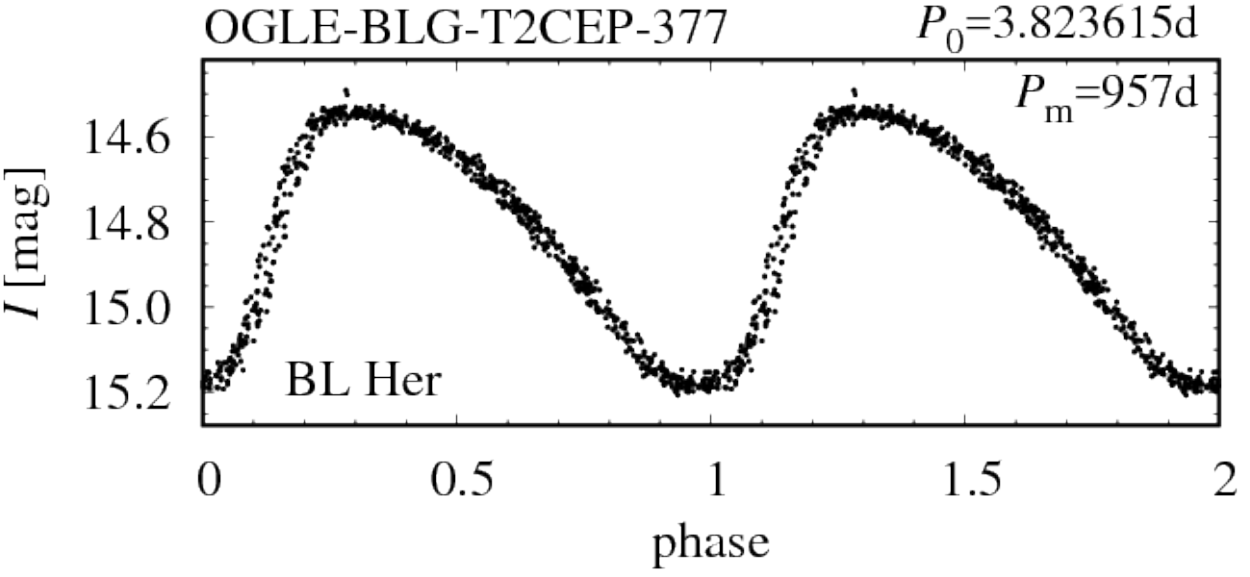}
\includegraphics[width=.66\columnwidth]{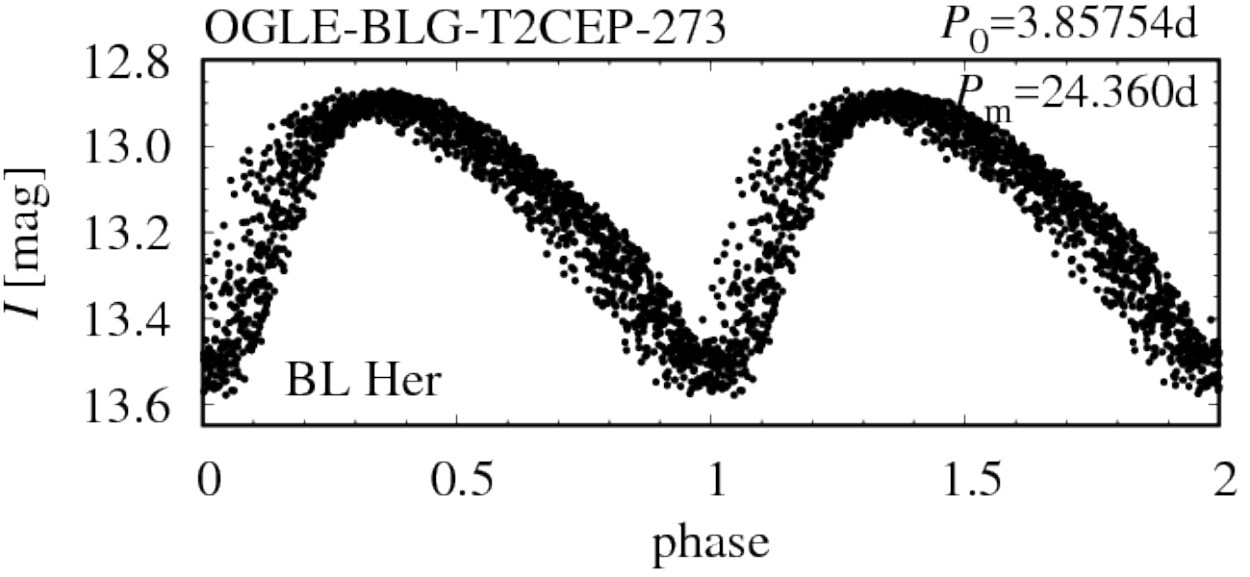}
\caption{Collection of phased light curves for modulated BL~Her stars. Star's ID, type, pulsation period and modulation period are given in each panel. Stars are sorted by the increasing pulsation period.}
\label{fig:t2cepBL_mod_lc}
\end{figure*}

Based on the analysis of frequency spectra (equidistant patterns) we identified 16 modulated BL~Her stars. Basic data for these stars are collected in the top section of Tab.~\ref{tab:mod}. In the consecutive columns we provide star's ID, pulsation period, modulation period, amplitude of the fundamental mode and amplitudes of the modulation peaks detected at $\fm$, $\fF-\fm$ and $\fF+\fm$. Information about additional modulation side-peaks and remarks are given in the last column. In several stars, ID's of which are marked with superscript `A', the amplitude of the modulation is large enough to visualise the effect with an animation. These animations can be found as a Supplementary Online Material. For all stars discussed in this section, in Fig.~\ref{fig:t2cepBL_mod_lc} we show the light curves folded with the pulsation period. It is immediately visible, that the modulations are of low amplitude. In none of the light curves the effect is immediately visible, as it is the case, e.g. for RR~Lyrae stars and the Blazhko effect. For all modulated stars the frequency spectra showing the most prominent modulation peaks are presented either below, or as Supplementary Online Material, Appendix~B.

\begin{table*}
\centering
\caption{Properties of modulated BL~Her and W~Vir variables. The consecutive columns give: star's ID (OGLE-BLG-T2CEP-NNN), pulsation period, $\PF$, modulation period, $\Pm$, pulsation amplitude, and amplitudes of the modulation side-peaks located at $\fm$, $\fF-\fm$ and $\fF+\fm$; $A_0$, $A_{\rm m}$, $A_{-}$ and $A_{+}$, respectively. The last column lists additional modulation side-peaks detected in the spectrum and remarks. Additional side-peaks: A -- $k\fF\pm\fmj$ for $k=2,\ldots,8$ without $5\fF-\fmj$ and $8\fF-\fmj$; (weak) $\fF+\fmj+\fmd$, $2\fF-\fmj+\fmd$; B -- $2\fF\pm\fm$; $3\fF\pm\fm$; $4\fF+\fm$; $\fF\pm2\fm$; $2\fF-2\fm$; $\fF-3\fm/2$, $2\fF-3\fm/2$, $\fm/2$ and $3\fm/2$. Remarks: `am' -- additional long-period modulation possible; `fX' -- additional single periodicity detected; `tdp' -- modulation becomes clear after time-dependent prewhitening.}
\label{tab:mod}
\begin{tabular}{llllllll}
ID  & $\PF$\,(d) & $\Pm$\,(d) & $A_{0}$\,(mag) & $A_{\rm m}$\,(mag) & $A_{-}$\,(mag) & $A_{+}$\,(mag) & additional peaks, remarks\\
\hline
\offi{075}$^{A}$\tech{BLG534.32.44546-EP}  & 1.3900567(4)& 443(2)     & 0.2527(2) & 0.0014(2) & 0.0031(2) & 0.0042(2) & A \\ 
                                          &             & 108.4(1)   &           & 0.0014(2) & 0.0028(2) & 0.0033(2) & \\ 
\offi{153}\tech{BLG501.02.68787-PM}       & 1.4843614(5)&  15.792(7) & 0.0899(1) &  -        & 0.0007(1) & 0.0007(1) & tdp, fX\\ 
\offi{251}$^{A}$\tech{BLG511.16.58981-RS}  & 1.5461499(2)&  89.86(4)  & 0.2362(1) & 0.0013(1) & 0.0040(1) & 0.0041(1) & $2\fm$, $2\fF\!\pm\!\fm$ \\ 
\offi{020}$^{A}$\tech{BLG611.21.132345-EP} & 1.6605306(3)& 257.3(4)   & 0.2231(2) & 0.0013(2) & 0.0028(2) & 0.0022(2) & $2\fF\!\pm\!2\fm$, (weak) $3\fF\!\pm\!2\fm$, $3\fF\!\pm\!\fm$\\ 
\offi{178}$^{A}$\tech{BLG505.15.201926-RS} & 1.8173612(3)& 167.3(1)   & 0.2233(1) & 0.0023(1) & 0.0028(1) & 0.0029(1) & tdp, am, $2\fF\!\pm\!\fm$ \\ 
\offi{159}\tech{BLG535.18.128627-RS}      & 1.8893494(8)& 146.3(5)   & 0.2272(4)  & 0.0025(4) & 0.0033(4) &  -       & am\\  
\offi{274}$^{A}$\tech{BLG516.13.89169-RS}  & 2.115251(2) & 175.4(5)   & 0.3234(8) & 0.0031(5) & 0.0036(5) & 0.0032(5) & tdp\\ 
\offi{214}\tech{BLG508.29.18238-PM}       & 2.2378475(8)& 142.9(2)   & 0.2760(3) & 0.0024(3) &  -        & 0.0056(3) & $2\fF+\fm$, am \\ 

\offi{361}$^{A}$\tech{BLG617.03.69359-EP}  & 2.337876(2) & 140.4(2)   & 0.2958(6) & 0.0049(6) & 0.0052(6) & 0.0048(6) & $2\fF\!\pm\!\fm$, am\\ 
\offi{805}$^{A}$\tech{BLG500.27.34197-RS}  & 2.3760435(7)& 865(3)     & 0.3008(3) &  -        & 0.0068(3) & 0.0048(3) & $k\fF\pm\fmj$ for $k=2,\ldots,7$, am\\ 
\offi{775}$^{A}$\tech{BLG534.13.6292-PM}   & 2.991630(4) &  79.25(4)  & 0.278(2)  & 0.0049(3) & 0.0064(3) & 0.0051(3) & \\ 
\offi{878}$^{A}$\tech{BLG513.10.81158-RS}  & 3.52009(1)  &  38.54(2)  & 0.1547(8) & 0.0126(8) & 0.0041(8) & 0.0098(8) & $2\fF\!+\!\fm$, am\\ 
\offi{543}$^{A}$\tech{BLG652.15.73587-PM}  & 3.54839(2)  &  39.70(3)  & 0.190(1)  & 0.013(1)  & 0.012(1)  &  -        & \\ 
\offi{122}$^{A}$\tech{BLG501.29.67488-RS}  & 3.694937(1) & 973(4)     & 0.3079(1) &  -        & 0.0054(1) & 0.0049(1) & $k\fF\pm\fmj$ for $k=2,\ldots,6$, $7\fF-\fm$\\ 
\offi{377}$^{A}$\tech{BLG613.32.38453-PM}  & 3.823615(7) & 957(5)     & 0.3121(8) &  -        & 0.0142(8) & 0.0180(8) & $k\fF\pm\fmj$ for $k=2,\ldots,4$, $5\fF+\fm$\\ 
\offi{273}$^{A}$\tech{BLG513.30.132494-PM} & 3.85754(2)  &  24.360(5) & 0.268(2)  & 0.0343(5) & 0.0133(5) & 0.0119(5) & B, tdp\\ 

\hline

\offi{783}\tech{BLG500.12.142517-EP}      & 5.247911(8) &  50.07(2)  & 0.2964(4) &  -        &  -        & 0.0072(4) & $2\fF+\fmj$, $3\fF+\fmj$; $5\fmj=6\fmd$ \\ 
                                          &             &  59.92(4)  &           &  -        & 0.0061(4) &  -        & $2\fF-\fmd$, $3\fF-\fmd$ \\                
\offi{696}$^{A}$\tech{BLG660.28.28062-EP}  & 5.87379(3)  &  50.86(2)  & 0.230(1)  & 0.026(1)  & 0.015(1)  &  -        & $2\fF-\fm$ \\                              
\offi{804}$^{A}$\tech{BLG500.27.89659-EP}  & 8.31937(8)  & 988(5)     & 0.0537(3) &  -        &  -        & 0.0137(3) &  $2\fF+\fmj$ \\ 
\offi{282}\tech{BLG512.03.173511-EP}      & 8.91565(3)  &  82.24(6)  & 0.1856(4) &  -        & 0.0106(4) &  -        & $2\fF-\fm$ \\
\offi{187}$^{A}$\tech{BLG510.31.74018-PM}  & 9.5937(1)   & 107.6(3)   & 0.270(2)  &  -        & 0.014(2)  &  -        & $2\fF-\fm$, peculiar W~Vir\\               
\offi{443}\tech{BLG667.07.151-PM}         & 10.6364(3)  & 124.5(4)   & 0.327(4)  & 0.010(2)  & 0.013(2)  & 0.013(2)  & tdp \\
\offi{594}\tech{BLG680.19.2882-RS}        & 11.396(2)   & 116(1)     & 0.254(4)  &  -        & 0.040(5)  &  -        & $2\fF-\fm$, poor data \\ 
\offi{764}$^{A}$\tech{BLG639.29.80329-RS}  & 12.4293(4)  & 115.6(2)   & 0.215(3)  & 0.014(3)  & 0.043(3)  &  -        & $2\fF-\fm$, am \\                          
\offi{230}\tech{RS}                       & 16.3985(3)  & 219(2)     & 0.451(1)  &   -       & 0.013(1)  & 0.007(1)  &  O-III data only\\   

 \hline  
\end{tabular}
\end{table*}

Modulated BL~Her stars cover a relatively wide range of pulsation periods ($\approx$1.4--3.9\,d) and modulation periods ($\approx$15.8--970\,d). In the majority of stars, modulation manifests through equidistant triplets centred at $\fF$ and by a signal at the modulation frequency, $\fm$. Strongly asymmetric modulation patterns (doublets) are less frequent. In many cases, modulation co-exists with irregular variation of amplitude and/or phase of the fundamental mode, which manifests as unresolved remnant power at $\fF$, and hampers the analysis of modulation. Modulations of pulsation amplitude, pulsation phase and of mean brightness are all present in the discussed sample.

Below we provide some details and illustrate the modulation for particular stars. 

\begin{figure}
\includegraphics[width=\columnwidth]{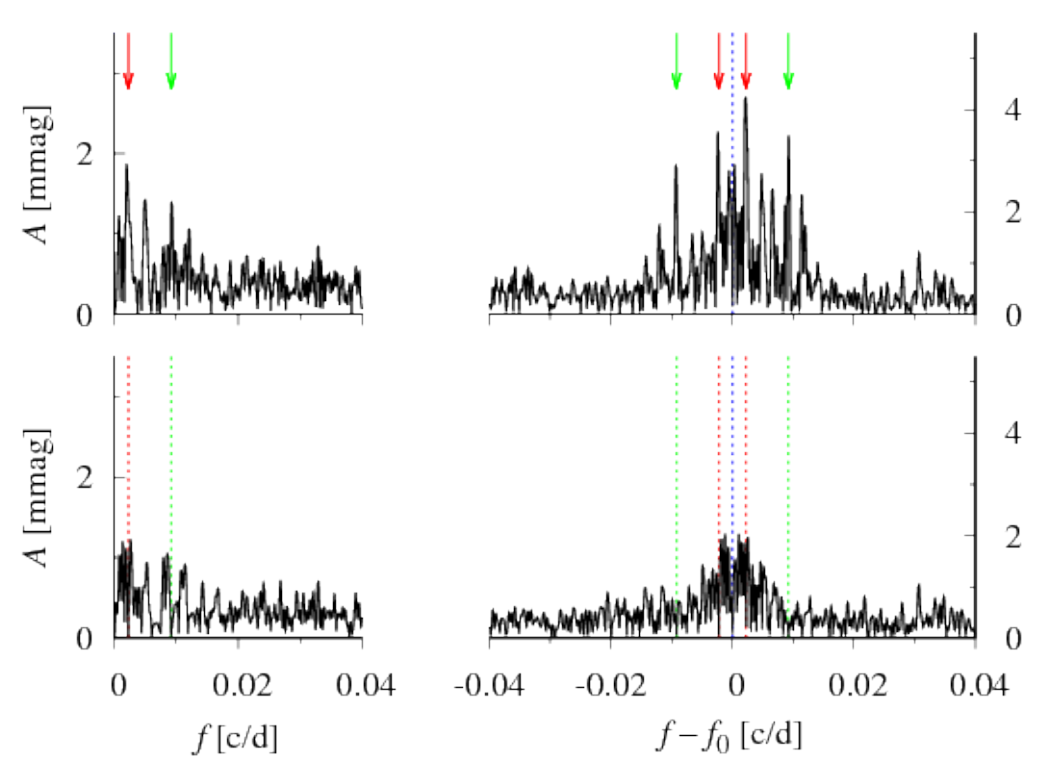}
\caption{Top panel: sections of the frequency spectrum for T2CEP-075 (BL~Her) at the low frequency range (left) and centred at $\fF$ (right), after prewhitening with the fundamental mode and its harmonics (blue dashed line). Modulation side-peaks are marked with arrows. Bottom panel: the corresponding sections of the frequency spectrum after prewhitening with the modulation side-peaks (red and green dashed lines).}
\label{fig:fspmod075}
\end{figure}

\begin{figure}
\includegraphics[width=\columnwidth]{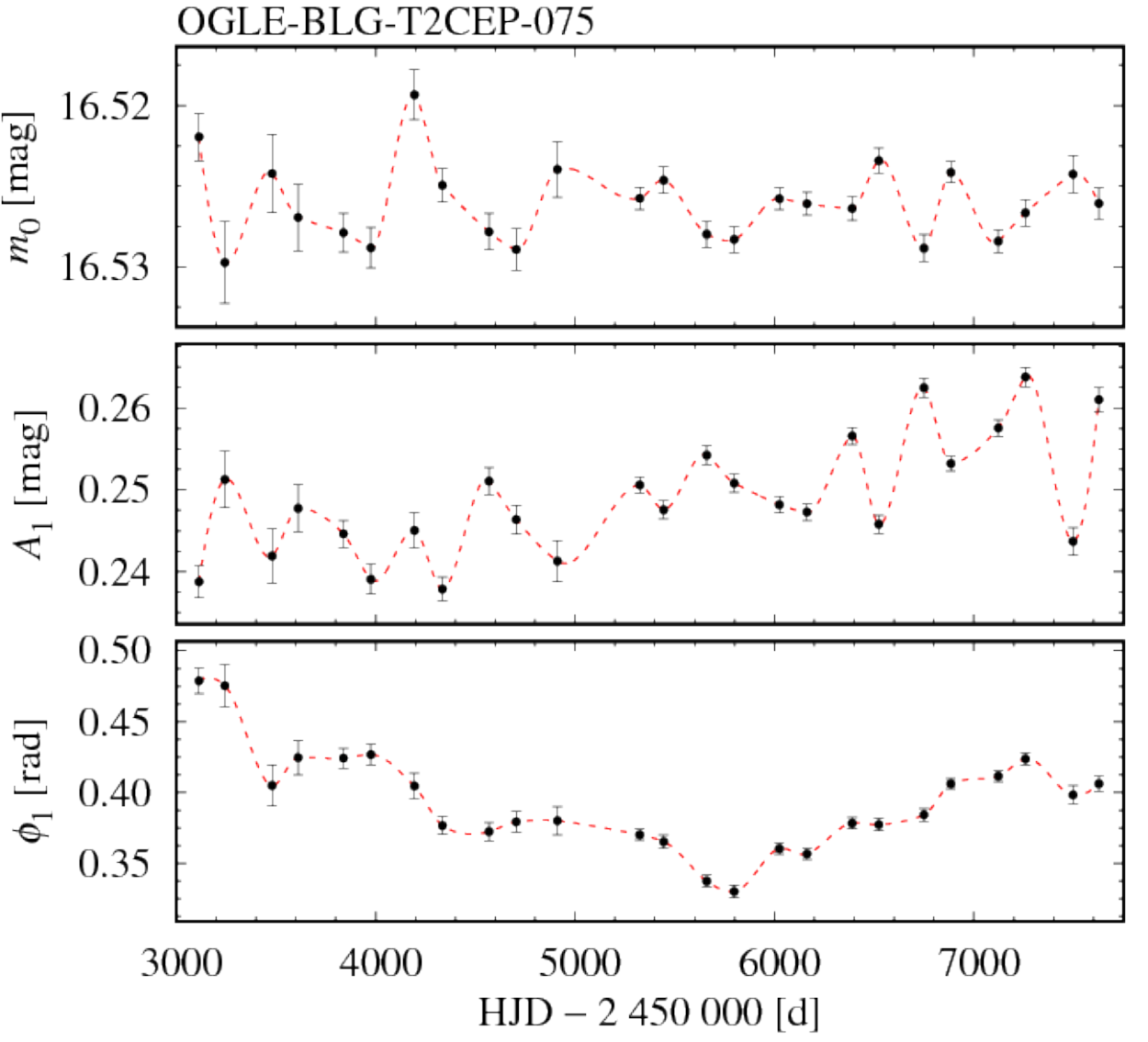}
\caption{Time-dependent Fourier analysis ($\Delta t\approx 125$\,d)  for T2CEP-075 (BL~Her): mean brightness (top), amplitude (middle) and phase (bottom) changes.}
\label{fig:tdfd075}
\end{figure}

{\bf \offi{T2CEP-075}\tech{BLG534.32.44546-EP}}. In this star, we clearly detect two modulation periods as illustrated in the frequency spectrum in Fig.~\ref{fig:fspmod075}. The primary modulation, of larger amplitude, has a period of $\Pmj\approx443$\,d. Triplets at $\fF$ and at the harmonics and a signal at the modulation frequency $\fmj$, are detected. Secondary modulation ($\Pmd\approx108.4$\,d; $\Pmj/\Pmd\approx4.1$) is evident; the modulation peaks are detected at $\fF\pm\fmd$ and at the modulation frequency $\fmd$. Two peaks at combination frequencies ($\fF\!+\!\fmj\!+\!\fmd$ and $2\fF\!+\!\fmj\!-\!\fmd$) are also present in the spectrum, but are weak. OGLE-III data for this star are of inferior quality, however, the last seasons of OGLE-III data, together with OGLE-IV data, allow the time-dependent analysis with a time resolution of about $\Delta t\!\approx\!130$\,d ($\sim$half of the season), which we present in Fig.~\ref{fig:tdfd075}. Amplitude modulation on the timescale of $\Pmj$ is obvious. It is clear that the mean brightness also varies on the same time-scale. These variations are anticorrelated: the lower the pulsation amplitude, the brighter the star. Phase changes, on the other hand, are much more smooth and dominated by a secular, roughly parabolic variation.

We also note a slight difference in amplitudes between OGLE-III and OGLE-IV observations which, as explained in Sect.~\ref{sec:data}, is likely an instrumental effect. OGLE-IV data also enable time-dependent Fourier analysis on a much shorter time scale which shows that also the secondary modulation is an amplitude modulation. Based on frequency solution, the two modulations were disentangled and illustrated with the animations (see Supplementary Online Material).

{\bf \offi{T2CEP-153}\tech{BLG501.02.68787-PM}}. The remnant, unresolved power at $\fF$ dominates the prewhitened frequency spectrum of this star. In addition, a strong signal at $\fx/\fF=0.451$ is present. We do not detect any linear frequency combinations of the two frequencies. Since $\px$ is longer than $\PF$, this additional signal is either a non-radial pulsation mode or arises due to contamination. After time-dependent prewhitening ($\Delta t\approx 90$\, d), two side-peaks, symmetrically placed around $\fF$ are firmly detected (at $\sn$ equal to 4.9 and 4.4) and indicate a modulation with $\Pmj\approx15.8$\,d\footnote{For all stars, figures with sections of the frequency spectrum, illustrating the modulation peaks, are either included in the main body of the paper, or in the Supplementary Online Material -- Appendix B.}.

\begin{figure}
\includegraphics[width=\columnwidth]{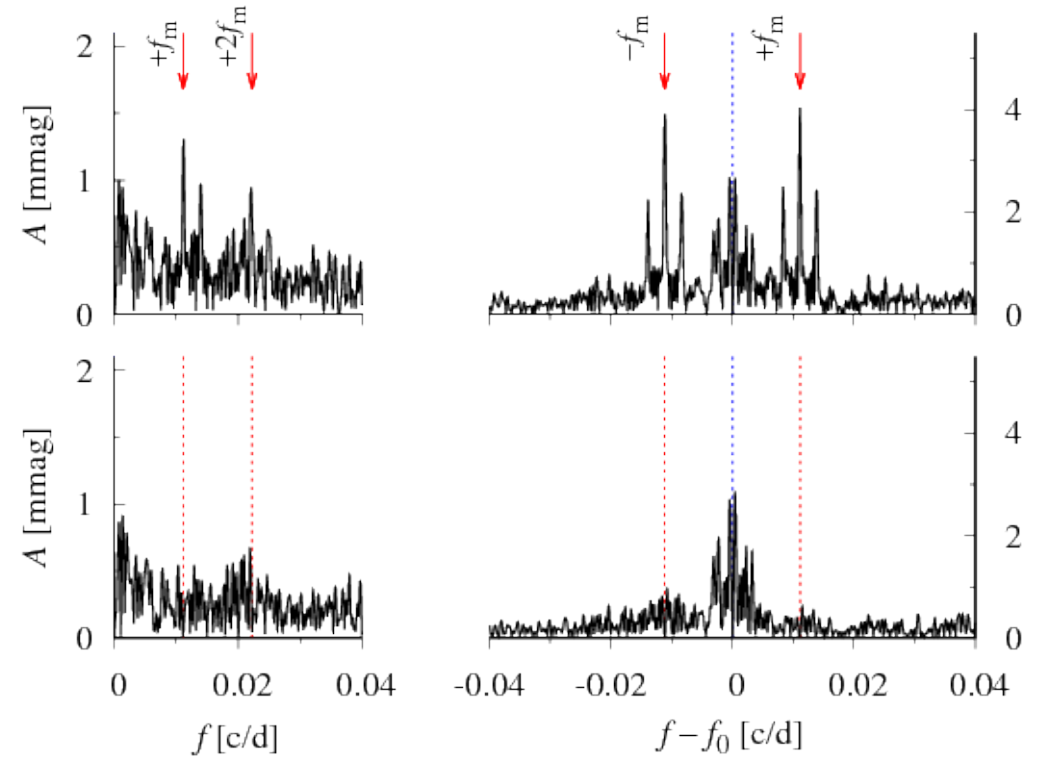}
\caption{Top panel: sections of the frequency spectrum for T2CEP-251 (BL~Her) at the low frequency range (left) and centred at $\fF$ (right), after prewhitening with the fundamental mode and its harmonics (blue dashed line). Modulation side-peaks are marked with arrows (their 1-yr aliases are also well visible). Bottom panel: the corresponding sections of the frequency spectrum after prewhitening with the modulation side-peaks (red dashed lines).}
\label{fig:fspmod251}
\end{figure}

{\bf \offi{T2CEP-251}\tech{BLG511.16.58981-RS}}. In the frequency spectrum we detect equidistant triplets centred at $\fF$ and $2\fF$ as well as a peak at the modulation frequency, $\fm$, and at its harmonic -- Fig.~\ref{fig:fspmod251}. Modulation period is nearly 90\,d. Amplitude modulation is clear and illustrated with an animation (see Supplementary Online Material).

\begin{figure}
\includegraphics[width=\columnwidth]{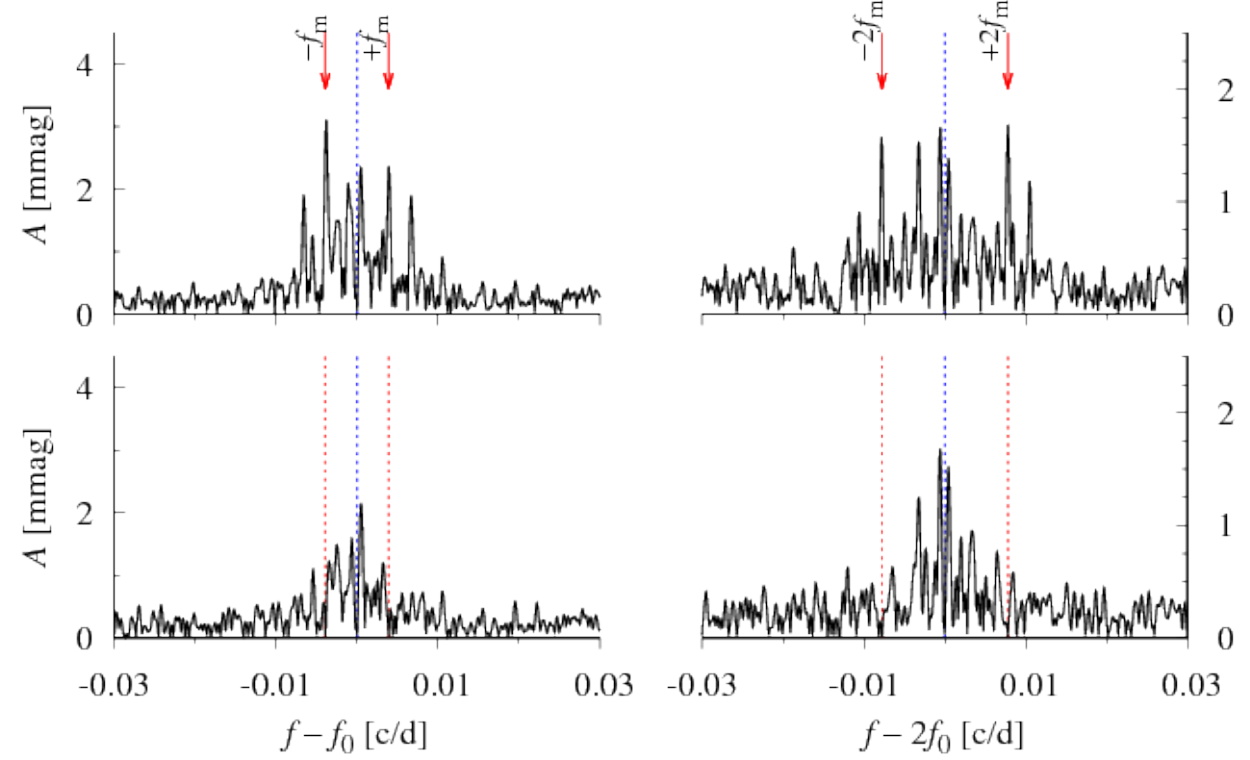}
\caption{Top panel: sections of the frequency spectrum for T2CEP-020 (BL~Her) centred at $\fF$ (left) and $2\fF$ (right), after prewhitening with the fundamental mode and its harmonics (blue dashed lines). Modulation side-peaks are marked with arrows. Bottom panel: The corresponding sections of the frequency spectrum after prewhitening with the modulation side-peaks (red dashed lines).}
\label{fig:fspmod020}
\end{figure}

{\bf \offi{T2CEP-020}\tech{BLG611.21.132345-EP}}. The star clearly shows a modulation of pulsation with a period of $257.3$\,d and with a very interesting frequency spectrum pattern. While at $\fF$ we detect triplet components, $\fF\pm\fm$, at the harmonic we detect the quintuplet components, $2\fF\pm 2\fm$, but not the $2\fF\pm\fm$ side-peaks -- Fig.~\ref{fig:fspmod020}. At $3\fF$, signature of all quintuplet components can be found, but the signals are weak. In the time domain we observe peculiar amplitude modulation: the amplitude is roughly constant for most of the modulation cycle and increases a little only during its relatively small fraction (see animation in Supplementary Online Material).

{\bf \offi{T2CEP-178}\tech{BLG505.15.201926-RS}}. After prewhitening with the fundamental mode and its harmonics, frequency spectrum of this star is dominated by a strong low-frequency signal (due to trend) and remnant power at $k\fF$. After time-dependent prewhitening on a season-to-season basis, modulation with a period of $\Pm=167.3$\,d is clear. Strong modulation peaks are detected at $\fm$ and at $\fF\pm\fm$. For animation, see Supplementary Online Material.


{\bf \offi{T2CEP-159}\tech{BLG535.18.128627-RS}}. Modulation is rather weak. We detect only two side-peaks, at $\fF\!-\!\fm$ and at $\fm$ that may correspond to a modulation with a period of $146.3$\,d. Weak ($\sn=4.4$) additional signal with frequency lower than $\fF$ is present in the spectrum after prewhitening. The corresponding period ratio is $\PF/\px\!\approx\!0.89$. This additional peak may arise, e.g. due to secondary modulation. 

\begin{figure}
\includegraphics[width=\columnwidth]{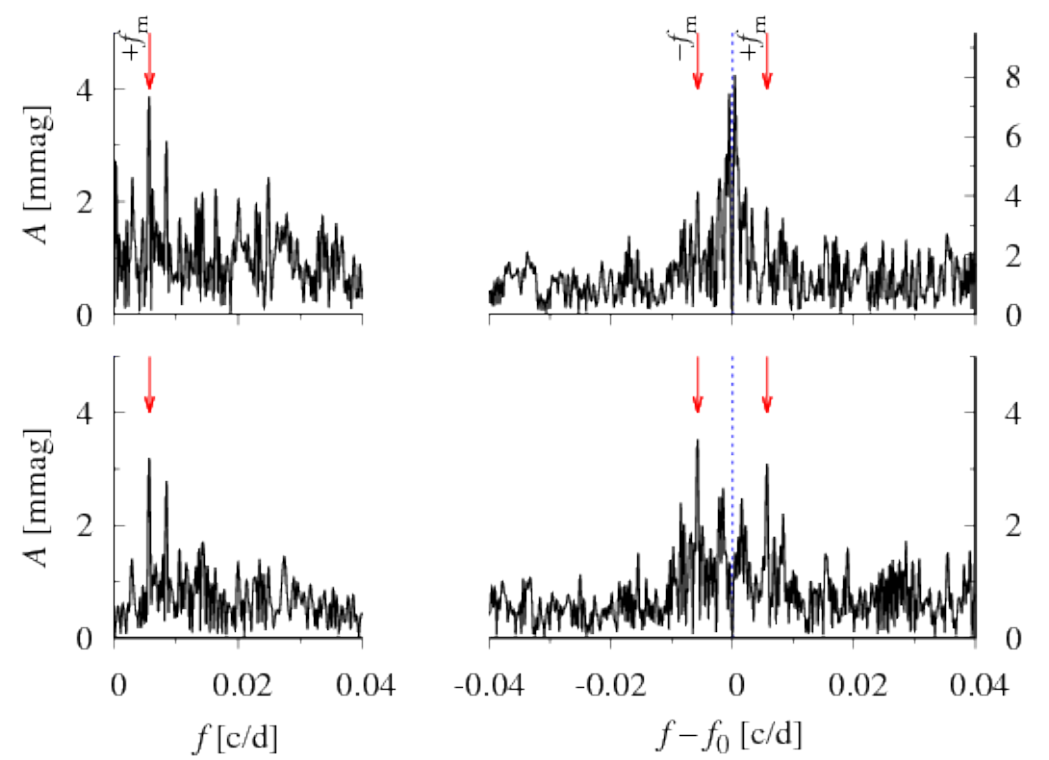}
\caption{Top panel: sections of the frequency spectrum for T2CEP-274 (BL~Her) at the low frequency range (left) and centred at $\fF$ (right), after prewhitening with the fundamental mode and its harmonics (blue dashed line). Modulation side-peaks are marked with arrows. Bottom panel: The corresponding sections of the frequency spectrum after time-dependent prewhitening ($\Delta t\approx 500$\,d). Only now the modulation side-peaks at $\fF$ become well visible.}
\label{fig:fspmod274}
\end{figure}

{\bf \offi{T2CEP-274}\tech{BLG516.13.89169-RS}}. Signal in the low frequency range is clearly detected (Fig.~\ref{fig:fspmod274}; top left) and suggests a modulation with a period of about $175$\,d. The corresponding side-peaks at $\fF$ are barely noticeable as the spectrum is dominated by residual power at $\fF$ (Fig.~\ref{fig:fspmod274}; top right). Hence, we applied the time-dependent prewhitening with a rather large $\Delta t\approx 500$\,d so the prewhitening does not cut the suspected modulation of relatively long period. As a result, the modulation side-peaks become clearly visible and significant (Fig.~\ref{fig:fspmod274}; bottom right). We also note that with $\Delta t\approx 500$\,d, the power at $\fF$ is not entirely removed. Except the peaks at $\fm$ and $\fF\pm\fm$, no other modulation peaks are detected. Modulation is dominated by phase variation, illustrated with an animation (see Supplementary Online Material).

\begin{figure}
\includegraphics[width=\columnwidth]{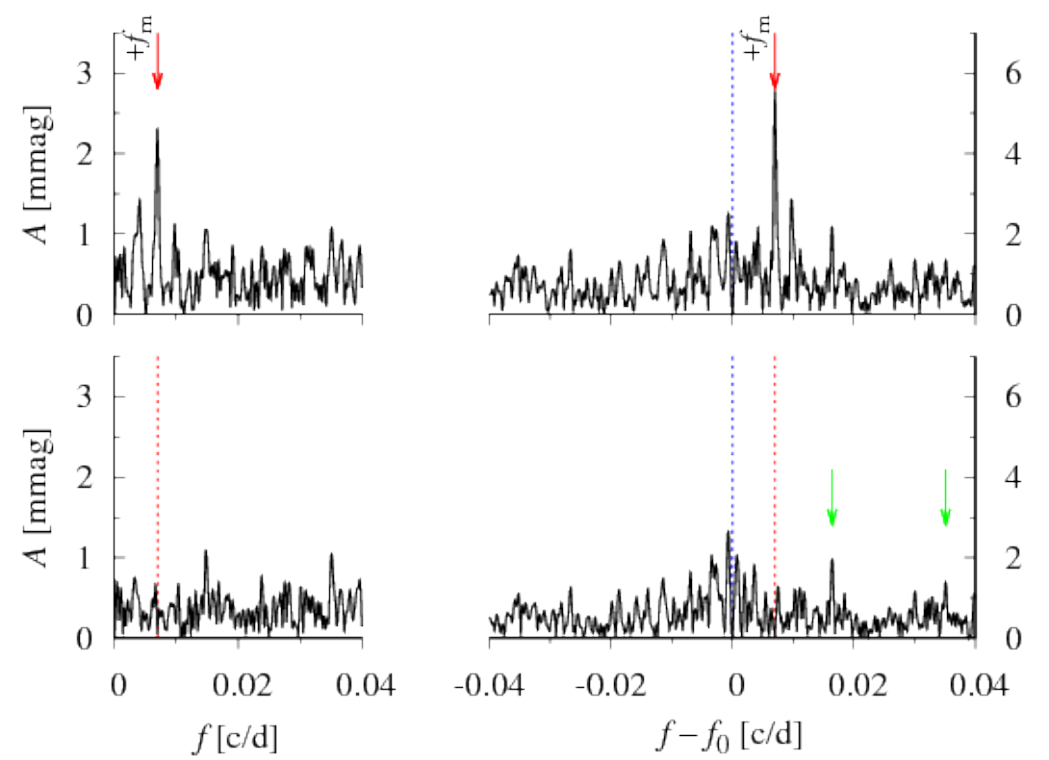}
\caption{Top panel: sections of the frequency spectrum for T2CEP-214 (BL~Her) at the low frequency range (left) and centred at $\fF$ (right), after prewhitening with the fundamental mode and its harmonics (blue dashed line). Modulation side-peaks are marked with arrows. Bottom panel: The corresponding sections of the frequency spectrum after prewhitening with the modulation side-peaks (red dashed lines).}
\label{fig:fspmod214}
\end{figure}

{\bf \offi{T2CEP-214}\tech{BLG508.29.18238-PM}}. In addition to remnant power at $\fF$ (phase change), side-peaks at $\fm$, $\fF+\fm$ and $2\fF+\fm$ are clearly detected (Fig.~\ref{fig:fspmod214}). After time-dependent prewhitening on a season-to-season basis two additional peaks located close to $\fF$ become significant (marked with arrows in the bottom-right panel of Fig.~\ref{fig:fspmod214}) and may correspond to secondary modulations (with periods of 60.8 and 28.46\,d).

{\bf \offi{T2CEP-361}\tech{BLG617.03.69359}}. The data are rather scarce, but modulation with $\Pm=140.4$\,d is clear and manifests with equidistant triplets at $\fF$, $2\fF$ and as a signal at modulation frequency, $\fm$. After prewhitening with the modulation components, a significant and resolved peak is detected at $\fF$, which may correspond to secondary modulation of longer period (more than $1000$\,d). In addition, a significant peak at $\fF-\fm/2$ is present. Despite poor data, modulation can be noticed in an animation (see Supplementary Online Material).


{\bf \offi{T2CEP-775}\tech{BLG534.13.6292}}. Analysis of this star is particularly difficult because of its close-to-integer pulsation period, $\PF=2.991630$\,d. As a consequence, the light curve is composed of three separate chunks, at the brightness maximum, at the descending branch and at the brightness minimum (Fig.~\ref{fig:t2cepBL_mod_lc}). In addition slow trends are present in the data which we removed by fitting two low-frequency sine functions. Still, the modulation with $\Pm=79.25$\,d is easily detectable. In the frequency spectrum it manifests through prominent modulation side-peaks located at $\fm$, and at $\fF\pm\fm$. Side-peaks at higher order harmonics are not detected after the prewhitening, which may be due to peculiar pulsation frequency: they are expected at the location of daily aliases of the peaks at $\fm$ and $\fF\pm\fm$. In the animation, modulation is best visible at the minimum brightness (see Supplementary Online Material).


{\bf \offi{T2CEP-878}\tech{BLG513.10.81158}}. The strongest modulation peak is detected at the modulation frequency, which means that the mean brightness of this BL~Her star is strongly modulated with a rather short period of $38.54$\,d. The side-peaks are also detected at $\fF\pm\fm$ and at $2\fF+\fm$. After prewhitening, additional significant and resolved peaks are detected at $\fF$ which may correspond to secondary modulations on longer time-scales. However, we do not detect other modulation side-peaks with the same separations, that would confirm the secondary modulations. Animation nicely confirms a significant modulation of the mean brightness (see Supplementary Online Material).


{\bf \offi{T2CEP-543}\tech{BLG652.15.73587-PM}}. Modulation period is rather short, $39.70$\,d. Only two side-peaks are firmly detected, at $\fm$ and at $\fF-\fm$. Modulation of mean brightness is clear and illustrated with an animation (see Supplementary Online Material). 

\begin{figure}
\includegraphics[width=\columnwidth]{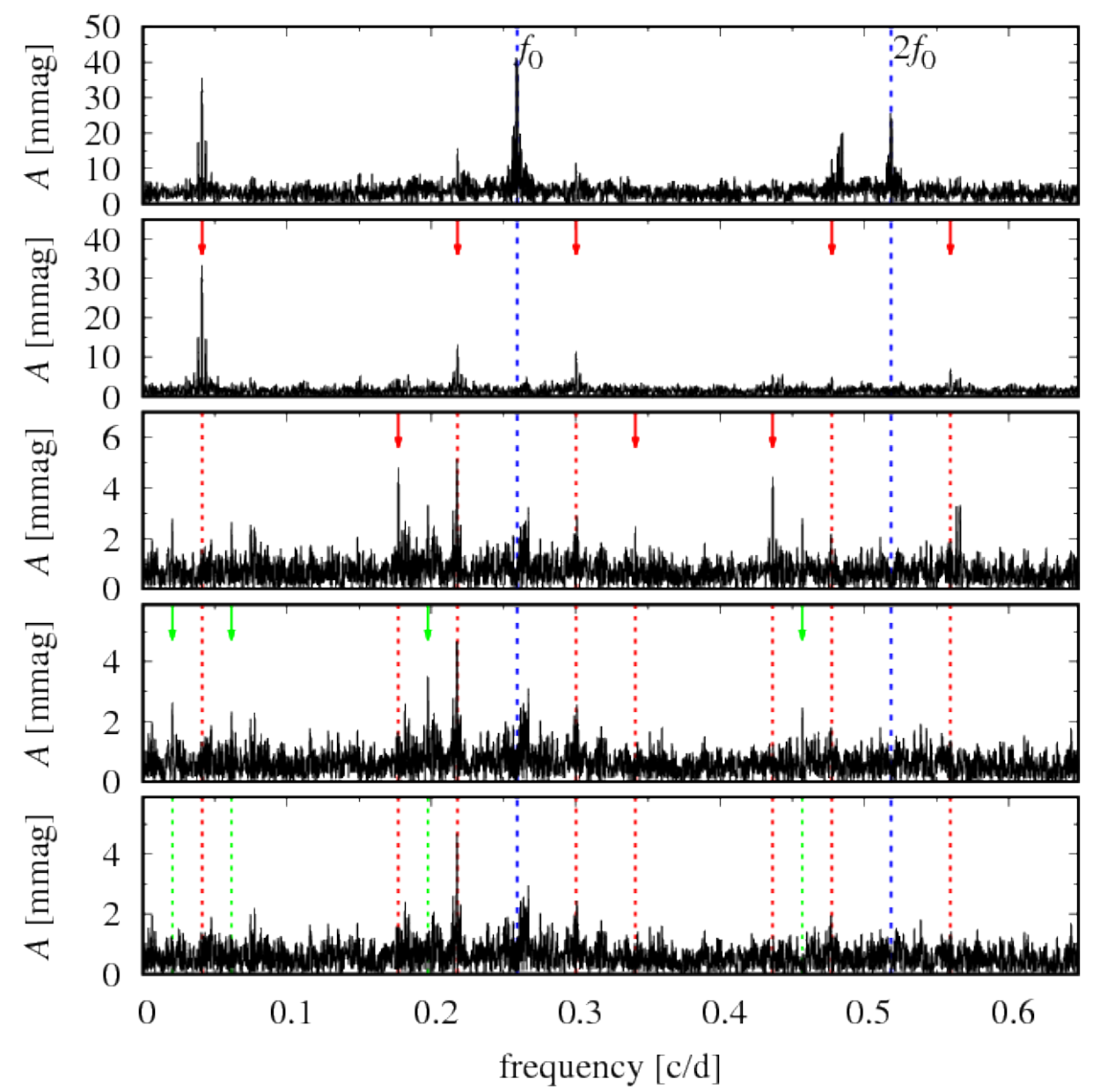}
\caption{Prewhitening sequence for modulated BL~Her star \offi{T2CEP-273}\tech{BLG513.30.132494}. Dashed lines indicate the location of prewhitened frequencies, arrows indicate newly detected peaks. After standard prewhitening with the fundamental mode and its harmonics strong unresolved remnant power was detected (top panel). Additional modulation side-peaks are detected after time-dependent prewhitening ($\Delta t\approx 120$\,d; panels 2--5).}
\label{fig:fspmod273}
\end{figure}

{\bf \offi{T2CEP-273}\tech{BLG513.30.132494-PM}}. After standard prewhitening with fundamental mode and its harmonics, remnant unresolved power dominates the frequency spectrum, as illustrated in the top panel of Fig.~\ref{fig:fspmod273}. As soon as this unresolved power is removed with time-dependent prewhitening ($\Delta t\approx 120$\,d; half of the season), a very rich spectrum of modulated side-peaks is revealed as demonstrated in the consecutive panels of Fig.~\ref{fig:fspmod273}. Full quintuplet is detected at $\fF$. Separation between side-peaks corresponds to modulation period of 24.361\,d. Incomplete quintuplet is also detected at $2\fF$. In the low frequency range, signal at modulation frequency is detected. Interestingly, after prewhitening with all the modulation side-peaks, another set of peaks, located at $\fF-3\fm/2$, $2\fF-3\fm/2$, $\fm/2$ and $3\fm/2$ becomes apparent, but these peaks are weak. The modulation is illustrated with an animation (see Supplementary Online Material).

In several stars we suspect a quasi-periodic phase modulation, on a time-scale of a few hundreds of days. Firm detection of such long-period modulation, with typically seven observing seasons, and for stars in which irregular phase and amplitude variations are common, is challenging. Unambiguous interpretation of the frequency spectra, due to aliasing problems and superposition of the quasi-periodic modulation and irregular changes, is sometimes difficult. In the stars discussed below, the modulation signature is detected in the frequency spectrum (triplet structures at $k\fF$), in the time-dependent Fourier analysis (phase changes on the same time-scale) and finally illustrated with animations. In one of these stars phase modulation may reflect the light time travel effect due to motion in a binary system, as we discuss in more detail in Sect.~\ref{ssec:binary}.

\begin{figure}
\includegraphics[width=\columnwidth]{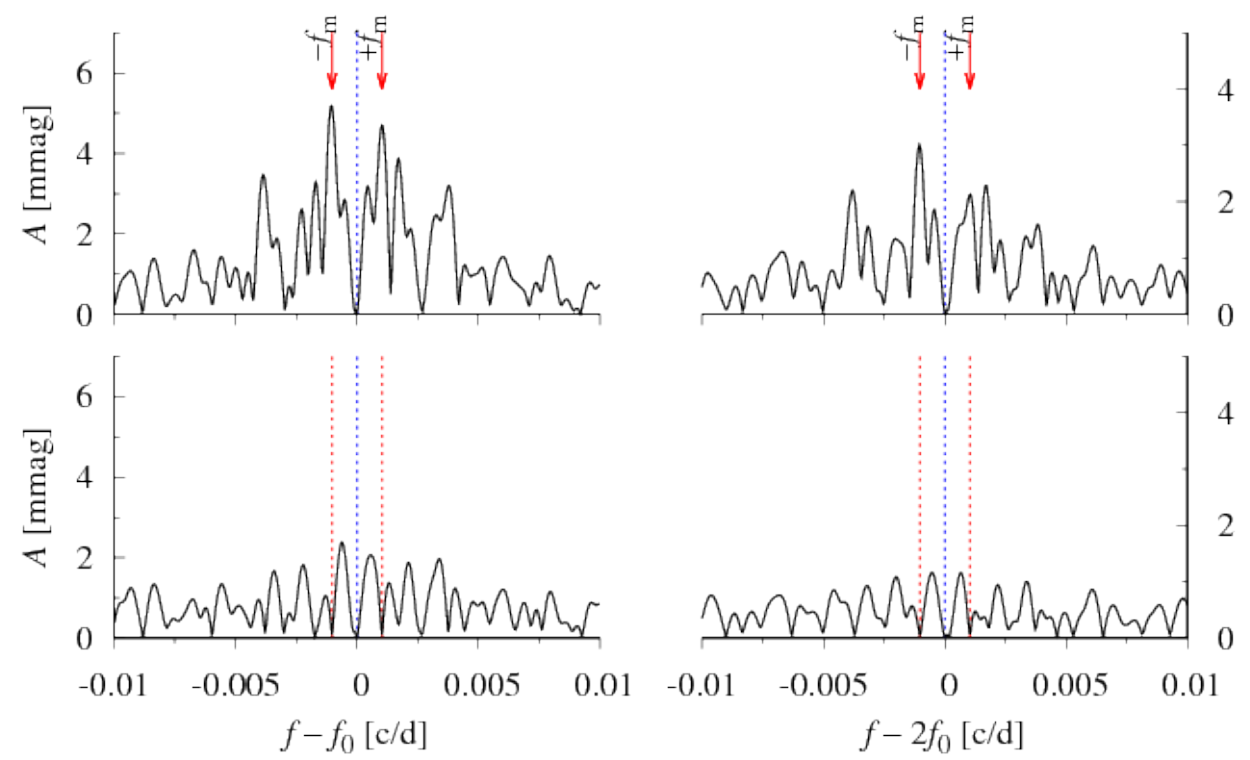}
\caption{Top panel: sections of the frequency spectrum for T2CEP-122 (BL~Her) centred at $\fF$ (left) and $2\fF$ (right), after prewhitening with the fundamental mode and its harmonics (blue dashed lines). Modulation side-peaks are marked with arrows. Bottom panel: the corresponding sections of the frequency spectrum after prewhitening with the modulation side-peaks (red dashed lines).}
\label{fig:fspmod122}
\end{figure}

\begin{figure}
\includegraphics[width=\columnwidth]{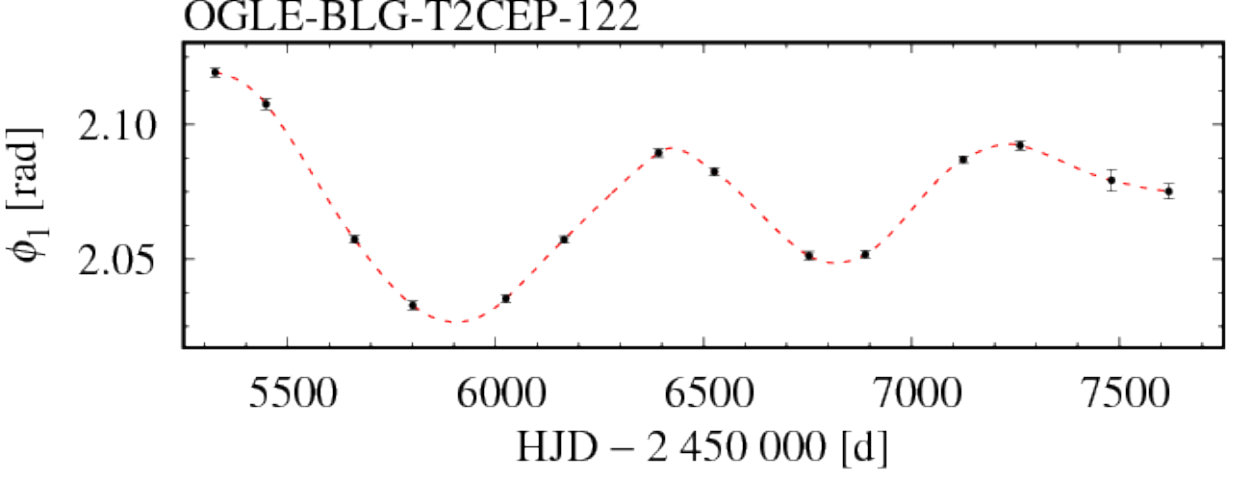}
\includegraphics[width=\columnwidth]{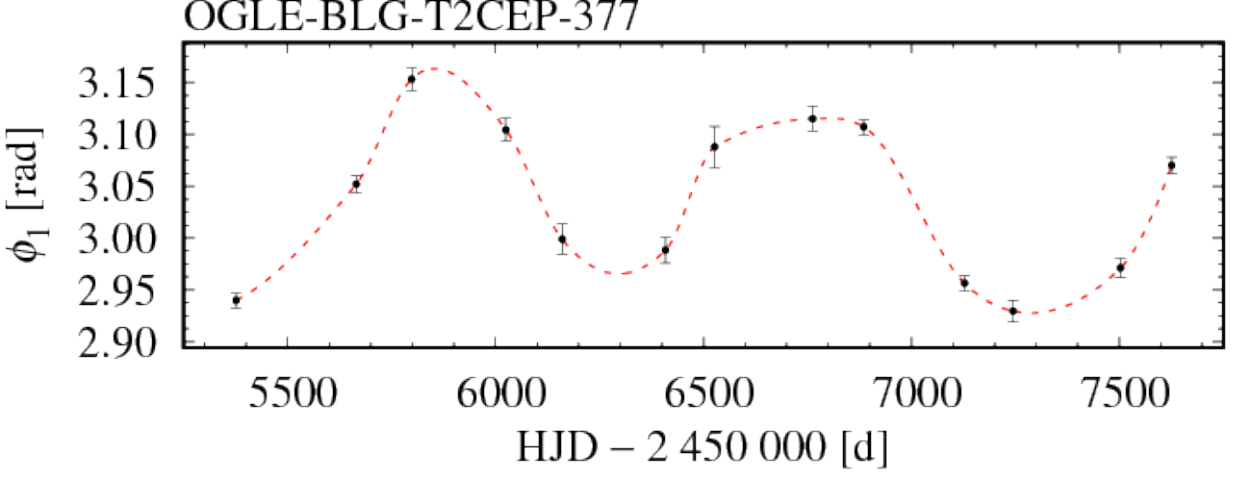}
\includegraphics[width=\columnwidth]{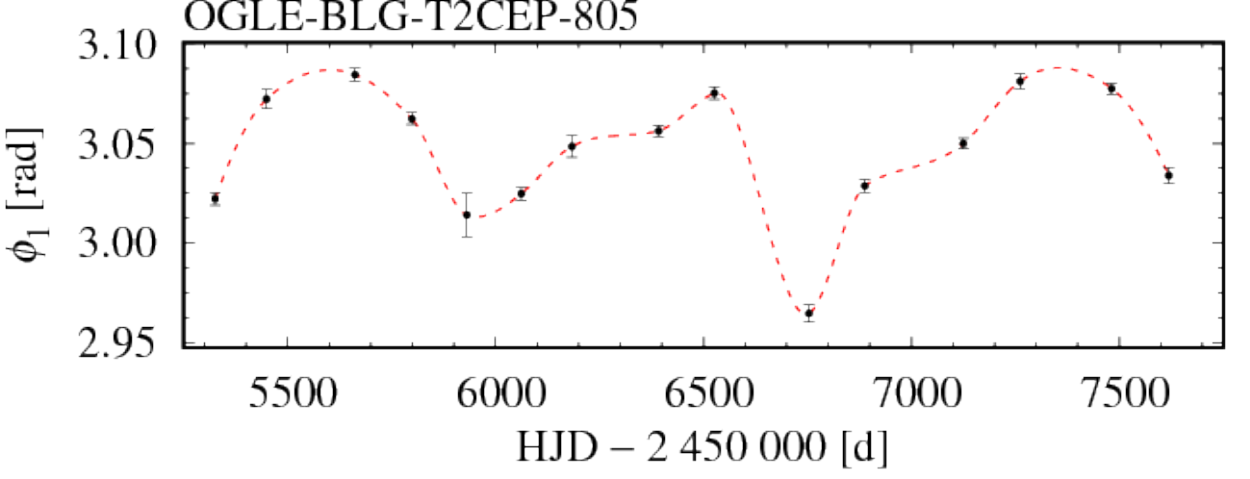}
\caption{Quasi-periodic phase modulation in three BL~Her-type stars illustrated with the help of time-dependent Fourier analysis (with a time resolution of $\sim$half of the observing season). Data were fitted with spline function to guide the eye. Stars' IDs are given in the top section of each panel.}
\label{fig:tdfd_logpmod}
\end{figure}

Frequency spectrum for \offi{T2CEP-122}\tech{BLG501.29.67488-RS}, centred at $\fF$ and $2\fF$, is illustrated in Fig.~\ref{fig:fspmod122}. Symmetric triplets are detected up to $6\fF$. Frequency separation suggests a modulation with a period of 973\,d. Time dependent Fourier analysis (Fig.~\ref{fig:tdfd_logpmod}, top panel) clearly indicates phase changes on the same time scale. Animation (Supplementary Online Material) nicely illustrates the phase modulation.

A very similar frequency spectra were obtained for \offi{T2CEP-377}\tech{BLG613.32.38453-PM} and \offi{T2CEP-805}\tech{BLG500.27.34197-RS}, and suggest the modulation with periods of 957\,d and 865\,d, respectively. Time-dependent Fourier analysis (Fig.~\ref{fig:tdfd_logpmod}; middle and bottom panels) indeed shows the phase modulation on the same time-scale. In T2CEP-805, after the prewhitening,  we detect a lot of remnant power and additional peaks close to the fundamental mode frequency and its harmonics; secondary modulations are possible. Modulations in T2CEP-377 and T2CEP-805 are illustrated with animations (see Supplementary Online Material).

We note that the above long-period phase modulations are irregular, as Fig.~\ref{fig:tdfd_logpmod} nicely illustrates. The determined modulation periods are rather modulation time-scales. Phase variations are often complex in type II Cepheids and may temporarily mimic the quasi-periodic variability. We cannot rule out such possibility for the discussed stars, as the data cover little more than two modulation cycles. More detailed discussion is postponed to Sect.~\ref{ssec:disc_mod}.

\subsection{Periodic modulation of pulsation: W~Vir variables}

\begin{figure*}
\includegraphics[width=.66\columnwidth]{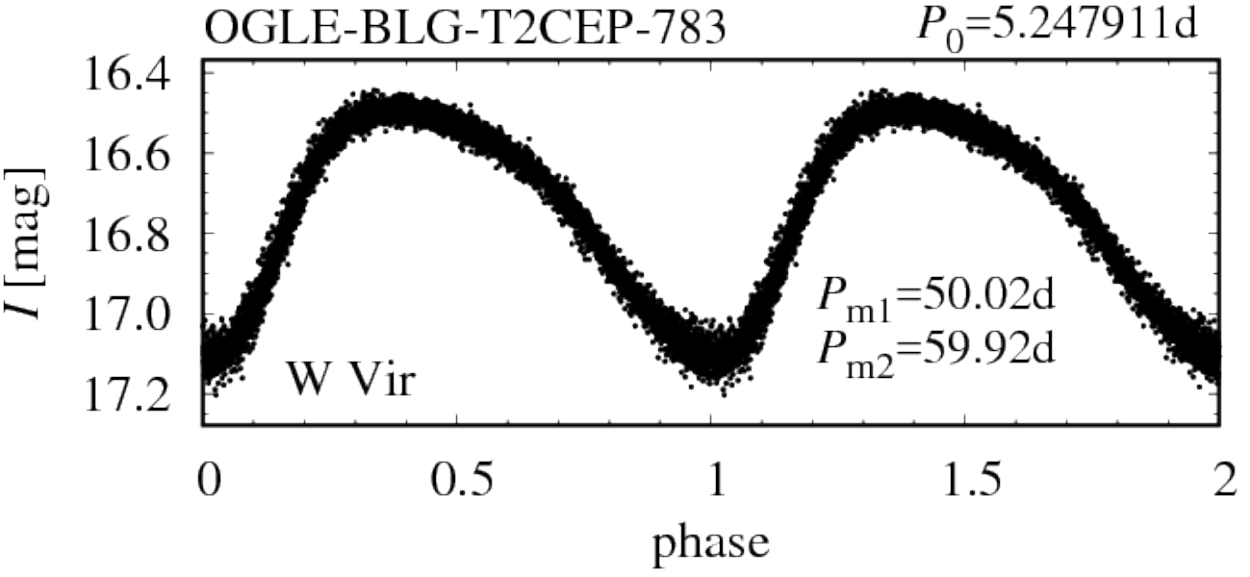}
\includegraphics[width=.66\columnwidth]{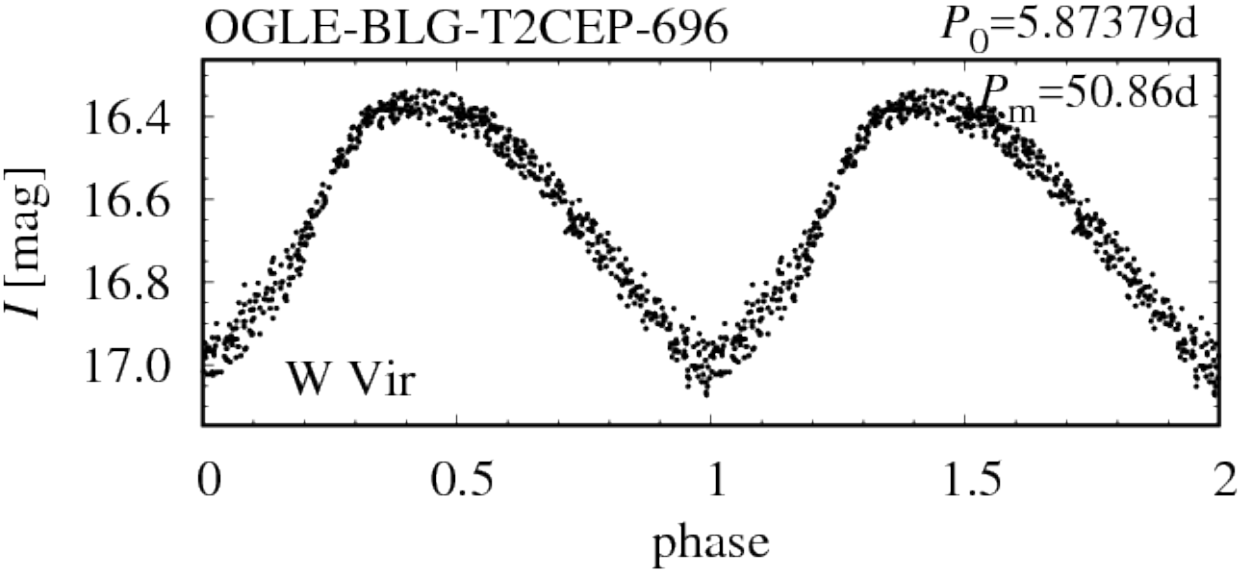}
\includegraphics[width=.66\columnwidth]{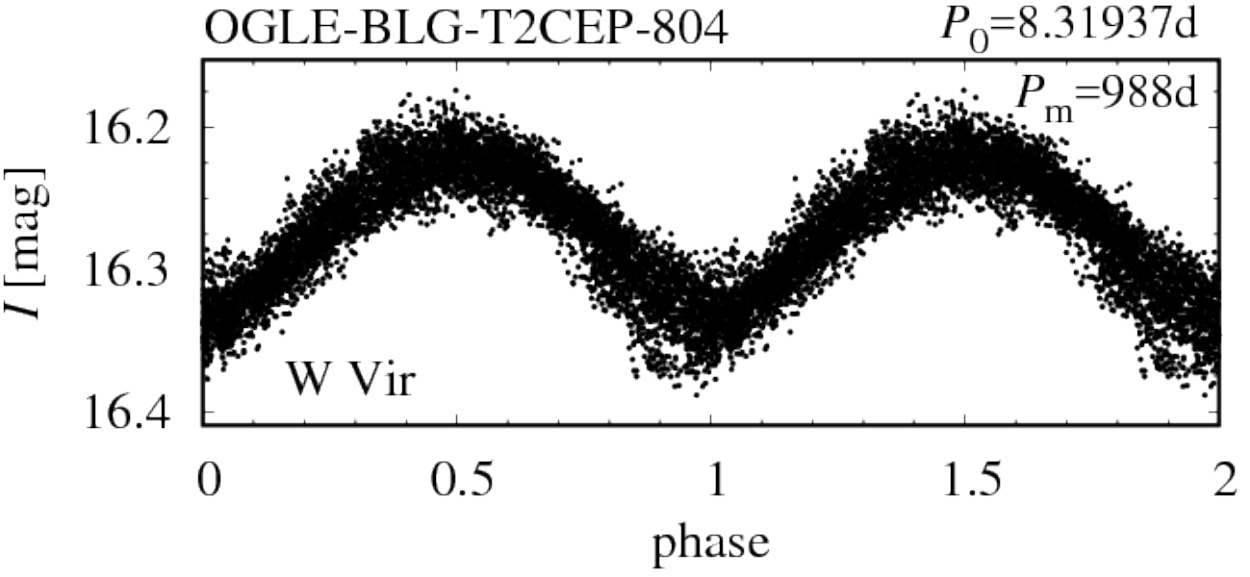}
\includegraphics[width=.66\columnwidth]{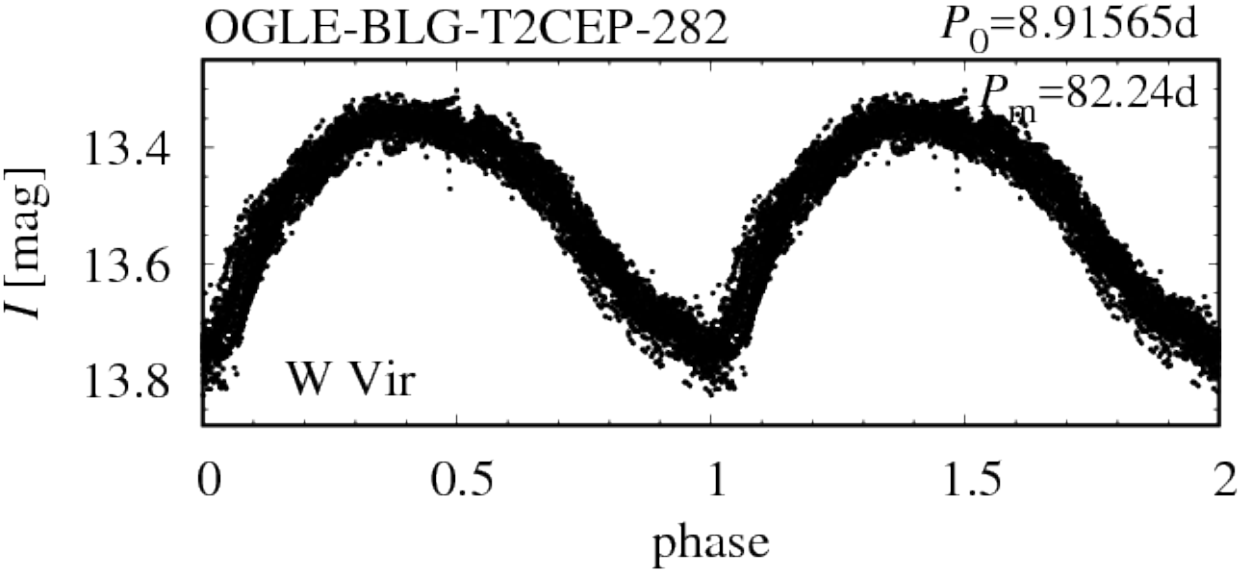}
\includegraphics[width=.66\columnwidth]{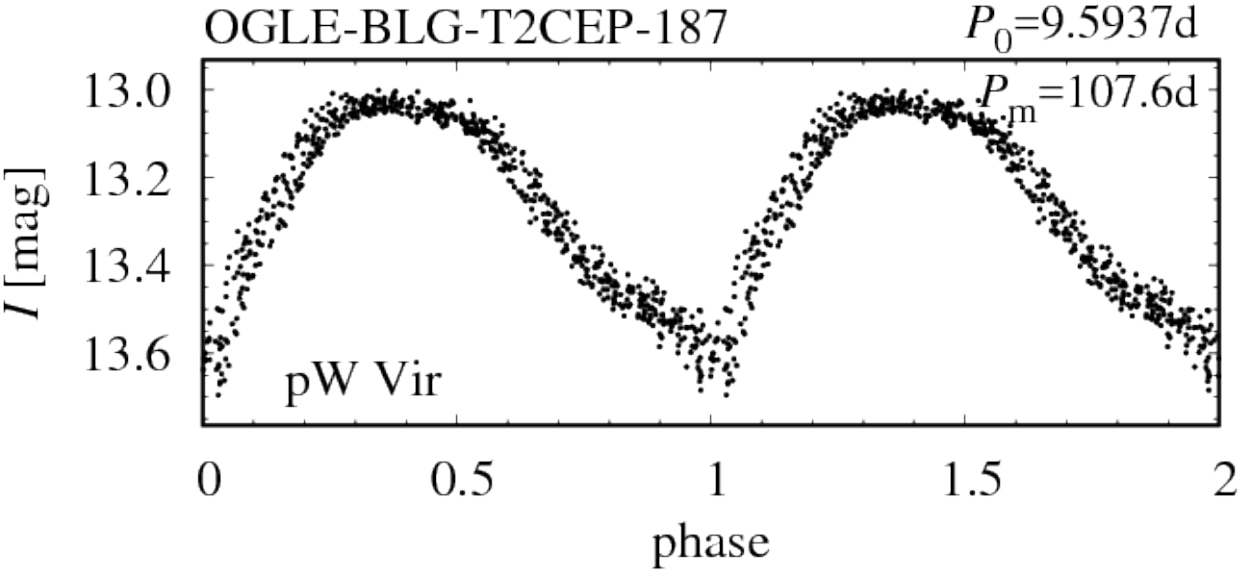}
\includegraphics[width=.66\columnwidth]{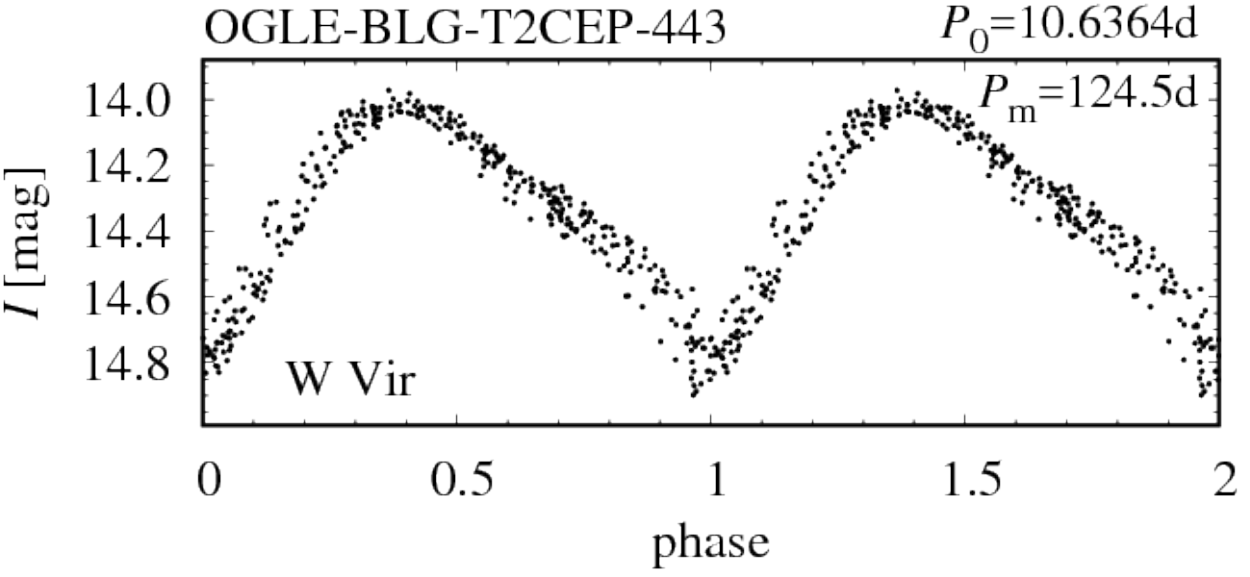}
\includegraphics[width=.66\columnwidth]{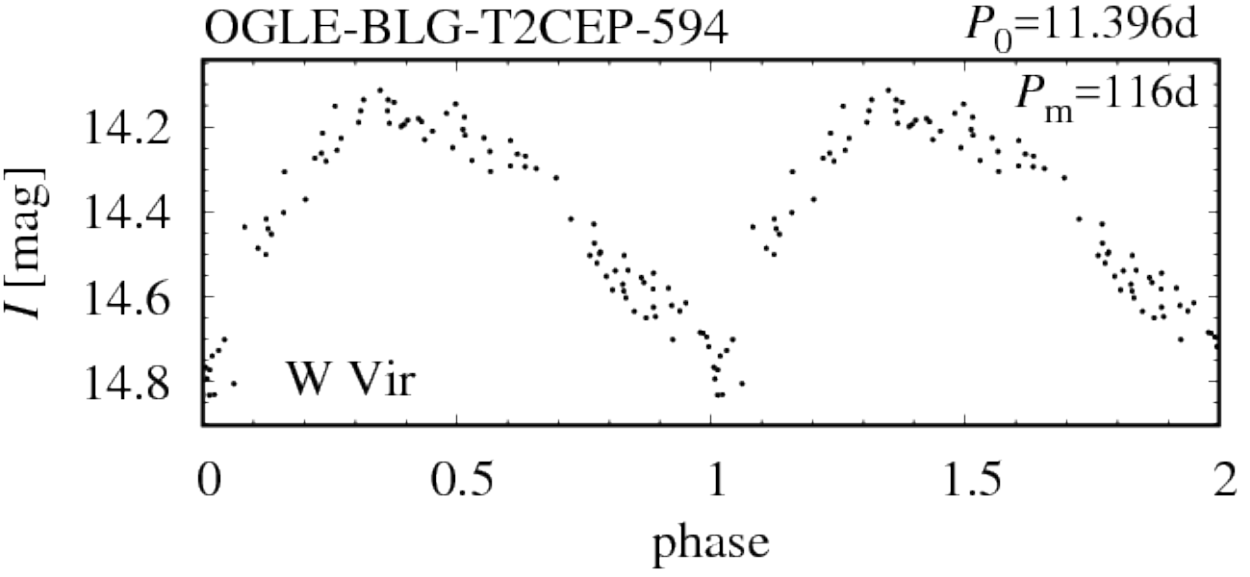}
\includegraphics[width=.66\columnwidth]{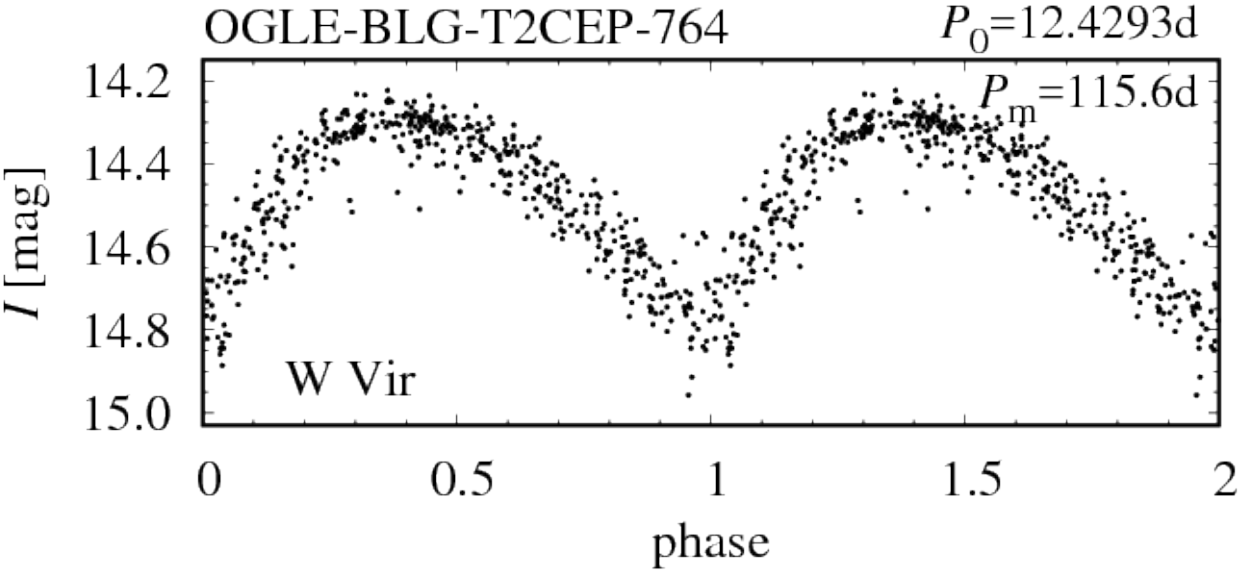}
\includegraphics[width=.66\columnwidth]{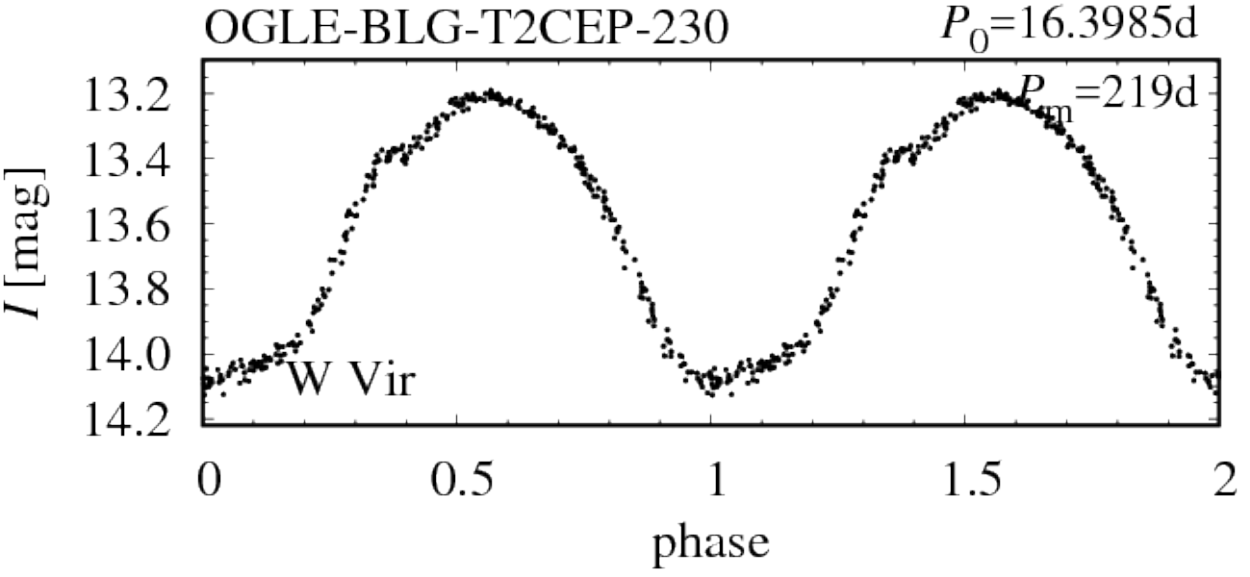}
\caption{Collection of phased light curves for modulated W~Vir stars. Star's ID, type, pulsation period and modulation period are given in each panel. Stars are sorted by the increasing pulsation period.}
\label{fig:t2cepWV_mod_lc}
\end{figure*}

Search for modulation is even more difficult for W~Vir stars, as irregular phase/amplitude changes, that produce power excess at $\fF$ after the prewhitening, are more common. Still, for a few W~Vir stars the detection of modulation is convincing. Their basic properties are given in the bottom section of Tab.~\ref{tab:mod} and the light curves folded with pulsation period are plotted in Fig.~\ref{fig:t2cepWV_mod_lc}. In contrast to BL~Her variables, we rarely observe equidistant triplets at $\fF$ and its harmonics. More commonly, doublets are present. Signal at the modulation frequency is detected in three stars. Pulsation periods cover a wide range; the shortest period is $\approx$5.25\,d, the longest is $\approx$16.4\,d. Modulation periods are rather short, except in one star, all below 220\,d. In one star we detect two modulation periods. Below we briefly discuss the modulated W~Vir stars.

\begin{figure}
\includegraphics[width=\columnwidth]{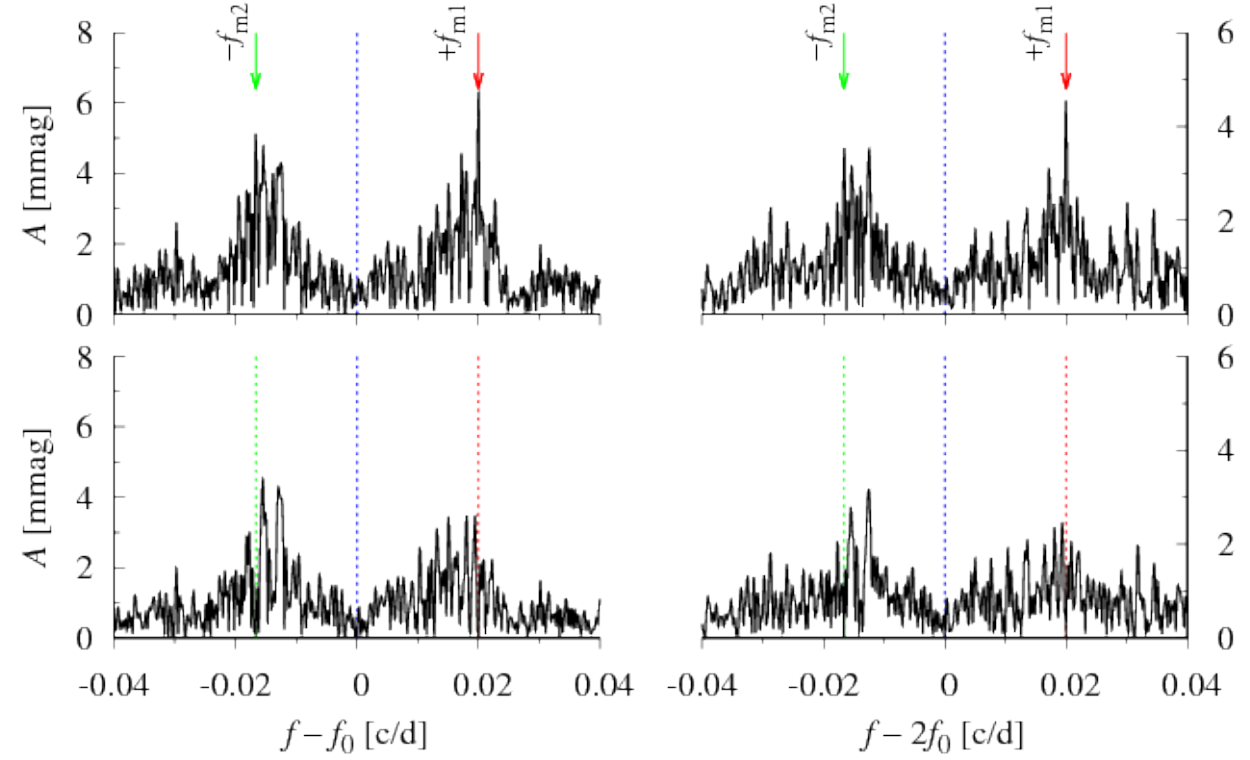}
\caption{Top panel: sections of the frequency spectrum for T2CEP-783 (W~Vir) centred at $\fF$ (left) and $2\fF$ (right), after time-dependent prewhitening with the fundamental mode and its harmonics ($\Delta t\approx 125$\,d, half of the season; blue dashed lines). Modulation side-peaks are marked with arrows. Bottom panel: The corresponding sections of the frequency spectrum after prewhitening  with the modulation side-peaks (red and green dashed lines).}
\label{fig:fspmod783}
\end{figure}

{\bf \offi{T2CEP-783}\tech{BLG500.12.142517-EP}}. Pulsation of this star is irregular, both amplitude and phase vary, which manifests in strong remnant power at $k\fF$ after prewhitening. In the top panel of Fig.~\ref{fig:fspmod783} we show the sections of frequency spectrum centred at $\fF$ and $2\fF$ after time-dependent prewhitening ($\Delta t\approx 125$\,d; half of the season). Two families of side-peaks are detected: one at $k\fF+\fmj$, the other at $k\fF-\fmd$ with $k=1,\ldots,3$. If these peaks are interpreted as due to modulation, the respective modulation periods are $\Pmj=50.07$\,d and $\Pmd=59.92$\,d. Interestingly, within frequency resolution of the data, $5\fmj=6\fmd$. The side-peaks are strongly non-stationary, as illustrated in the bottom panel of Fig.~\ref{fig:fspmod783}.

{\bf \offi{T2CEP-696}\tech{BLG660.28.28062-EP}}. Among pronounced  modulation side-peaks detected at $\fm$, $\fF-\fm$ and $2\fF-\fm$, the one at the modulation frequency is the highest, and corresponds to strong modulation of the mean brightness with a period of $50.86$\,d -- Fig.~\ref{fig:fspmod696} (for animation, see Supplementary Online Material).

\begin{figure}
\includegraphics[width=\columnwidth]{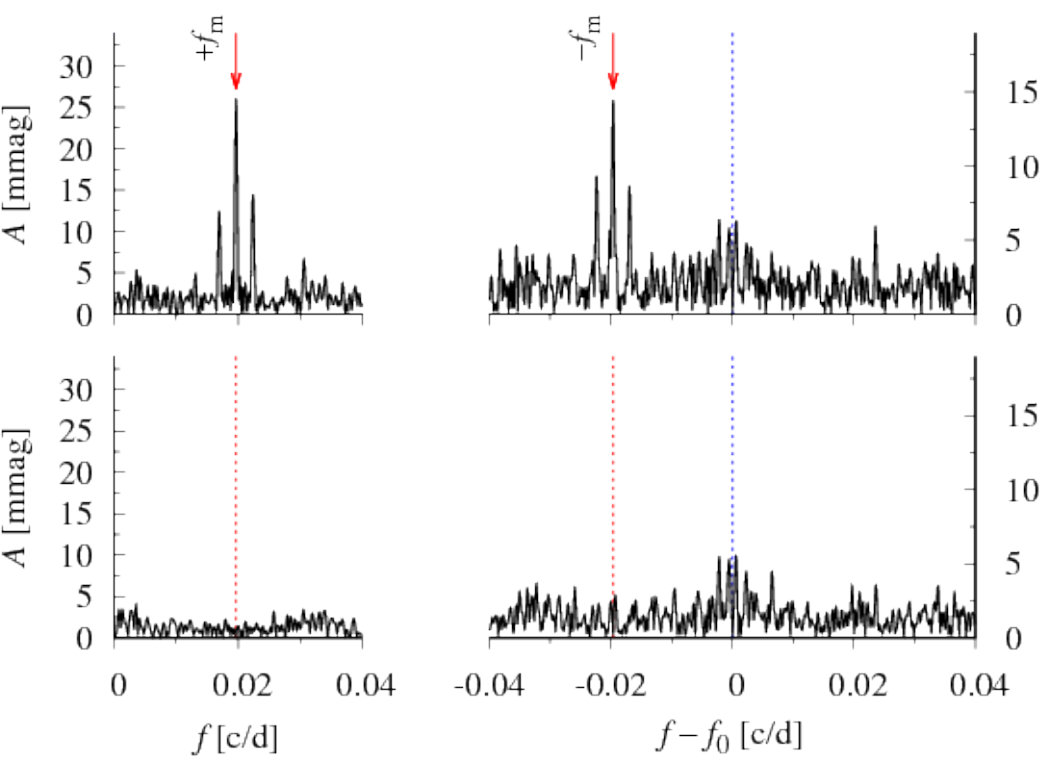}
\caption{Top panel: sections of the frequency spectrum for T2CEP-696 (W~Vir) at the low frequency range (left) and centred at $\fF$ (right), after prewhitening with the fundamental mode and its harmonics (blue dashed line). Modulation side-peaks are marked with arrows (their 1-yr aliases are also well visible). Bottom panel: The corresponding sections of the frequency spectrum after prewhitening with the modulation side-peaks (red dashed lines).}
\label{fig:fspmod696}
\end{figure}

{\bf \offi{T2CEP-804}\tech{BLG500.27.89659-EP}}. Long-period ($\Pm=988$\,d) modulation manifests through side-peaks located at $\fF+\fm$ and $2\fF+\fm$. Amplitude modulation is pronounced and accompanied with phase variation. After prewhitening with the fundamental mode, its harmonic and modulation side-peaks, remnant power at $\fF$ is present.

{\bf \offi{T2CEP-282}\tech{BLG512.03.173511-EP}}. Fundamental mode and its harmonic are non-stationary. Modulation side-peaks are well resolved and located at $\fF-\fm$ and $2\fF-\fm$ ($\Pm=82.24$\,d). The peaks are wide and strong remnant remains after the prewhitening, which indicates that we observe non-coherent modulation.


{\bf \offi{T2CEP-187}\tech{BLG510.31.74018-PM}}. Peculiar W~Vir star. Side-peaks are detected at $\fF-\fm$ and $2\fF-\fm$, and may correspond to a modulation with $\Pm=107.6$\,d. After prewhitening, a remnant power at $k\fF$ is detected and is due to pronounced phase change (for animation, see Supplementary Online Material). 

{\bf \offi{T2CEP-443}\tech{BLG667.07.151-PM}}. Well resolved modulation peaks become prominent only after time-dependent prewhitening ($\Delta t\approx 120$\,d, half of the season). They are located at $\fm$, $\fF-\fm$ and $\fF+\fm$ and correspond to a modulation with a period of $\Pm=124.5$\,d. 

{\bf \offi{T2CEP-594}\tech{BLG680.19.28827-RS}}. Despite poor data -- two seasons are only available -- the amplitude modulation is clear in the light curve. We detect strong peak at $\fF-\fm$ ($\Pm=116$\,d) and much weaker modulation peak at the harmonic, $2\fF-\fm$ ($\sn=3.6$). 


{\bf \offi{T2CEP-764}\tech{BLG639.29.80329-RS}}. Modulation peaks are detected at $\fm$, $\fF+\fm$ and $2\fF+\fm$, and correspond to a modulation with $\Pm=115.6$\,d. Secondary long-period modulation is also possible but more data (longer time base) are needed to confirm this (for animation, see Supplementary Online Material).


{\bf \offi{T2CEP-230}\tech{OIII-RS}}. Only OGLE-III data are available for this long-period ($\PF=16.3985$\,d) pulsator. The side-peak detected at $\fF-\fm$ is strong and indicates a modulation with $\Pm=219$\,d. After prewhitening, the other side-peak at $\fF+\fm$ is the most prominent in the spectrum; nevertheless, with $\sn=3.8$, it is weak. No other modulation peaks were detected in the spectrum.

\subsection{RV~Tau stars with modulation periods close to $4\PF$?}\label{ssec:rvtp4mod}

\begin{figure}
\includegraphics[width=\columnwidth]{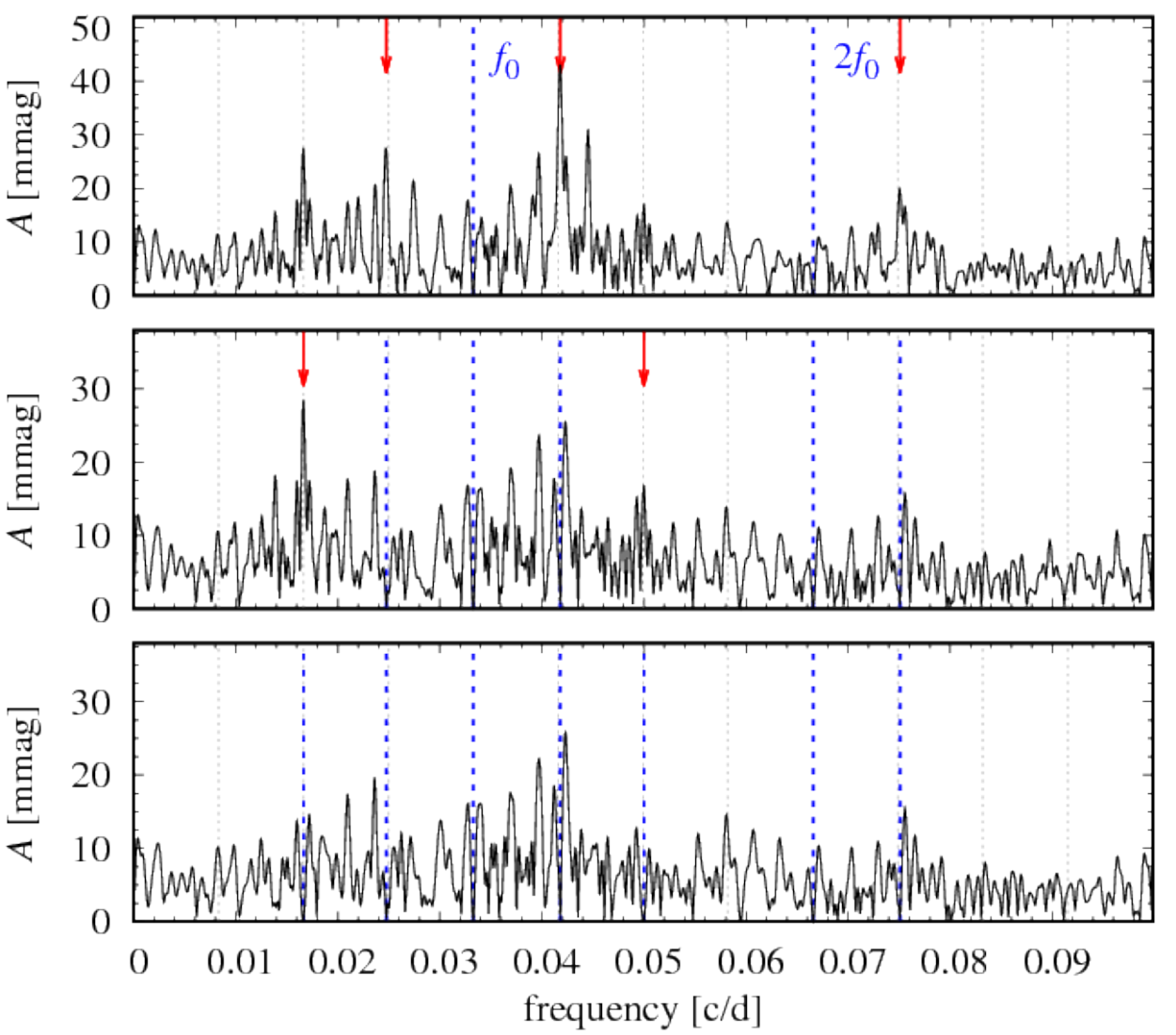}
\includegraphics[width=\columnwidth]{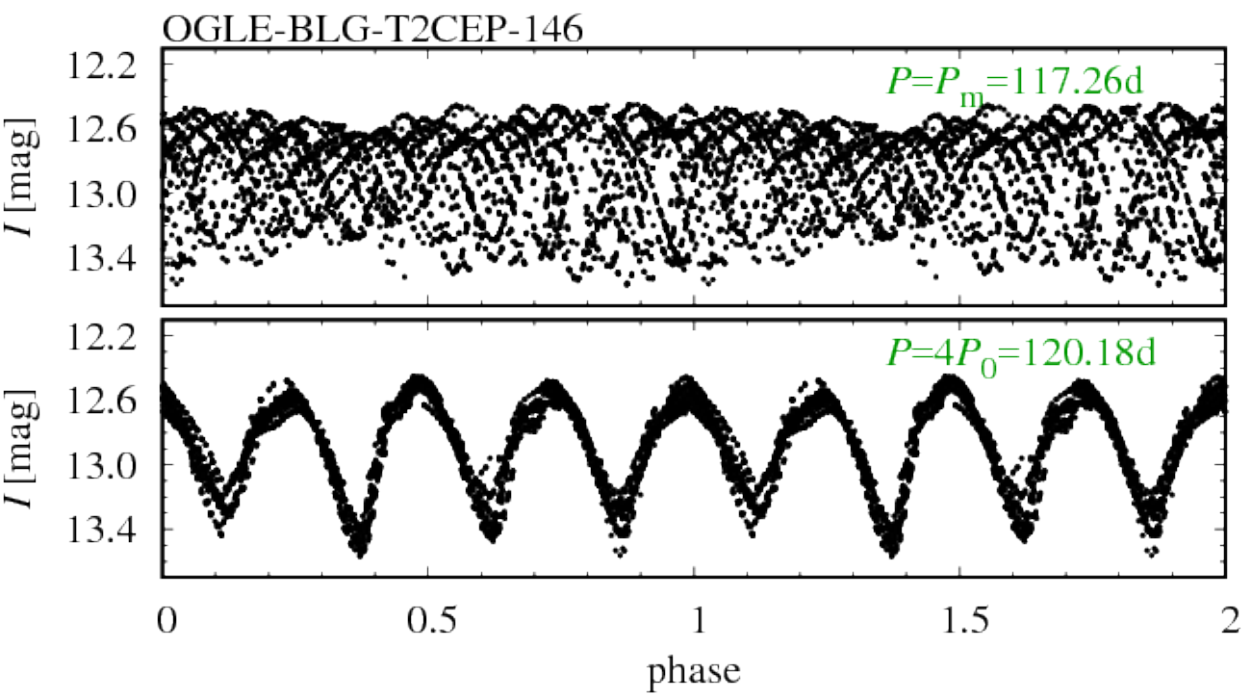}
\caption{Top panels: frequency spectra during consecutive prewhitening steps for RV~Tau star \offi{T2CEP-146}\tech{BLG534.27.120371-PM}. Thin grey vertical lines form a grid with a separation of $\fjc$. Prewhitened frequencies are marked with thick dashed blue lines, new identified frequencies are marked with red arrows. Fundamental mode and its harmonics are prewhitened in all panels. Bottom panels: light curves for \offi{T2CEP-146}\tech{BLG534.27.120371-PM} folded either with suspected modulation period, or with $4\PF$.}
\label{fig:146_p4mod}
\end{figure}

\begin{figure}
\includegraphics[width=\columnwidth]{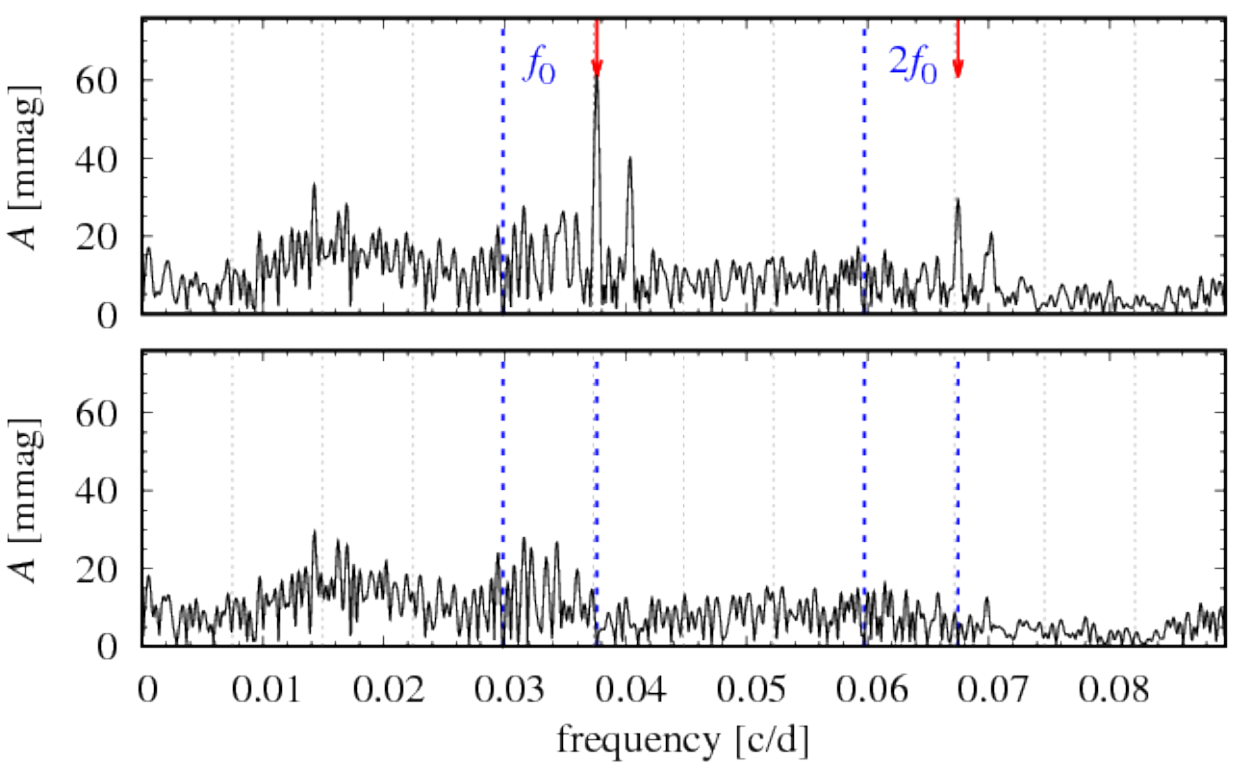}
\includegraphics[width=\columnwidth]{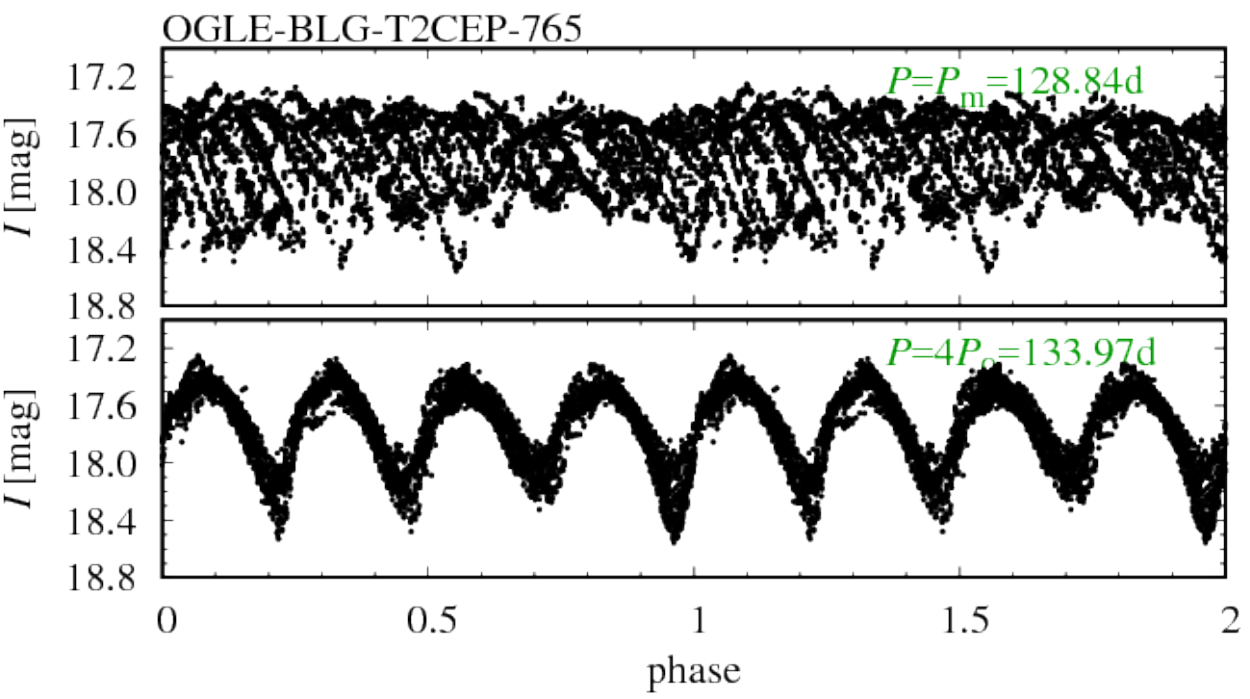}
\caption{The same as Fig.~\ref{fig:146_p4mod} but for \offi{T2CEP-765}\tech{BLG500.31.16917-EP}.}
\label{fig:765_p4mod}
\end{figure}

In the frequency spectra of seven RV~Tau stars, properties of which are listed in Tab.~\ref{tab:rvtp4mod}, we have detected very similar patterns: additional signals located very close to $\fpc$, $\fdc$ (in all stars) and to $\ftrzc$ (in five stars; see sixth column of Tab.~\ref{tab:rvtp4mod}), besides the fundamental mode, its harmonics, and signals centred at sub-harmonic frequencies, $\fjd$ and $\ftd$ (last column of Tab.~\ref{tab:rvtp4mod}). All these stars have periods in the $30<\PF<35$\,d range. Frequency spectra for two stars, T2CEP-146 and T2CEP-765, are illustrated in the top panels of Figs~\ref{fig:146_p4mod} and \ref{fig:765_p4mod}, respectively\footnote{Analogous figures for all stars discussed in this section as available as Supplementary Online Material, Appendix C.}. For these stars, the signals located close to $\fpc$ dominate the spectrum after prewhitening with the fundamental mode and its harmonics. The signals close to $\fdc$ and $\ftc$ (T2CEP-146) are also noticeable. Signal at $\fjd$ is well visible in T2CEP-146, while in T2CEP-765 only a broad and rather insignificant power excess, centred at $\fjd$, is present.

\begin{table*}
\centering
\caption{Properties of RV~Tau stars with suspected modulation such that $\Pm\approx4\PF$. Consecutive columns contain: star's ID, fundamental mode period, $\PF$, modulation period, $\Pm$, and, for comparison, $4\PF$, parameter $r$ defined as $r=(\fm\!-\!\fjc)/(1/\Delta T)$, the list of detected modulation peaks and the list of peaks detected at sub-harmonic frequencies.}
\label{tab:rvtp4mod}
\begin{tabular}{lllllll}
  star ID & $\PF$\,(d) & $\Pm$\,(d) & $4\PF$\,(d) & $r$ & additional peaks & sub-harmonics\\
\hline
T2CEP-135 & 23.2681(4) &  93.07 & 93.07  & 0.002& $\fF\!+\!\fm$, $2\fF\!+\!\fm$                                 &$\fjd$, $\ftd$        \\
\hline
T2CEP-146 & 30.0451(6) & 117.26 & 120.18 & 0.50 & $\fF\!+\!\fm$, $2\fF\!+\!\fm$, $3\fF\!+\!\fm$, $\fF\!-\!\fm$  & $\fjd$, $\ftd$        \\ 
T2CEP-364 & 31.413(2)  & 118.5  & 125.7  & 1.17 & $\fF\!+\!\fm$, $2\fF\!+\!\fm$, $3\fF\!+\!\fm$,                & $\fjd$, $\ftd$, $\fjd\!+\!\fm$\\
T2CEP-401 & 32.815(2)  & 115.7  & 131.3  & 2.48 & $\fF\!+\!\fm$, $2\fF\!+\!\fm$,                                & $\fjd$, $\ftd$, $\fpc$ \\
T2CEP-765 & 33.4931(9) & 128.84 & 133.97 & 0.72 & $\fF\!+\!\fm$, $2\fF\!+\!\fm$, $3\fF\!+\!\fm$,                &    ($\fjd$) \\ 
T2CEP-513 & 33.851(3)  & 123.0  & 135.4  & 1.78 & $\fF\!+\!\fm$, $2\fF\!+\!\fm$, $3\fF\!+\!\fm$ (weak)          &  $\fjd$, $\ftd$ (weak), $\fF\!-\!\fmd$ (weak)\\
T2CEP-868 & 34.545(3)  & 125.4  & 138.2  & 1.76 & $\fF\!+\!\fm$, $2\fF\!+\!\fm$ (weak)                          &  $\fjd$, $\ftd$\\
T2CEP-019 & 34.881(1)  & 135.2  & 139.5  & 0.56 & $\fF\!+\!\fm$, $2\fF\!+\!\fm$, $3\fF\!+\!\fm$,                & $\fjd$, $\ftd$ \\
 \hline  
\end{tabular}
\end{table*}

\begin{figure}
\includegraphics[width=\columnwidth]{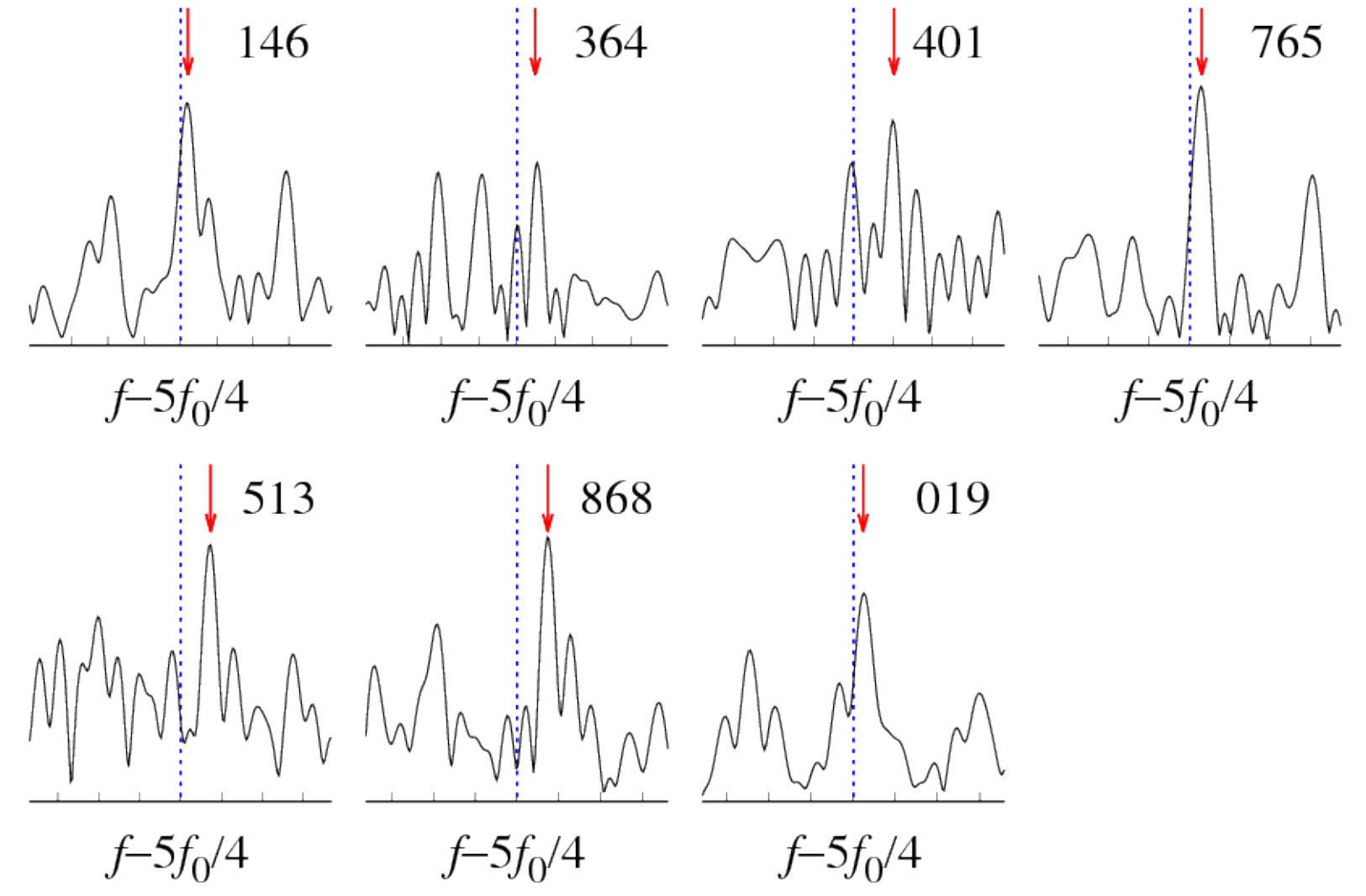}
\caption{Zooms into frequency spectra of seven RV~Tau stars, centred at $\fpc$ (dashed line). Frequency of the highest nearby peak, as inferred from the sine series fit, is marked with arrow. Stars' IDs are given in top right section of each panel.}
\label{fig:resolved}
\end{figure}

Interestingly, in all discussed stars, the additional peaks are not centred on  $\fpc$, $\fdc$, or $\ftrzc$, but have a bit higher frequency. This is illustrated in the seven panels of Fig.~\ref{fig:resolved}, in which we show small sections of the frequency spectra centred at $\fpc$ (dashed line), with the total width of the section equal to $\fjc$. The arrows are placed at the locations of the additional peaks inferred from the sine series fits. The offsets are very small, only in the case of T2CEP-401 the two frequencies are well resolved. In fact, in this case a weak peak centred exactly at $\fpc$ is visible. For a given star, the offsets are the same at $\fpc$, $\fdc$ and $\ftrzc$.

There are three possible interpretations of the discussed frequency patterns. The peaks detected at  $\fpc$, $\fdc$ and $\ftrzc$ may be denoted as $\fx$, $\fF+\fx$ and $2\fF+\fx$, and may correspond to {\it excitation of additional mode} and its linear frequency combinations. Alternatively, the peaks may be denoted as $\fF+\fm$, $2\fF+\fm$ and $3\fF+\fm$ (see the penultimate column of Tab.~\ref{tab:rvtp4mod}) and may be interpreted as due to {\it modulation of the fundamental mode} with a frequency $\fm$, slightly larger than $\fjc$. Admittedly, the offset between the additional peaks and exact harmonics of $\fjc$, i.e. $\fm-\fjc$ (or $\fx-\fpc$) is very small. It can be expressed in the units of formal frequency resolution of the data (which is $1/\Delta T$; $\Delta T$ -- data length), $r=(\fm-\fjc)/(1/\Delta T)$. This parameter is given in the fifth column of Tab.~\ref{tab:rvtp4mod}. Within our conservative approach, only in T2CEP-401 the two frequencies are resolved ($r>2$). Consequently, the third interpretation is {\it period-4 pulsation}. In the first row of Tab.~\ref{tab:rvtp4mod} we have included the data for T2CEP-135, a period-4 candidate discussed in Sect.~\ref{ssec:rvt-p4}. This star has a bit shorter period than the other seven stars, and peaks interpreted as due to period-4 pulsation are exactly centred at $\fpc$ and its harmonics ($r$$\approx$0).

Which of the three scenarios is more likely? In the two bottom panels of Figs~\ref{fig:146_p4mod} and \ref{fig:765_p4mod} we show the light curves for T2CEP-146 and T2CEP-765 folded either with the suspected modulation period, $\Pm=1/\fm$ or with $4\PF$. Since the two periods are very close, both plots are telling. Data folded with $\Pm$ suggest that we deal with periodic modulation of pulsation; amplitude modulation is clear. Data folded with $4\PF$  suggest a period-4 pulsation. Beating of the two modes also cannot be excluded. 

Excitation of additional mode seem the least probable scenario. Assuming that the highest additional peak corresponds to the additional mode, i.e. that $\fx$ is a bit higher than $\fpc$, the period ratios, $\px/\PF$, would be just a bit lower than 0.8, and would be confined in a narrow range ($\sim$0.78--0.80). For the radial first overtone ($\fx\!=\!\fO$), the expected period ratios are lower \citep[see, e.g.][]{s16}, consequently, the additional modes cannot be radial. Non-radial mode calculations are scarce for giant stars. Calculations for RR~Lyrae stars (which are less luminous) show that non-radial low degree modes may be excited, but with frequencies rather close to the frequencies of the radial modes. Their growth rates are at least an order of magnitude lower than for radial modes \citep{vdk98}. Calculations for classical Cepheids (which are more massive) indicate that non-radial low degree modes are not excited \citep{mulet}.

Period-4 interpretation also faces difficulties. First, the peaks are not centred at $\fpc$, $\fdc$, or $\ftrzc$, but are {\em systematically} off. This cannot be interpreted as due to irregular effect or as due to switching of the deep and shallow minima. By analogy with period doubling (Sect.~\ref{ssec:rvt-switch}), these phenomena should lead to more complex structures in the frequency spectra, but still centred at $\fjc$ and its harmonics. Second, in the discussed stars we detect sub-harmonic frequencies as well (Tab.~\ref{tab:rvtp4mod}) -- a signature of period-doubling effect. Sub-harmonic frequencies are obviously also expected for period-4 pulsation. These peaks, however, are well centred at $\fjd$ and $\ftd$ in contrast to peaks located $\fpc$, $\fdc$ or $\ftrzc$.

Modulation of pulsation, with periods slightly shorter than $4\PF$ seems the most likely interpretation. Admittedly, only in one star, T2CEP-146 (Fig.~\ref{fig:146_p4mod}) we detect a weak peak located at $\fF-\fm$ and consequently we detect an equidistant triplet, the most convincing signature of modulation. However, as discussed in Sect.~\ref{sec:data}, doublets are a common signature of modulation in ground-based observations. In Addition, the modulation side-peaks may be intrinsically highly asymmetric as a result of specific phase relation between amplitude and phase modulation \citep[e.g.][]{benko11}. No doubt, the modulation in seven RV~Tau stars is very similar --  the modulation side-peaks are located on higher frequency side of $\fF$ and its harmonics, modulation periods are always a bit shorter than $4\PF$ and pulsation periods of the modulated stars are in a relatively narrow range of $30<\PF<35$\,d. It may indicate a specific mechanism behind the suspected modulation.

In a few other RV~Tau stars we detect hints for periodic modulation of different nature than just discussed. In the majority of cases, the hints are single, resolved peaks located near $\fF$. Two cases are worth mentioning. T2CEP-258 ($\PF=32.762$\,d) has a very simple frequency spectrum: significant peaks are detected at $\fF$ and $2\fF$, and then two peaks located near $\fF$, at $\fF+\fmj$ ($\Pmj=514$\,d) and at $\fF-\fmd$ ($\Pmd=200$\,d). No other peaks that could confirm the above modulations are detected in the spectrum. Incidentally, the peak at $\fF-\fmd$ corresponds to $5\fF/6$ and the light curve folded with $6\PF$ is telling. Interestingly, sub-harmonics are hardly present in the spectrum of this star. Seasonal analysis reveals that period doubling alternations are transient and strongly irregular. Another interesting case is T2CEP-347 ($\PF=25.5951$\,d) which shows pronounced alternations that might be modulated. In the frequency spectrum we detect possible modulation peaks in the low frequency range, at $\fm$, and at $\fjd-\fm$. Unfortunately, 1-year aliases are of similar height, consequently, we cannot resolve what is the modulation period: 679\,d or 236\,d.

\section{Discussion}\label{sec:discussion}

\subsection{Period doubling(s) in type II Cepheids}

Period doubling effect is present in non-linear pulsation models of BL~Her, W~Vir and RV~Tau stars. It is also observed in all three classes of variables. The agreement between the models and the observations is only qualitative, however.  

The existence of PD in BL~Her star was predicted by \cite{bm92} based on radiative hydrodynamic models. However, only recently the phenomenon was detected in T2CEP-279 (firm detection) and in T2CEP-257 (candidate star) based on OGLE-III data \citep{blher_ogle,blherPD}. Here we confirm both detections and add one more BL~Her star to the sample, T2CEP-820 (Sect.~\ref{ssec:pdinblher}). Only in T2CEP-279 the effect is strong and well visible in the folded light curve. In T2CEP-257 and T2CEP-820 the effect is of very low amplitude and is present in some of the observing seasons only. These three stars constitute less than 0.3 per cent of the OGLE sample which counts 372 BL~Her stars. No doubt the effect is very rare and is of very low amplitude (see Tab.~\ref{tab:blherpd}).

On the other hand, the models predict that the effect should be much more common. \cite{blherPD} considered a grid of non-linear convective models aimed at reproducing the pulsation of T2CEP-279 best, while models computed by \cite{s16} covered the full extent of the BL~Her domain. The extent of the computed PD domain depends on the parameters of the convection model used. In particular, large eddy-viscous dissipation leads to the vanishing of the PD domain. However, with parameters that reproduce the pulsation of T2CEP-279 best, the PD domain in the HR diagram is large and covers a significant part of the BL~Her domain and of the low-luminosity W~Vir domain -- see fig.~5 in \cite{s16}. The possible pulsation periods within the PD domain cover a wide range, from about 2\,d to about 6.5\,d, and predicted alternation amplitudes are large, typically much larger than that observed in T2CEP-279 -- see figs.~13 and 15 in \cite{blherPD}. This is all in contrast with the observations, in which, as pointed above, the effect is very rare and of low amplitude. The observations also indicate that the actual PD domain must be very small and, most likely, so are the ranges of physical parameters of period doubled stars. This conclusion follows from similarity of the three period doubled BL~Her stars. First, they all have very close pulsation periods (see Tab.~\ref{tab:blherpd}). In T2CEP-279 and in T2CEP-820 the period is nearly the same, $\approx$2.4\,d, and in T2CEP-257 it is slightly shorter ($\approx$2.25\,d). Second, all three stars have a very similar light curve shape -- cf. Figs~\ref{fig:pd279}, \ref{fig:257} and \ref{fig:820} --  with characteristic bump at minimum light and with wide rounded maximum with standstill on the ascending branch. Considering the whole BL~Her domain, the light curve morphology varies significantly, see, e.g. fig.~2 in \cite{blher_ogle}. Since it is the 3:2 resonance between the fundamental mode and the first overtone ($\fO/\fF=3:2$) that underlies the PD effect \citep{mb90,blherPD}, we may speculate that, e.g. the resonant mismatch, $\fO/\fF-1.5$, must be in a very narrow range or, that some other very specific conditions are needed for the PD effect to occur.

We have not detected PD effect in stars with pulsation periods in between $\sim$2.4\,d and $\sim$15\,d, although models predict the effect also in this period range. Again we refer the reader to fig.~5 in \cite{s16}. Two PD domains were identified theoretically in the models. The low luminosity one, that was just discussed and that encompasses the three period-doubled BL~Her stars, and the higher luminosity one, that starts at $\approx$500\,\LS\ ($0.6\,\MS$ models), and extends towards higher luminosities. This domain was also present in early radiative calculations of \cite{kb88}. There is no upper boundary of this PD domain; at least it was not determined with non-linear calculations, as numerical problems do not allow computation of more luminous, long-period models \citep[see sect.~2.1 in][]{s16}. This domain encompasses the RV~Tau variables, in which period doubling is common, but clearly the effect starts at too low luminosities and consequently at too short pulsation periods ($\approx10$\,d) as compared with observations. Our analysis shows however, that the disagreement is not as severe as thought before. Till recently, the transition between W~Vir stars towards period doubled RV~Tau stars was placed at a rather sharp boundary at $\PF\approx20$\,d, with a few shorter period PD stars detected only recently. Here, for the first time, we have shown (Sec.~\ref{ssec:pdinwvir}, Tab.~\ref{tab:transition}) that the transition is smooth and starts at pulsation periods just above 15\,d. In between 15\,and 20\,d, period doubled W~Vir stars are quite common.

In radiative non-linear model sequences of luminous type II Cepheids, the period doubling is only a first step in a period doubling cascade that finally leads to chaotic behaviour \citep{bk87,kb88}. Along a cascade, period-4, period-8, \ldots, period-$2k$, pulsation are present. The higher the $k$, the narrower the domains of period-$2k$ pulsation; ratios of the domains' extents are dictated by the Feigenbaum constant \citep[$\approx$4.67; see, e.g.][]{cvitanovic}. Consequently, period-4 pulsation should be more scarce than period-doubled pulsation, period-8 more scarce than period-4, etc. If this scenario is valid, one would expect to finally detect period-4 or in general period-$k$ pulsations in a numerous sample of type II Cepheids. Another possible route to chaos, that does not involve period-$k$ pulsations, is through tangent bifurcation \citep{kb88}. While irregular/chaotic pulsation is rather common at the long-period/high-luminosity RV~Tau and SRd stars (although firm confirmation of chaotic dynamics is challenging), the transition from periodic to irregular/chaotic pulsation remains elusive.

The first candidate for period-4 pulsation was reported by \cite{pollard} in MACHO observations of the LMC. For one RV~Tau star ($\PF=30.408$\,d) they show the light curve folded with $4\PF$. Unfortunately, frequency spectrum was not analysed\footnote{The star was also observed by OGLE \citep{ogle3_lmc_t2}; OGLE-LMC-T2CEP-162. Although the OGLE light curve folded with $4\PF$ may indeed suggest period-4 pulsation, the effect relies on a few dimmer data points at one of the minimums. No convincing signature of period-4 pulsation is detected in the frequency spectrum.}. Here we reported another candidate, T2CEP-135. Admittedly, we do not detect a permanent and regular period-4 pulsation. The effect is well visible during four observing seasons only (and confirmed with the analysis of the frequency spectrum). Otherwise, we observe a regular PD effect. Period-4 pulsation may be a short-lived transient. Alternatively, we may also witness another phenomenon related to chaotic dynamics, type-III intermittency, in which intermittent switching between period-$k$ and period-$2k$ occurs \citep{PM80}. But a series of such switchings must be observed and hence much longer monitoring of T2CEP-135 is needed to make any definite claims.

No doubt, in OGLE observations we find no support for period-doubling route to chaos. The observational picture is apparently more complex, or rather more obscured, than non-linear models predict. The models are essentially either periodic (including period-$k$ pulsation), or chaotic. In type II Cepheids we observe strong irregularities superimposed on singly periodic pulsation (majority of BL~Her and W~Vir stars), or on period-doubled pulsation (RV~Tau stars). The nature of irregularities remains to be investigated, but as OGLE data contain seasonal gaps and are irregularly sampled, it will be challenging to prove the chaotic nature of these irregularities.

\subsection{Modulation of pulsation in type II Cepheids}\label{ssec:disc_mod}

Pronounced amplitude and phase modulations were also detected in classical Cepheids pulsating in radial overtones, including second overtone pulsator V473~Lyr \citep[for recent analysis see][]{v473lyr}, a sizeable sample of first overtone Cepheids \citep{ccmod}, and many double-mode, first plus second overtone pulsators \citep{mk09}. Small amplitude ($\sim$mmag level) periodic modulations were recently detected also in fundamental mode classical Cepheids \citep{derekas,s17}.

Detection of modulation in type II Cepheids is particularly challenging due to non-stationary nature of these stars: irregular changes of pulsation amplitude, pulsation period and of mean brightness are commonly present, which we could verify with the time-dependent Fourier analysis. Such changes often generate wide and pronounced bands of power centred, or located close to the pulsation frequency, its harmonics and in the low frequency range. Based on such frequency spectra, tens of stars were initially identified as modulation candidates, but were false verified in the subsequent, detailed analysis. Most of these stars can be divided into two classes. In the first, the pulsation amplitude/phase/brightness vary on a time-scale of $\sim$1000\,d. As typical data length in the sample is $\sim$2400\,d, in the frequency spectra we can identify a set of (marginally) resolved peaks that in principle can be attributed to a periodic modulation with a frequency $\fm$ (e.g., peaks at $\fm$, $\fF+\fm$ and $2\fF+\fm$). Prewhitening turns inefficient however: the power is only somewhat reduced; strong remnant remains in the spectrum. Time-dependent analysis indicates that $\fm^{-1}$ corresponds to a characteristic time-scale of irregular changes, rather than to a period of cyclic variation. Among such false-verified candidates are, e.g. T2CEP-328, T2CEP-470, T2CEP-561, T2CEP-817, T2CEP-894 (W~Vir stars) T2CEP-385 (RV~Tau). In Sect.~\ref{sec:results}, we have identified only a few stars in which long-period modulation was convincing. Still, barely more than two cycles are covered and long-term-monitoring of their brightness is necessary to confirm, whether the changes we observe are indeed periodic.

In the second class of stars identified as modulation candidates, changes of pulsation amplitude/phase/brightness are faster and limited frequency resolution is not an issue. In the frequency spectrum we observe wide bands of power typically located on one side of the pulsation frequency and its harmonics. Within such power excess there are several well resolved peaks of similar height; at $\fF$ and at $2\fF$ one can easily point the peaks with the same separation, $\fmj$, with respect to $k\fF$. Then, after prewhitening, another set of peaks with slightly different separation, $\fmd$, emerges, and so on. Such procedure is obviously subjective and pointless. We do not observe a manifestation of multi-periodic modulation, but complex changes on a variety of time-scales. A few examples of such stars are, e.g. T2CEP-200, T2CEP-269, T2CEP-447, T2CEP-815 (W~Vir stars).

The described complex variations may originate from chaotic dynamics as briefly discussed in the previous Section and in the Introduction. Standard consecutive prewhitening with periodic sine terms is not adequate to study such variations in detail. Time-frequency techniques need to be employed, but it is beyond the scope of the present analysis.

Fortunately, in several stars we could convincingly identify quasi-periodic modulation. As in the case of period doubling, phenomenon is present both in pulsation models of type II Cepheids and, as reported here for the first time, in observations. It is however difficult to find a direct correspondence between the models and observations.

\cite{blher_mod} found periodic and quasi-periodic modulation of pulsation in several sequences of BL~Her models covering a small domain in the HR diagram (see their fig.~1). These models were computed with strongly decreased eddy viscosity parameter however, and are far from realistic (no red edge of the instability strip, excessive pulsation amplitudes, peculiar light variations). They are interesting from dynamical point of view as they demonstrate that the half-integer resonances between pulsation modes (specifically the $\fO/\fF=3:2$ resonance) may drive the periodic modulation of pulsation, as suggested in one of the models explaining the Blazhko effect \citep{bk11}. Modulation in these models is present on top of the pronounced period doubling effect and modulation properties (period, amplitude) are a strong function of the resonance mismatch parameter. As soon as eddy viscosity parameter is increased, the modulation of pulsation vanishes in BL~Her-type models \citep{s16}. Hence, periodic modulation of pulsation, comparable to the observed phenomenon, has not been reproduced with pulsation models yet.

Contrary to period-doubled BL~Her stars, the modulated BL~Her stars seem not confined to any particular parameter range (Tab.~\ref{tab:mod}, Fig.~\ref{fig:t2cepBL_mod_lc}). They cover a wide period range, from $\approx$1.4\,d to $\approx$3.9\,d and display a variety of light curve shapes. The modulation properties (modulation periods, amplitudes) also widely differ. Both amplitude and phase modulations are observed. Patterns in the frequency spectra are rather simple: a few modulation side-peaks are observed, mostly in the vicinity of $\fF$ and sometimes close to the harmonics. Both symmetric patterns (triplets) and asymmetric patterns (doublets) are detected. Interestingly, in the majority of stars, a peak at the modulation frequency is also observed. Very similar frequency spectrum patterns were detected in fundamental mode classical Cepheids by \cite{s17}.

Periodic modulation of pulsation was found in a few isolated W~Vir-type models computed by \cite{s16}; no large modulation domain was found. Independent of mass and metallicity, these models have roughly the same location on the HR diagram, have pulsation periods close to 9.5\,d, and have very similar modulation properties \citep[see tab.~2 of][]{s16}, which suggested a resonant origin of the modulation in the models.

The modulated W~Vir stars (Tab.~\ref{tab:mod}) are not restricted to any specific parameter range. Their periods are in the $\approx$5.2\,d to $\approx$16.4\,d range. Modulation periods are rather short; except one star, all shorter than $\approx$220\,d. For three stars: T2CEP-282 ($\PF=8.91565$\,d, $\Pm=82.24$\,d), T2CEP-187 ($\PF=9.5937$\,d, $\Pm=107.6$\,d) and T2CEP-443 ($\PF=10.6364$\,d, $\Pm=124.5$\,d) we can find a relatively close match among the models published by \cite{s16}, as both the pulsation and modulation periods are close to that seen in the models. In addition, in all theoretical models the lower frequency side-peaks are of higher amplitude. In T2CEP-282 and T2CEP-187 only the lower frequency side-peaks are detected, while in T2CEP-443 symmetric triplet centered on $\fF$ is present. We note that T2CEP-187 is a peculiar W~Vir star, consequently it may be a member of the binary system \citep[see][]{ogle4_gb_t2}. The asymmetric pattern observed in its frequency spectrum suggests however, that the observed modulation is intrinsic to the star (see next section).

\subsection{T2CEP-377: modulation due to binarity?}\label{ssec:binary}

Motion of a Cepheid in a binary system leads to periodic  phase modulation ($\Pm=P_{\rm orb}$) due to light time travel effect (LTTE). In the frequency spectrum of pulsating star, LTTE manifests as equidistant multiplets centred at pulsation frequency and its harmonics. In addition, the amplitude ratio between the side-peaks located at $k\fF\pm\fm$, and the central component of each multiplet, $k\fF$, is proportional to $k\fF$, and the phase difference between the central component and the $k\fF\pm\fm$ side-peaks of each multiplet is $\uppi/2$. Details of the frequency spectrum depend on the orbital parameters of the system \citep{shiba1, shiba2}. As demonstrated by \cite{shibaEPJ}, frequency spectrum may be used to discriminate between phase modulation due to Blazhko effect and phase modulation due to binarity in RR~Lyrae stars.

\begin{figure}
\includegraphics[width=\columnwidth]{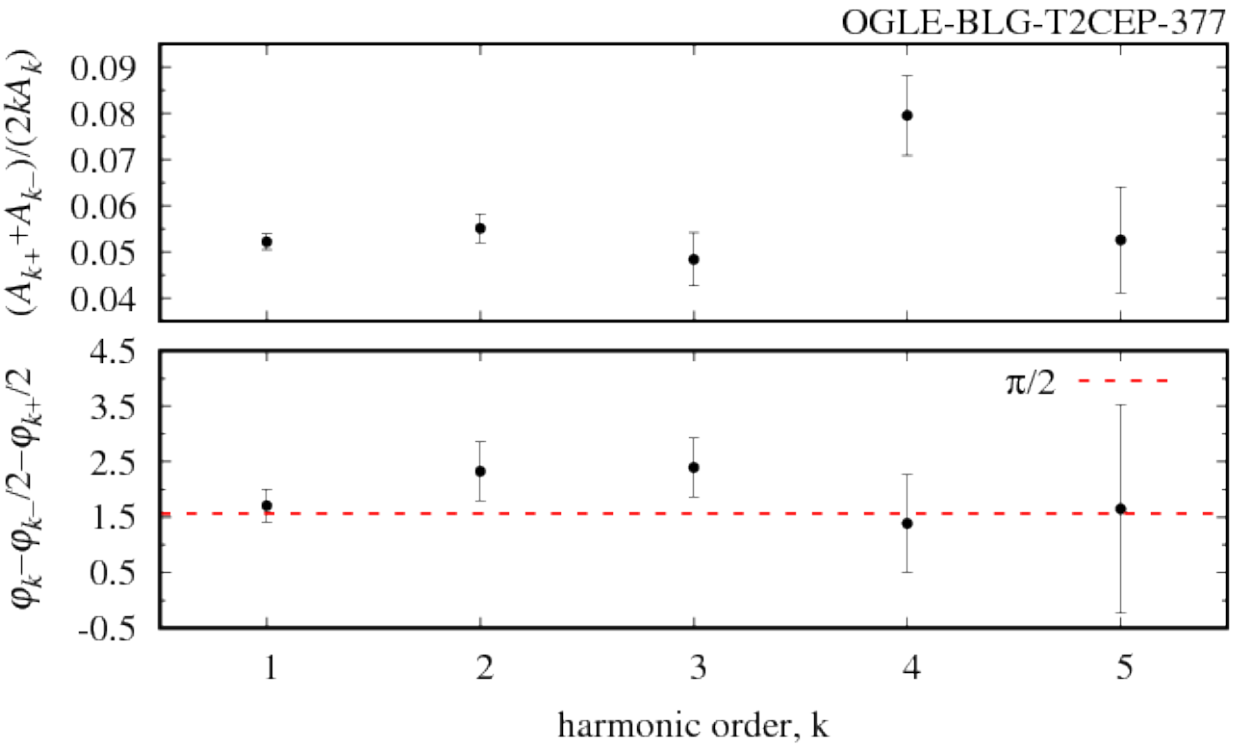}
\caption{Amplitude ratios and phase differences between the side-peaks and central components of the triplets for T2CEP-377, a candidate for BL~Her star in a binary system.}
\label{fig:shiba}
\end{figure}

In a few type II Cepheids we observe long-period phase modulations. In Fig.~\ref{fig:tdfd_logpmod} we have illustrated phase changes for three BL~Her stars. Of these, frequency spectrum of T2CEP-377 ($\PF=3.823615$\,d, $\Pm=957$\,d) indicates that LTTE may be responsible for the modulation. Although the phase changes displayed in Fig.~\ref{fig:tdfd_logpmod} are not strictly repetitive, in this star the phase modulation is superimposed on the irregular phase changes on longer time scale, that are most likely intrinsic to the star. Amplitude is, within $2\,\sigma$, constant. In the frequency spectrum we have detected equidistant and roughly symmetric triplets up to $5\fF$. In Fig.~\ref{fig:shiba} we show the relevant amplitude ratios, $(A_{k+}+A_{k-})/(2kA_k)$ (top panel) and phase differences, $\phi_k-\phi_{k-}/2-\phi_{k+}/2$ (bottom panel), where $A_{k\pm}$ and $\phi_{k\pm}$ refer to amplitudes and phases of the side-peaks located at $k\fF\pm\fm$. For this plot, all side-peaks were fitted as independent frequencies. Except at $4\fF$, the amplitude ratios (here divided by the harmonic order, $k$) are constant within 1\,$\sigma$, as expected due to LTTE. The phase differences are within 2\,$\sigma$ equal to $\uppi/2$, again supporting LTTE as origin of the modulation.

As long-term irregular phase changes are also present in T2CEP-377, spectroscopic monitoring is needed to verify the binary nature of the observed modulation.

\section{Summary}

We have analysed $I$-band photometry of 924 Galactic bulge type II Cepheids observed by the OGLE project. Seven seasons of the OGLE-IV photometry are the basic data set we have considered for each star. We have also analysed OGLE-III data for stars not observed in the fourth phase of the project, and analysed both data sets for the most interesting stars, if these were available. Our most important findings are the following.

\begin{itemize}
  
\item We identify first radial double-mode type II Cepheids of BL~Her type, T2CEP-749 ($\PF=1.041572$\,d, $\PO=0.73442$\,d, $\PO/\PF=0.7051$) and T2CEP-209 ($\PF=1.1812857$\,d, $\PO=0.832986$\,d, $\PO/\PF=0.7052$) pulsating simultaneously in the radial fundamental and radial first overtone modes. Although these are the first double-mode BL~Her stars, they are also members of the group of double-mode variables identified previously among long-period RR~Lyrae stars \citep{eRRd}. These stars are characterised by long fundamental mode periods ($\PF>0.6$\,d) as compared to majority of RRd stars, lower period ratios, low amplitudes of the radial first overtone and characteristic light curve shape corresponding to the dominant fundamental mode. This class now spans across RR~Lyrae and BL~Her classes.
  
\item We confirm period doubling in BL~Her-type stars T2CEP-279 and T2CEP-257 \citep{blherPD} and add one more star to the sample, T2CEP-850. Only in T2CEP-279 the effect is of relatively large amplitude and is stable ($\sim$coherent pulsations are followed for more than 2300 pulsation cycles now). In the other two stars the effect is of very low amplitude and is a transient phenomenon, present in some observing seasons only. In addition, in T2CEP-820 switching of the deep and shallow pulsation cycles occurs.

\item The three period doubled BL~Her stars constitute less than 0.3 per cent of the OGLE BL~Her sample, have periods in a narrow range ($\approx2.25-2.4$\,d) and very similar light curve shapes, which suggests that the period doubling domain in the HR diagram is very small. This finding is in conflict with non-linear pulsation models, which predict that the PD phenomenon should be much more common and that amplitude of the effect should be large.

\item Twenty five period-doubled W~Vir stars with pulsation periods between 15\,d and 20\,d are reported. The longer the pulsation period the higher the incidence rate of period-doubled W~Vir stars and the higher the amplitude of the effect. The transition towards period doubling behaviour is smooth and starts at pulsation periods slightly above 15\,d. Non-linear pulsation models are again in conflict with these observations, as they predict the transition at even shorter pulsation periods, just above 10\,d.

\item In many RV~Tau stars the period doubling effect is irregular. In some stars the depth of the deep and shallow maxima varies. Particularly interesting are tens of stars in which deep and shallow pulsation cycles interchange.

\item One star, T2CEP-135 ($\PF=23.2681$\,d), is a candidate for period-4 pulsation. The effect is not permanent however. It is present during four observing seasons and is revealed both in the folded light curve and in the frequency spectrum. In other seasons, standard period-doubled pulsation is observed. The effect may be a transient. The other possibility is intermittent switching between period-2 and period-4 pulsation (type-III intermittency) but much longer monitoring is needed to prove that this is the case.
  
\item For the first time, periodic modulation of pulsation was detected in all subgroups of type II Cepheids.

\item The most numerous are modulated BL~Her stars. Sixteen cases were identified with pulsation periods in the $\approx$1.4-3.9\,d range, characterized by a variety of light curve shapes. In the frequency spectra, modulation patterns are rather simple and indicate both amplitude and phase modulations. In the majority of stars modulation of the mean stellar brightness is also detected. Although some modulated non-linear BL~Her-type models were already published, they do not reproduce the observed stars.

\item In BL~Her star, T2CEP-377, pulsation amplitude is constant (within $2\,\sigma$) and observed phase modulation ($\PF=3.823615$\,d, $\Pm=957$\,d) is most likely due to light time travel effect, as deduced from the analysis of its frequency spectrum. Spectroscopic monitoring is needed to confirm that T2CEP-377 is a member of the binary system.
  
\item Periodic modulation of pulsation was detected in 9 W~Vir stars. Pulsation periods are in the $\approx\!5.2-16.4$\,d range. Except one star, modulation periods are below $\approx\!220$\,d. As compared to modulated BL~Her stars, equidistant triplets are rarely detected in the frequency spectrum. Modulation of the mean brightness is also less frequent; it was detected in three W~Vir stars.

\item In seven RV~Tau stars confined to a relatively narrow period range, $\approx30-35$\,d, similar frequency patterns were identified, namely additional peaks (beyond fundamental mode frequency, its harmonics and sub-harmonics) located at frequencies which are always slightly larger than $\fpc$, $\fdc$ and $\ftrzc$, are detected. As discussed in Sect.~\ref{ssec:rvtp4mod} the most likely interpretation is periodic modulation of pulsation with periods slightly shorter than $4\PF$. The less likely interpretations are period-4 pulsation or excitation of additional non-radial modes.

\item In the majority of type II Cepheids power excess is detected at the location of the fundamental mode after the prewhitening. Time-dependent analysis shows that irregular and complex changes of pulsation amplitude and of pulsation phase are common in these variables.
  
\end{itemize}

\section*{Acknowledgements}
This research is supported by the Polish National Science Centre, grant agreement DEC-2015/17/B/ST9/03421. EP was supported by the NKFIH PD-121203 and K-115709 grants, the J\'anos Bolyai Research Scholarship and the LP2014-17 and LP2018-7/2018 Programs of the HAS. IS is also supported by the MAESTRO grant 2016/22/A/ST9/00009. The OGLE project has received funding from the National Science Centre, Poland, grant MAESTRO 2014/14/A/ST9/00121 to AU.


\section*{SUPPORTING~~INFORMATION}

Additional Supporting Information may be found in the online version of this article:

{\bf Animations:} A collection of animations illustrating periodic modulation of pulsation in BL~Her and W~Vir stars.

{\bf Appendix A:} A collection of figures illustrating light curves and frequency spectra of RV~Tau stars with period doubling switching events.

{\bf Appendix B:} A collection of figures illustrating frequency spectra for modulated BL Her and W Vir stars.

{\bf Appendix C:} A collection of figures illustrating light curves and frequency spectra of modulated RV~Tau stars.

\bsp	
\label{lastpage}
\end{document}